\documentclass[12pt]{article}

\usepackage{frankstyle}

\title{Identifying Causal Effects in Experiments with Spillovers and Non-compliance\thanks{The views expressed in this article are those of the authors and do not necessarily reflect the position of the Federal Reserve Bank of Chicago or the Federal Reserve System.} \thanks{We thank Esther Duflo, Roland Rathelot, and Philippe Zamora for their help securing our access to the experimental data set we use in this paper. We also thank Steve Bond, Christina Goldschmidt, Luojia Hu, seminar participants at The Philadelphia Fed, the 2018 IAAE Annual Conference, UPenn, Oxford, the 2018 SEA Annual Meetings, and the 2020 Econometric Society World Congress for helpful comments and suggestions.}}

\author[1]{Francis J.\ DiTraglia\thanks{Corresponding Author: \href{mailto:francis.ditraglia@economics.ox.ax.uk}{francis.ditraglia@economics.ox.ac.uk}, Manor Road, Oxford OX1 3UQ, UK.}}
\author[2]{Camilo Garc\'{i}a-Jimeno}
\author[1]{Rossa O'Keeffe-O'Donovan}
\author[3]{Alejandro S\'{a}nchez-Becerra}
\affil[1]{\normalsize Department of Economics, University of Oxford}
\affil[2]{\normalsize Federal Reserve Bank of Chicago}
\affil[3]{\normalsize Department of Quantitative Theory and Methods, Emory University}

\date{\small First Version: September 19, 2019 \quad Final Version: December 11, 2022} 

\begin{document}

\clearpage
\maketitle
\thispagestyle{empty}

\vspace{-0.8cm}
\begin{abstract}
  \singlespacing
\vspace{-0.3cm}
This paper shows how to use a randomized saturation experimental design to identify and estimate causal effects in the presence of spillovers--one person's treatment may affect another's outcome--and one-sided non-compliance--subjects can only be offered treatment, not compelled to take it up.
Two distinct causal effects are of interest in this setting: direct effects quantify how a person's own treatment changes her outcome, while indirect effects quantify how her peers' treatments change her outcome. 
We consider the case in which spillovers occur within known groups, and take-up decisions are invariant to peers' realized offers.
In this setting we point identify the effects of treatment-on-the-treated, both direct and indirect, in a flexible random coefficients model that allows for heterogeneous treatment effects and endogenous selection into treatment.
We go on to propose a feasible estimator that is consistent and asymptotically normal as the number and size of groups increases.
We apply our estimator to data from a large-scale job placement services experiment, and find negative indirect treatment effects on the likelihood of employment for those willing to take up the program.
These negative spillovers are offset by positive direct treatment effects from own take-up.


\end{abstract}

\onehalfspacing


\section{Introduction}
Random saturation experiments provide a powerful tool for estimating causal effects in the presence of spillovers---also known as interference---by generating exogenous variation in both individuals' own treatment offers and the fraction of their peers who are offered treatment \citep{hudgens2008}.
These two sources of variation allow researchers to study both direct causal effects---the effect of Alice's treatment on her own outcome---and indirect causal effects---the effect of Bob's treatment on Alice's outcome. 
A complete understanding of both direct and indirect effects is crucial for program evaluation in settings with spillovers.
When considering a national job placement program, for example, policymakers may worry that the indirect effects of the program could completely offset the direct effects: in a slack labor market, job placement could merely change who is employed without affecting the overall employment rate \citep{crepon2013}.

In this paper we provide methods that use data from a randomized saturation design to identify and estimate direct and indirect causal effects in the presence of spillovers and one-sided non-compliance.
In real-world experiments non-compliance is the norm rather than the exception.
In their study of the French labor market, \cite{crepon2013} found that only 35\% of workers offered job placement services took them up.
Despite pervasive non-compliance in practice, most of the existing literature on randomized saturation designs either assumes perfect compliance---all subjects adhere to their experimentally-assigned treatment allocation---or identifies only intent-to-treat-effects---the effect of being \emph{offered} treatment.
Intent-to-treat effects are generally insufficient for policy analysis: comparing costs and benefits requires an estimate of the average effect of treatment on those who experience it.
For this reason, we go beyond intent-to-treat effects.
In particular, we use the randomized saturation design as a source of instrumental variables to estimate treatment-on-the-treated and treatment-on-the-untreated effects when subjects endogenously select into treatment on the basis of their experimental offers. 

In a world of homogeneous treatment effects, a simple instrumental variables (IV) regression using individual treatment offers and group saturations as instruments would identify both direct and indirect effects.
In most if not all real-world settings, however, treatment effects vary across individuals.
In the presence of heterogeneity, this ``na\"{i}ve'' IV approach will not in general recover interpretable causal effects.
To allow for realistic patterns of heterogeneity in a tractable framework, we study a flexible random coefficients model in which causal effects may depend on an individual's treatment take-up as well as that of her peers.\footnote{As we discuss further below, the random coefficients model is not itself restrictive, but the flexibility of the models that one can identify in practice is constrained by the design of the experiment. See Appendix \ref{sec:BasisFunctionsRank} for a detailed discussion of this point.}

Our approach relies on four key assumptions.
First is \emph{partial interference}: we assume that each subject belongs to a
single, known group and that spillovers occur only within groups. 
This is reasonable in many experimental settings where, for example, groups correspond to villages, and spillovers across them are negligible.
Second is \emph{anonymous interactions}: we assume that individuals' potential outcome functions depend on their peers' treatment take-up only through the \emph{average} take-up in their group.
Under this assumption only the number of treated neighbors matters, not their identities \citep{Manski2013}.
In the absence of detailed network data, the assumption of anonymous interactions is a natural starting point and is likely to be reasonable in settings such as the labor market example described above.
Third is \emph{one-sided non-compliance}: we assume that the only individuals who can take up treatment are those to whom treatment was offered via the experimental design.
One-sided non-compliance is relatively common in practice, for example when an ``encouragement design'' is used to introduce a new program, product or technology that is otherwise unavailable \citep[e.g.][]{miguel2004, crepon2013}.

We refer to our fourth key assumption as \emph{individualized offer response}, or IOR for short.
IOR requires that each subject's treatment take-up decision is invariant to the realized treatment offers made to her peers. 
While IOR is a strong assumption, it is \emph{a priori} reasonable in many contexts, for example in online settings where other subjects' treatment offers are unobserved by others \citep{bond2012,anderson2014,eckles2016}  confidential \citep{yi2015}, or observed with a delay.
IOR limits but does not rule out strategic behavior.
For example, it holds when agents act strategically on their own beliefs about others' actions provided that they are unaware of their peers' offers when making their own take-up decisions.
(\cite{bhattacharya2021demand} call this an ``incomplete information equilibrium.'')
Most importantly, IOR has testable implications and we find no evidence against it in our empirical example.\footnote{See \autoref{sec:IOR} for details.}

When combined with one-sided non-compliance, IOR allows us to divide the population into never-takers and compliers, two of the traditional LATE strata.\footnote{One-sided non-compliance rules out always-takers and defiers.}
Under the randomized saturation design and a standard exclusion restriction, we show how to construct valid and relevant instruments that identify the average causal effects of interest. 
The key to our approach is a result showing that conditioning on group size $n$ and the share of compliers $\bar{c}$ in a group breaks any dependence between peers' average take-up and an individual's random coefficients.\footnote{Our identification approach relates to a large literature on random coefficients models, e.g.\ \cite{wooldridge2004}, \cite{masten2016}, and \cite{graham2022semiparametrically}, and to the literature that uses control functions to identify structural effects \citep{altonji2005cross,imbens2009identification}.}
Under the randomized saturation design, the share of Alice's neighbors who are offered treatment is exogenous. 
Under IOR, their average take-up depends only on how many of them are compliers and whether they are offered treatment.
Thus, conditional on $n$ and $\bar{c}$, any residual variation in the take-up of Alice's neighbors comes solely from the experimental design.
Although group size is observed, the share of compliers in a given group is not.
In a large group, however, the rate of take-up among those offered treatment, call it $\widehat{c}$, closely approximates $\bar{c}$.
Using this insight, we provide feasible estimators of direct and indirect causal effects that are consistent and asymptotically normal in the limit as group size grows at an appropriate rate relative to the number of groups.
After constructing the appropriate instruments, our estimators can be implemented as simple IV regressions without the need for non-parametric estimation.
In a series of simulations we demonstrate that our estimator works well at reasonable sample sizes.

We apply our methods to experimental data from \cite{crepon2013}, a large-scale randomized saturation experiment carried out across French labor markets that offered job-placement services to young adults.
In particular, we estimate direct and indirect treatment effects of program take-up for compliers (the treated) and spillovers for never-takers (the untreated).
We find large negative indirect effects for compliers, a more vulnerable sub-population than never-takers based on their observed characteristics at baseline.
Take-up of the program by these individuals, however, shields them from the negative spillovers induced by the increased take-up of job-placement services by others in their city.
The never-taker sub-population, in contrast, is unaffected by such negative spillover effects.
Our results go beyond the intent-to-treat effects estimated by \cite{crepon2013}.
Whereas they estimate the spillovers from \emph{offering} job placement services, we estimate the labor market displacement effects of \emph{providing} them.
While we do not consider additional applications here, we believe that the methods developed in this paper can be usefully applied in a variety of other settings.\footnote{In \autoref{sec:MoreApplications} we discuss a number of recent studies with non-compliance that appear to satisfy the assumptions and data requirements of our estimators.}

This paper relates most closely to recent work by \cite{Kang2016} and \cite{Imai2018}, who also study randomized saturation experiments with social interactions under non-compliance.
\cite{Imai2018} identify a ``complier average direct effect'' (CADE), in essence a Wald estimand calculated for all groups with the same share of offers (saturation).
While it is identified under a weaker condition than IOR, the CADE is a hybrid of direct and indirect effects unless one is willing to impose IOR.
Under IOR, the CADE quantifies the effect of an individual's own treatment take-up, given that her group has been assigned a particular saturation.
In contrast, the direct effects that we recover below quantify the effect of an individual's own treatment take-up given that a certain share of her neighbors have \emph{taken up} treatment.
\cite{Kang2016} identify effects similar to those of \cite{Imai2018} using an assumption they call ``personalized encouragement,'' the equivalent of our IOR assumption.
Both \cite{Kang2016} and \cite{Imai2018} identify well-defined effects while placing limited structure on the potential outcome functions.
The cost of this generality is that the effects they recover have a ``reduced form'' flavor, and are only defined relative to the specific saturations used in the experiment.
While our assumption of anonymous interactions places more restrictions on the potential outcome functions, we recover ``fully structural'' causal effects that are not specific to the design of the experiment.

In a recently and closely related paper, \cite{VazquezBare} uses instrumental variables to identify spillovers without relying on a particular experimental design. 
\cite{VazquezBare} focuses on settings with pairs of people, for example roommates or couples, and considers spillovers both in outcomes and take-up.
Under one-sided non-compliance and a novel monotonicity restriction, he identifies two causal effects without invoking the IOR assumption: a direct effect for compliers whose partner is untreated, and an indirect effect for untreated individuals whose partner is a complier.
This identification result does not extend to groups of more than two people.
In larger groups, \cite{VazquezBare} identifies average potential outcomes under anonymous interactions without IOR, instead assuming that individuals' potential outcomes are independent of their peers' compliance types.
While our results rely on IOR, we do not invoke his latter assumption because in many applied settings a person's potential outcomes may be related to the characteristics of her peers. 

Our paper also relates to the applied literature that estimates spillover effects, including ``partial population'' studies in which a subset of subjects in the treatment group are left untreated and their outcomes are compared to those of subjects in a control group \citep{duflo2003,bobonis2009,angelucci2009,barrera2011,haushofer2016}.
It also includes cluster-randomized trials where groups are defined by a spatial radius within which spillovers may arise \citep{miguel2004,bobba2014}.
In general, the applied literature focused on spillovers estimates intent-to-treat (ITT) effects. Two notable exceptions are \cite{crepon2013} and \cite{Akram2018} who estimate effects that are similar in spirit to the CADE of \cite{Imai2018}.

The remainder of the paper is organized as follows.
Section \ref{sec:notationAssumptions} details our notation and assumptions, \autoref{sec:identification} presents our identification results, and \autoref{sec:estimationinference} provides consistent and asymptotically normal estimators of the effects identified in \autoref{sec:identification}.
In \autoref{sec:application} we implement our estimator on data from a well-known labor market experiment, and discuss our findings.
In \autoref{sec:simulations} we present a brief simulation study illustrating the behavior of our estimator.
Section \ref{sec:conclusion} concludes.
Proofs and additional results appear in the appendix.

\section{Notation and Assumptions}
\label{sec:notationAssumptions}

We observe $N$ individuals divided between $G$ groups.
We assume throughout the paper that each group has at least two members so there is scope for spillovers.
Let $g =1, \dots, G$ index groups and $i = 1, \dots, N_g$ index individuals within a given group $g$.
Using this notation, $N = \sum_g N_g$.
For each individual $(i,g)$ we observe a binary treatment offer $Z_{ig}$, an indicator of treatment take-up $D_{ig}$, and an outcome $Y_{ig}$.
For each group $g$ we observe a saturation $S_g \in [0,1]$ that determines the fraction of individuals offered treatment in that group.
A bold letter indicates a vector and a $g$-subscript shows that this vector is restricted to members of a particular group.
For example $\boldsymbol{Z}$ is the $N$-vector of all treatment offers $Z_{ig}$ while $\boldsymbol{Z}_g$ is the $N_g$-vector obtained by restricting $\boldsymbol{Z}$ to group $g$.
Define $\boldsymbol{D}$ and $\boldsymbol{D}_g$ analogously and let $\boldsymbol{S}$ denote the $G$-vector of all $S_g$.
At various points in our discussion we will need to refer to the average value of a variable for everyone in a group \emph{besides} person $(i,g)$.
As shorthand, we refer to these other individuals as person $(i,g)$'s \emph{neighbors}.
To indicate such an average, we use a bar along with an $(i,g)$ subscript.
For instance, $\bar{D}_{ig}$ denotes the treatment take-up rate in group $g$ excluding $(i,g)$, while $\bar{Z}_{ig}$ is the analogous treatment offer rate:
\begin{equation}
  \bar{D}_{ig} \equiv \frac{1}{N_g - 1} \sum_{j\neq i} D_{jg}, \quad
  \bar{Z}_{ig} \equiv \frac{1}{N_g - 1} \sum_{j\neq i} Z_{jg}.
  \label{eq:barDef}
\end{equation}
Under this definition, $\bar{D}_{ig}$ and $\bar{Z}_{ig}$ vary across individuals in the same group depending on their values of $D_{ig}$ or $Z_{ig}$.
For example in a group of eleven people, of whom five take up treatment, $\bar{D}_{ig}=0.5$ if $D_{ig} = 0$ and $0.4$ if $D_{ig} = 1$.
We now introduce our basic assumptions, beginning with the experimental design.

\begin{figure}[t]
  \centering
\tikzset{
    man0/.pic={    
      \begin{scope}[black!60]
\node[transform shape, circle,fill,minimum size=4.5mm] (head) at (0,0) {};
\node[transform shape, draw, fill, trapezium, trapezium angle=90, trapezium stretches=true, rounded corners=2pt, minimum width=0.4cm, minimum height=0.8cm, text = white,below = 0pt of head, inner sep=1pt] (body) {0};
\draw[line width=1.25mm,round cap-round cap, rounded corners] ([yshift=-.5\pgflinewidth]body.north) --++(2.5mm,0)--++(-75:6mm);
\draw[line width=1.25mm,round cap-round cap, rounded corners] ([yshift=-.5\pgflinewidth]body.north) --++(-2.5mm,0)--++(255:6mm);
\end{scope}
}}

\tikzset{
    man1/.pic={    
      \begin{scope}[red!80]
\node[transform shape, circle,fill,minimum size=4.5mm] (head) at (0,0) {};
\node[transform shape, draw, fill, trapezium, trapezium angle=90, trapezium stretches=true, rounded corners=2pt, minimum width=0.4cm, minimum height=0.8cm, text = white,below = 0pt of head, inner sep=1pt] (body) {1};
\draw[line width=1.25mm,round cap-round cap, rounded corners] ([yshift=-.5\pgflinewidth]body.north) --++(2.5mm,0)--++(-75:6mm);
\draw[line width=1.25mm,round cap-round cap, rounded corners] ([yshift=-.5\pgflinewidth]body.north) --++(-2.5mm,0)--++(255:6mm);
\end{scope}
}}

\tikzset{
  urn/.pic={
\draw[line join=round,line cap=round,ultra thick](0,1)--++(0,-.5) arc[start angle=180, end angle=360, radius=0.5]--++(0,0.5);
}}

\tikzset{
  balls/.pic={
  \begin{scope}[radius= 0.1, fill=blue!50, thick]
    \draw[fill] (0.26,0.58) circle; 
    \draw[fill] (0.5,0.58) circle; 
    \draw[fill] (0.74,0.58) circle; 
    \draw[fill] (0.39,0.38) circle; 
    \draw[fill] (0.61,0.38) circle; 
    \draw[fill] (0.15,0.38) circle; 
    \draw[fill] (0.85,0.38) circle; 
    \draw[fill] (0.29,0.19) circle; 
    \draw[fill] (0.71,0.19) circle; 
    \draw[fill] (0.5,0.13) circle; 
  \end{scope}
}}

\tikzset{
  group25/.pic={
  \draw[radius=2.5, thick, fill=blue!50, fill opacity=0.25](0,0) circle ;
  \pic at (-0.5,1.9) {man0};
  \pic at (0.5,1.9) {man0};
  \pic at (-1.5,0.4) {man1};
  \pic at (-0.5,0.4) {man0};
  \pic at (0.5,0.4) {man0};
  \pic at (1.5,0.4) {man0};
  \pic at (-0.5,-1.1) {man0};
  \pic at (0.5,-1.1) {man1};
  }

}

\begin{tikzpicture}

  \pic[scale = 2] at (0,2) {urn};
  \pic[scale = 2] at (3,2) {urn};
  \pic[scale = 2] at (6,2) {urn};
  \pic[scale = 2] at (9,2) {urn};
  \pic[scale = 2] at (12,2) {urn};

  \pic[scale = 2,fill opacity = 0] at (0,2) {balls};
  \pic[scale = 2,fill opacity = 0.25] at (3,2) {balls};
  \pic[scale = 2, fill opacity = 0.5] at (6,2) {balls};
  \pic[scale = 2, fill opacity = 0.75] at (9,2) {balls};
  \pic[scale = 2, fill opacity = 1] at (12,2) {balls};

  \node at (1,1.5) {$0\%$};
  \node at (4,1.5) {$25\%$};
  \node at (7,1.5) {$50\%$};
  \node at (10,1.5) {$75\%$};
  \node at (13,1.5) {$100\%$};

  \draw[->,very thick] (4.75,2.15) to [bend left = 15] (6.3,-0.5);
  \pic at (6.8,-3.1) {group25};

\end{tikzpicture}
\caption{Randomized Saturation Design. In the first stage groups (balls) are randomly assigned to saturations (urns). In the second stage, individuals within a group are randomly assigned treatment offers at the saturation selected in the first stage. The figure zooms in on a group of size eight that has been assigned to a 25\% saturation: two individuals are offered treatment.}
\label{fig:randsat}
\end{figure}
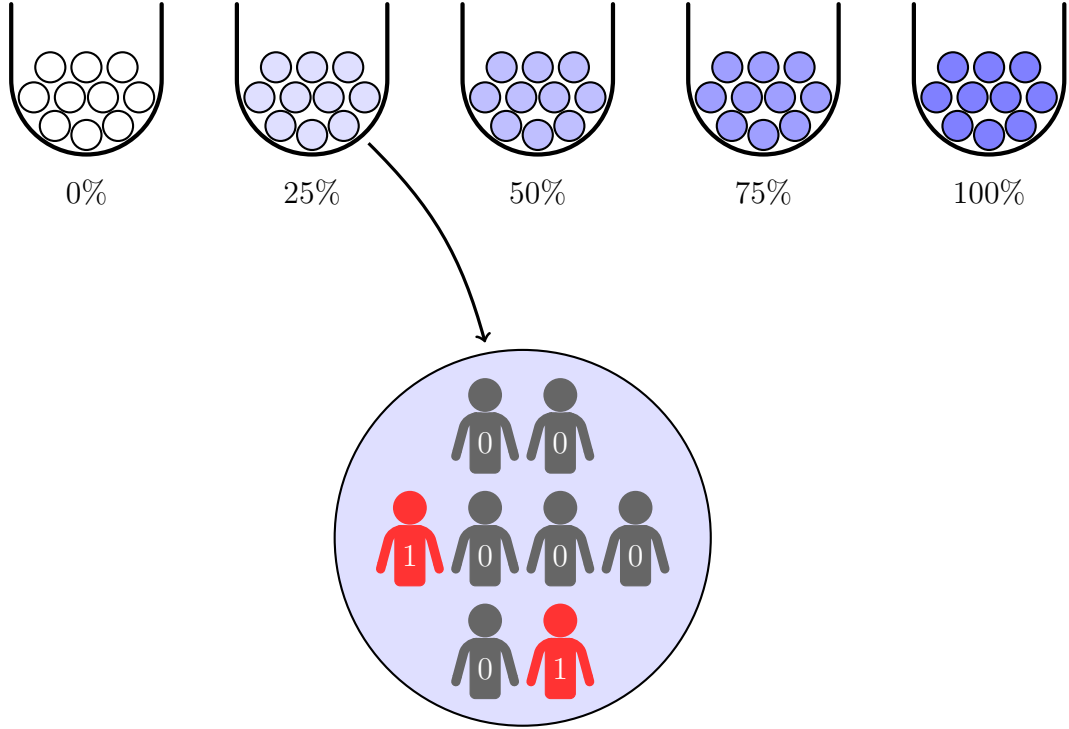

\begin{assump}[Assignment of Saturations]
  \label{assump:saturations}
  Let $\mathcal{S} = \{s_1, s_2, \dots, s_J\}$ where $s_j \in [0,1]$ for all $j$.
  Saturations are assigned to groups completely at random from $\mathcal{S}$ such that a fixed number $m_j$ of groups are assigned to saturation $s_j$, and $\sum_{j=1}^J m_j = G$.
  In other words,
  \[
    \mathbbm{P}(S_g = s_j) = \left\{\begin{array}{ll} m_j/G & \text{for } j = 1, \dots, J\\
        0 & \text{otherwise}
    \end{array}\right.
  \]
\end{assump}

\autoref{assump:saturations} details the first stage of the randomized saturation design.
In this stage, each group $g$ is assigned a saturation $S_g$ drawn completely at random from a set $\mathcal{S}$. 
In the example from \autoref{fig:randsat}, fifty groups (balls) are divided equally between five saturations (urns), namely $\mathcal{S} = \{0, 0.25, 0.5, 0.75, 1\}$.
The saturation drawn in this first stage determines the fraction of individuals in the group that will be offered treatment in the second stage.
\autoref{fig:randsat}, for example, depicts a group of eight individuals that has been assigned to the 25\% saturation: two are offered treatment and six are not.
For simplicity we assume that treatment offers in the second stage follow a \emph{Bernoulli design}, in which $S_g$ determines the probability of treatment rather than the number of treatment offers.
With minor modifications, our results can be extended to a completely randomized design, in which the number of treatment offers made to a given group is fixed conditional on the saturation.\footnote{For details see \autoref{sec:CompletelyRandomized}.}

\begin{assump}[Bernoulli Offers]
  \[
  \mathbbm{P}(\boldsymbol{Z}_g = \boldsymbol{z}|S_g = s, N_g = n) = \prod_{i=1}^n s^{z_i}(1 - s)^{1-z_i}.
  \]
  \label{assump:Bernoulli}
\end{assump}
\vspace{-0.8cm}
The randomized saturation design creates exogenous variation at the individual and group levels.
Within a group some individuals are offered while others are not.
Between groups, some have a large number of individuals offered treatment---a high saturation---while others do not.
Many randomized saturation experiments, like the illustration in \autoref{fig:randsat}, feature a 0\% saturation or even a 100\% saturation.
We refer to 0\% and 100\% saturations as \emph{corner saturations} to distinguish them from all other saturations, which we call \emph{interior}.
There is no variation in treatment offers between individuals in a group assigned a corner saturation.
For this reason, as we discuss in \autoref{subsec:identification_example} below, the number of interior saturations in the design will determine the flexibility with which we can model potential outcome functions.

Assumptions \ref{assump:saturations}--\ref{assump:Bernoulli} concern the design of the experiment.
Our remaining assumptions, in contrast, concern the potential outcome and treatment functions.
Without imposing any restrictions, an individual's potential outcome function $Y_{ig}(\cdot)$ could in principle depend on the treatment take-up of all individuals in the sample.
We denote this unrestricted potential outcome function by $Y_{ig}(\boldsymbol{D})$. 
\autoref{assump:randcoef} restricts $Y_{ig}(\cdot)$ to depend only on $D_{ig}$ and $\bar{D}_{ig}$ via a random coefficients model.

\begin{assump}[Random Coefficients Model]
  Let $\mathbf{f}(\cdot)$ be a $K$-vector of known functions $f_k\colon [0,1]\mapsto \mathbbm{R}$, each of which satisfies $\sup_{x\in [0,1]}|f_k(x)|<\infty$.
We assume that
\[
Y_{ig}(\boldsymbol{D})=Y_{ig}(\boldsymbol{D}_g) = Y_{ig}(D_{ig}, \bar{D}_{ig}) = \mathbf{f}(\bar{D}_{ig})' \left[(1 - D_{ig})\boldsymbol{\theta}_{ig} + D_{ig} \boldsymbol{\psi}_{ig}\right]
\]
where $\boldsymbol{\theta}_{ig}$ and $\boldsymbol{\psi}_{ig}$ are $K$-dimensional random vectors that may be dependent on $(D_{ig},\bar{D}_{ig})$.
\label{assump:randcoef}
\end{assump}

The first equality in \autoref{assump:randcoef} is the so-called \emph{partial interference} assumption, used widely in the literature on spillover effects.
This assumption states that there are no spillovers between people in different groups: only the treatment take-up of individuals in group $g$ affects the potential outcome of person $(i,g)$. 
The second equality in \autoref{assump:randcoef} states that person $(i,g)$'s potential outcome is only affected by the treatment take-up the others in her group through the \emph{aggregate} $\bar{D}_{ig}$.\footnote{Recall that $\bar{D}_{ig}$ is defined to exclude person $(i,g)$.}
This is effectively identical to the \emph{anonymous interactions} assumption from the network literature  \citep{Manski2013}.\footnote{In particular, because our treatment is binary, assuming that $\boldsymbol{D}_g$ only affects $Y_{ig}$ through $\bar{D}_{ig}$ is equivalent to assuming \emph{exchangeability}: only the number of $(i,g)$'s neighbors who take up treatment matters for her outcome; their identities are irrelevant. 
When researchers do not observe the social network within groups, as in \cite{crepon2013}, exchangeability is a natural assumption.} 
While we maintain this assumption throughout, \autoref{sec:conclusion} discusses some potential ways of relaxing it.

The third equality in \autoref{assump:randcoef} posits a finite basis function expansion for the potential outcome functions $Y_{ig}(0,\bar{D}_{ig})$ and $Y_{ig}(1,\bar{D}_{ig})$, namely
\[
  Y_{ig}(0, \bar{D}_{ig}) = \sum_{k=1}^K \theta_{ig}^{(k)} f_k(\bar{D}_{ig}), \quad 
  Y_{ig}(1, \bar{D}_{ig}) = \sum_{k=1}^K \psi_{ig}^{(k)} f_k(\bar{D}_{ig}) \quad 
\]
or, written more compactly in matrix form,
\begin{equation}
  Y_{ig} = \mathbf{X}_{ig}' \mathbf{B}_{ig}, \quad \mathbf{X}_{ig} \equiv \begin{bmatrix} 1 \\ D_{ig} \end{bmatrix} \otimes \mathbf{f}(\bar{D}_{ig}), \quad
  \mathbf{B}_{ig} \equiv \begin{bmatrix} \boldsymbol{\theta}_{ig} \\ \boldsymbol{\psi}_{ig} - \boldsymbol{\theta}_{ig} \end{bmatrix}
  \label{eq:XBdef}
\end{equation}
where the coefficient vectors $\boldsymbol{\theta}_{ig}$ and $\boldsymbol{\psi}_{ig}$, and hence $\mathbf{B}_{ig}$, are allowed to vary arbitrarily across groups and individuals. 
If, for example, person $(i,g)$ has some prior knowledge of her potential outcome function $Y_{ig}(\cdot,\cdot)$, her take-up decision may depend on $\boldsymbol{\theta}_{ig}$ and $\boldsymbol{\psi}_{ig}$.
More generally, the same unobserved characteristics that determine a person's decision to take up treatment could affect her potential outcomes. 
To account for these possibilities, we allow arbitrary statistical dependence between $(D_{ig},\bar{D}_{ig})$ and $\mathbf{B}_{ig}$. 
Our assumption of a random coefficients model is not in itself restrictive.
In principle one could even consider adapting the choice of $K$ to the data at hand using non-parametric series methods.
In practice, however, the design of the randomized saturation experiment limits the number of basis functions that can be used in practice.
To satisfy the rank condition introduced below, $K$ should not exceed the number of saturations.\footnote{See Appendix \ref{sec:BasisFunctionsRank} for details.}
For this reason, we treat $K$ as fixed throughout. 

Ideally, our goal would be to identify the average direct and indirect causal effects of the binary treatment $D_{ig}$.
Under \autoref{assump:randcoef}, we define these as follows, building on the definitions of \cite{hudgens2008}.
The direct treatment effect, $\text{DE}$, gives the average effect of exogenously changing an individual's own treatment $D_{ig}$ from 0 to 1 while holding the share of her treated neighbors $\bar{D}_{ig}$ fixed at $\bar{d}$, namely
\begin{equation}
  \text{DE}(\bar{d}) \equiv \mathbbm{E}\left[ Y_{ig}(1,\bar{d}) - Y_{ig}(0,\bar{d}) \right] = \mathbf{f}(\bar{d})'\mathbbm{E}\left[ \boldsymbol{\psi}_{ig} - \boldsymbol{\theta}_{ig} \right]
  \label{eq:DE}
\end{equation}
where the expectations are taken over all individuals in the population from which our experimental subjects were drawn.
Recall that $\bar{D}_{ig}$ excludes person $(i,g)$, ensuring that $\text{DE}(\bar{d})$ is well-defined.
An indirect treatment effect, in contrast, gives the average effect of exogenously increasing a person's share of treated neighbors $\bar{D}_{ig}$ from $\bar{d}$ to $\bar{d} + \Delta$ while holding her own treatment $D_{ig}$ fixed at $d$, in other words,
\begin{equation}
  \begin{split}
  \text{IE}_d(\bar{d},\Delta) &\equiv \mathbbm{E}\left[ Y_{ig}(d,\bar{d} + \Delta) - Y_{ig}(d,\bar{d}) \right]\\
  &= \left[ \mathbf{f}(\bar{d} + \Delta) - \mathbf{f}(\bar{d}) \right]' \left\{ (1-d)\mathbbm{E}\left[ \boldsymbol{\theta}_{ig} \right] + d \mathbbm{E}\left[\boldsymbol{\psi}_{ig} \right]\right\}
\end{split}
  \label{eq:IE}
\end{equation}
where $\Delta$ is a positive increment. 
There are two indirect treatment effect functions, $\text{IE}_0$ and $\text{IE}_1$, corresponding to the two possible values at which we could hold $D_{ig}$ fixed: a spillover on the untreated, and a spillover on the treated.
Because the direct and indirect causal effects are fully determined by $\mathbbm{E}[\mathbf{B}_{ig}]$ under \autoref{assump:randcoef}, this is our object of interest below. 
For example, if $\mathbf{f}(x)' = \left[1 \quad  x\right]$ we obtain a linear model of the form 
\begin{equation}
  Y_{ig} = \alpha_{ig} + \beta_{ig} D_{ig} + \gamma_{ig} \bar{D}_{ig} + \delta_{ig} D_{ig} \bar{D}_{ig}.
  \label{eq:LinearModel}
\end{equation}
In this case the direct effect is $\text{DE}(\bar{d}) = \mathbbm{E}[\beta_{ig}] + \mathbbm{E}[\delta_{ig}] \bar{d}$ while the indirect effects are
\[
  \text{IE}_0(\bar{d},\Delta) = \Delta \times \mathbbm{E}[\gamma_{ig}], \quad
  \text{IE}_1(\bar{d},\Delta) = \Delta \times \mathbbm{E}[\gamma_{ig} + \delta_{ig}].
\]
Notice that in the linear model, $\text{IE}_0$ and $\text{IE}_1$ do not depend on $\bar{d}$.
While all of our theoretical results apply to arbitrary random coefficients models, we focus on the linear model from \eqref{eq:LinearModel} in our empirical example and simulation study below.

\begin{figure}[t]
    \centering
    \begin{tikzpicture}[scale = 2.7]
      \draw [thick, <->] (0,2) node [above] {$Y_{ig}$} -- (0,0) -- (3,0) node [below]{$\bar{D}_{ig}$}; 
      \draw [thick] (0,1.5) node [left] {$\alpha_{ig} + \beta_{ig}$} -- (2.3,1.35) node [above right] {$Y_{ig}(1,\bar{D}_{ig})$};
      \draw [dashed] (1,1.43) -- (1,1.37) node [below] {$\gamma_{ig} + \delta_{ig}$} -- (1.75, 1.37);
      \draw [thick] (0,1) node [left] {$\alpha_{ig}$} -- (2.3,0.1) node [above right] {$Y_{ig}(0,\bar{D}_{ig})$};
      \node [below] at (0,0) {0};
      \draw [dashed] (1,0.6) -- (1,0.3) node [below] {$\gamma_{ig}$} -- (1.75, 0.3);
    \end{tikzpicture}
    \caption{A hypothetical example of the linear potential outcomes model from \eqref{eq:LinearModel}. The slope of the bottom line, $\gamma_{ig}$, is the indirect marginal effect when untreated while that of the top line, $\gamma_{ig} + \delta_{ig}$, is the marginal indirect effect when treated. The distance between the two lines is the direct treatment effect.}
    \label{fig:linear}
  \end{figure}
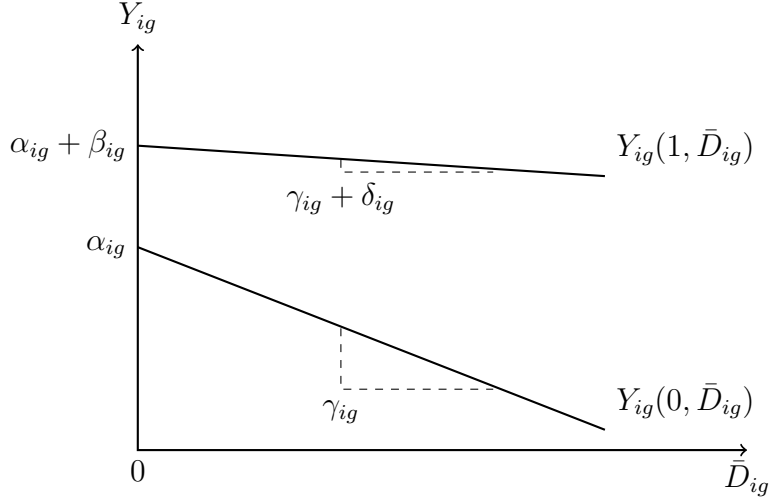

\autoref{fig:linear} presents a hypothetical example of \eqref{eq:LinearModel} in a setting with employment displacement effects.
Suppose that $Y_{ig}$ is Alice's probability of long-term employment.
Both $Y_{ig}(1, \bar{d})$ and $Y_{ig}(0, \bar{d})$ have a negative slope.
This means that Alice's probability of long-term employment \emph{decreases} if more of her neighbors obtain job placement services.
But since $\delta_{ig}$ is positive, the spillover is more harmful if Alice is untreated.
Alice's direct effect of treatment $Y_{ig}(1, \bar{d}) - Y_{ig}(0, \bar{d})$ is positive for all $\bar{d}$ in this example and increases as $\bar{d}$ does: job placement services are more valuable to Alice when more of her neighbors obtain them.
By averaging these effects for everyone in the population, we obtain $\text{IE}_0, \text{IE}_1$, and $\text{DE}$.

Under perfect compliance $D_{ig}$ would simply equal $Z_{ig}$, making both $D_{ig}$ and $\bar{D}_{ig}$ exogenous.
In this case a sample analogue of $\mathbbm{E}[Y_{ig}(d,\bar{d})]$ could be used to recover all of the treatment effects discussed above, at least at values of $\bar{d}$ that arise in the experimental design.
Unfortunately non-compliance is pervasive in real-world experiments, greatly complicating the identification of causal effects.
In a large-scale experiment carried out in France, for example, only 35\% of unemployed workers offered job placement services took them up \citep{crepon2013}.
Those who did take up treatment likely differ in myriad ways from those who did not: they may, for example, be more conscientious.
One way to to avoid this problem of self-selection is to carry out an intent-to-treat (ITT) analysis, conditioning on $Z_{ig}$ and $S_g$ rather than $D_{ig}$ and $\bar{D}_{ig}$.
But with take-up rates as low as 35\%, ITT estimates could be very far from the causal effects of interest.
In this paper we adopt a different approach.
Following the tradition in the local average treatment effect (LATE) literature, we provide conditions under which direct and indirect causal effects--rather than ITT effects--can be identified for well-defined sub-populations of individuals.
We focus on the case of \emph{one-sided noncompliance}, in which only those offered treatment can take it up.
One-sided non-compliance is common in practice and simplifies the analysis.\footnote{We suggest an avenue for extending our results to two-sided non-compliance in \autoref{sec:conclusion}.}

\begin{assump}[One-sided Non-compliance]
  If $Z_{ig} = 0$ then $D_{ig} = 0$.
  \label{assump:onesided}
\end{assump}

To account for endogenous treatment take-up, we define potential treatment functions $D_{ig}(\cdot)$. 
In principle these could depend on the treatment offers of every individual, $\boldsymbol{Z}$ in the experiment.
The following assumption restricts $D_{ig}(\cdot)$ to permit identification of the direct and indirect causal effects described above.

\begin{assump}[IOR]
  \label{assump:IOR}
  $D_{ig}(\boldsymbol{Z})= D_{ig}(\boldsymbol{Z}_g)= D_{ig}(Z_{ig},\bar{Z}_{ig}) = D_{ig}(Z_{ig})$.
\end{assump}

The first equality of \autoref{assump:IOR} is a partial interference assumption: it requires that person $(i,g)$'s take-up decision is invariant to the realized treatment offers made to people in \emph{different groups}.  
The second equality of \autoref{assump:IOR} states that person $(i,g)$'s take-up decision depends on the realized treatment offers of others in her group only through the fraction $\bar{Z}_{ig}$ of treatment offers made to the others in her group.
These first two equalities are not in general sufficient to point identify direct and indirect causal effects.
The third equality, which we call \emph{individualistic offer response} or IOR for short, imposes the further restriction that each person's take-up decision is invariant to the realized offers made to her peers.
Assumptions analogous or equivalent to IOR have appeared in the existing literature.
\cite{Kang2016}, for example, employ an assumption equivalent to IOR, which they call ``personalized encouragement.''
While \cite{Imai2018} derive their  ``complier average direct effect (CADE)'' under a weaker condition, this effect is a hybrid of direct and indirect effects unless one is willing to impose IOR.

IOR is a reasonable assumption in some but not all applications.
In settings where participants observe neither the saturation assigned to their group nor the treatment offers made to their neighbors, for example, IOR clearly holds.
IOR restricts but does not rule out strategic take-up.
For example, it also holds when agents act strategically on their beliefs about others' actions, provided that they are unaware of their peers' offers when making their own take-up decisions. 
In the introduction we list a number of recent randomized saturation experiments in which we consider IOR to be a reasonable assumption.
Moreover, as we discuss further in \autoref{sec:IOR}, IOR has testable implications in a randomized saturation experiment.
If the take-up rate among individuals who are offered treatment varies with saturation, this indicates a violation of IOR.

Under IOR and one-sided non-compliance (Assumptions \ref{assump:onesided} and \ref{assump:IOR}), we can divide individuals into never-takers and compliers, two of the principal strata from the LATE literature.
Never-takers are defined as those for whom $D_{ig}(0)=D_{ig}(1) = 0$, while
compliers are those for whom $D_{ig}(z) = z$ for all $z$.\footnote{Under
  one-sided non-compliance, \autoref{assump:onesided}, there are no
always-takers.} 
Defining $C_{ig}$ to be the indicator that person $(i,g)$ is a complier, Assumptions \ref{assump:onesided}--\ref{assump:IOR} imply that
\begin{equation}
  D_{ig} = C_{ig} Z_{ig}, 
  \quad \bar{D}_{ig} = \frac{1}{N_g - 1} \sum_{j\neq i} C_{jg} Z_{jg}. 
  \label{eq:IORimplication}
\end{equation}
By analogy to $\bar{Z}_{ig}$ and $\bar{D}_{ig}$, we define $\bar{C}_{ig}$ to be the share of compliers among person $(i,g)$'s neighbors in group $g$, namely 
\begin{equation}
\bar{C}_{ig} = \frac{1}{N_g - 1} \sum_{j\neq i} C_{jg}.
\label{eq:Cbar}
\end{equation}
Note that $\bar{C}_{ig}$ varies across individuals in the same group, depending on their values of $C_{ig}$.
Finally, let $\boldsymbol{C}_g$ denote the vector of $C_{ig}$ for all individuals in group $g$.

Our final assumption is an exclusion restriction for the treatment offers $\boldsymbol{Z}_g$ and saturation $S_g$.
To state it we require two additional pieces of notation.
First, let $\mathbf{B}_g$ denote the vector that stacks $\mathbf{B}_{ig}$ for all individuals in group $g$.
Second, following \cite{Dawid1979}, let ``$\indep$'' denote (conditional) independence so that $X \indep Y$ indicates that $X$ is statistically independent of $Y$ while $X \indep Y|Z$ indicates that $X$ is \emph{conditionally} independent of $Y$ given $Z$.
Using this notation, the exclusion restriction is as follows.


\begin{assump}[Exclusion Restriction]
  \label{assump:exclusion}
  \mbox{}
  \begin{enumerate}[(i)]
    \item $S_g \indep (\boldsymbol{C}_g, \mathbf{B}_g, N_g)$
    \item $\boldsymbol{Z}_g \indep (\boldsymbol{C}_g,\mathbf{B}_g)|(S_g, N_g)$ 
  \end{enumerate}
\end{assump}

Intuitively, \autoref{assump:exclusion} states that $(\boldsymbol{C}_g, \mathbf{B}_g, N_g)$ are ``predetermined'' with respect to the treatment offers and saturations. 
In a traditional LATE setting, its counterparts are the ``unconfounded type'' assumption and the independence of potential outcomes and treatment offers.
\autoref{assump:exclusion} could be violated in a number of ways.
If, for example, individuals chose their group membership based on knowledge of their group's saturation, $N_g$ would not be independent of $S_g$.
Similarly, if some individuals decided to comply with their treatment offers only because their group was assigned a high saturation, $\boldsymbol{C}_g$ would not be independent of $S_g$.
This latter possibility illustrates that \autoref{assump:exclusion} partially embeds IOR by ruling out ``selection into compliance.'' 
As discussed in \autoref{sec:IOR}, it also yields testable implications of the IOR assumption.
More prosaically, \autoref{assump:exclusion} would be violated if either $S_g$ or $Z_{ig}$ had a direct effect on the random coefficients $\mathbf{B}_g$.
Notice that part (ii) of \autoref{assump:exclusion} conditions on $(S_g, N_g)$.
This is because the second stage of the randomized saturation experiment assigns $\boldsymbol{Z}_g$ conditional on this information: see \autoref{assump:Bernoulli}.

\section{Identification}
\label{sec:identification}

\subsection{Conditioning on the Share of Compliers}
\label{subsec:compliers}

Under \autoref{assump:randcoef}, the functional form of the random coefficients model is known.
So why not simply use $(Z_{ig},S_g)$ as instrumental variables for $D_{ig}$ and $\mathbf{f}(\bar{D}_{ig})$?
As shown in a number of papers from the literature on random coefficients models
\citep{Wooldridge1997,heckman1998,Wooldridge2003,Wooldridge2016}, two-stage least squares identifies average effects when the causal effect of the instruments on the endogenous regressors is homogeneous. 
In our setting, however, this result does not apply because the conditional distribution of $\bar{D}_{ig}$ given $S_g$ varies with $(\bar{C}_{ig},N_g)$, as the following lemma shows.

\begin{lem}
  \label{lem:DbarDist}
  Let $\bar{c}$ be a value in $[0,1]$ such that $(n-1)\bar{c}$ is a non-negative integer.  
  Under Assumptions \ref{assump:saturations}--\ref{assump:Bernoulli} and \ref{assump:onesided}--\ref{assump:exclusion} and conditional on $(N_g = n, S_g = s, \boldsymbol{C}_g = \boldsymbol{c}, \bar{C}_{ig} = \bar{c}, Z_{ig} = z)$, $(n-1)\bar{D}_{ig}$ follows a $\text{Binomial}\left( (n-1)\bar{c}, s \right)$ distribution.
\end{lem}

Intuitively, the problem presented by \autoref{lem:DbarDist} is as follows.
Although $S_g$ is randomly assigned, the variation that it induces in $\bar{D}_{ig}$ is mediated by the share of compliers $\bar{C}_{ig}$.
Accordingly if $\bar{C}_{ig}$---a source of first-stage heterogeneity---is correlated with the random coefficients in the second stage, the IV estimator will not identify the effects of interest.
To make this problem more concrete, consider the linear potential outcomes model from \eqref{eq:LinearModel} and let $\boldsymbol{\vartheta}_{\text{IV}}$ be the IV estimand using instruments $(1, Z_{ig}, S_g, Z_{ig} S_g)$.
Throughout, we will refer to it as the ``na\"ive IV".
In this example $\boldsymbol{\vartheta}_{\text{IV}}$ takes a particularly simple form, as shown in the following lemma.

\begin{lem} 
  Let $\boldsymbol{\vartheta}_{\text{IV}}$ be the IV estimand from a regression of $Y_{ig}$ on  $\mathbf{X}_{ig} \equiv (1, D_{ig}, \bar{D}_{ig}, D_{ig} \bar{D}_{ig})'$ with instruments $\boldsymbol{\mathcal{Z}}_{ig} \equiv (1, Z_{ig}, S_g, Z_{ig}S_g)'$, namely
  \[
  \boldsymbol{\vartheta}_{\text{IV}} \equiv \begin{bmatrix} \alpha_{\text{IV}} & \beta_{\text{IV}}& \gamma_{\text{IV}}& \delta_{\text{IV}} \end{bmatrix}' = \mathbbm{E}\left[ \boldsymbol{\mathcal{Z}}_{ig}\mathbf{X}_{ig}' \right]^{-1} \mathbbm{E}\left[ \boldsymbol{\mathcal{Z}}_{ig} Y_{ig} \right].
  \]
  assuming that $\mathbbm{E}[\boldsymbol{\mathcal{Z}}_{ig} \mathbf{X}_{ig}']$ is invertible. 
  Then, under \eqref{eq:LinearModel} and Assumptions \ref{assump:saturations}--\ref{assump:Bernoulli} and \ref{assump:onesided}--\ref{assump:exclusion}, 
  \begin{align*}
    \alpha_{\text{IV}} &= \mathbbm{E}\left[ \alpha_{ig} \right] &
    \beta_{\text{IV}} &= \mathbbm{E}\left[ \beta_{ig}|C_{ig}=1 \right]\\
    \gamma_{\text{IV}} &= \mathbbm{E}\left[ \gamma_{ig} \right] + \frac{\text{Cov}(\bar{C}_{ig}, \gamma_{ig})}{\mathbbm{E}(\bar{C}_{ig})} & 
    \delta_{\text{IV}} &= \mathbbm{E}\left[ \delta_{ig}|C_{ig} = 1 \right] + \frac{\text{Cov}(\bar{C}_{ig}, \delta_{ig}|C_{ig} = 1)}{\mathbbm{E}(\bar{C}_{ig}|C_{ig} = 1)}.
  \end{align*}
  \label{lem:linear}
\end{lem}
\vspace{-0.6cm}
As we see from \autoref{lem:linear}, IV identifies the population average of $\alpha_{ig}$, along with the population average of $\beta_{ig}$ for the subset of individuals who select into treatment.
Neither of these, however, is itself a causal effect. 
In general, IV recovers neither direct nor indirect causal effects for any well-defined group of individuals.
Specializing \eqref{eq:IE} to the linear model from \eqref{eq:LinearModel} gives $\text{IE}_0(\bar{d}, \Delta) = \mathbbm{E}[\gamma_{ig}] \Delta$.
In other words, $\mathbbm{E}[\gamma_{ig}]$ is an average \emph{spillover}.
\autoref{lem:linear} shows that IV fails to identify this quantity unless the individual-specific spillovers $\gamma_{ig}$ are uncorrelated with the share of compliers $\bar{C}_{ig}$.
This condition could easily fail in practice.
In the labor market example from the introduction, cities with a particularly depressed labor market might be expected to contain a large share of compliers.
If negative spillovers are more intense in such cities, IV will not recover the average indirect effect. 
A similar problem hampers the interpretation of $\delta_{\text{IV}}$. 
Under \eqref{eq:LinearModel} the average direct effect for compliers, as a function of $\bar{d}$, is given by $\mathbbm{E}[\beta_{ig}|C_{ig}=1] + \mathbbm{E}[\delta_{ig}|C_{ig}=1] \bar{d}$.
While IV identifies the intercept of this function, it only identifies the slope if $\delta_{ig}$ is uncorrelated with $\bar{C}_{ig}$ for compliers.

As this example illustrates, identifying direct and indirect causal effects requires us to correct for possible dependence between individual-specific coefficients and group-level take-up that arises from the first-stage relationship in \autoref{lem:DbarDist}. 
The key to our approach, as shown in the following theorem, is to condition on $\bar{C}_{ig}$ and $N_g$. 

\begin{thm}
  \label{thm:conditional_indep}
  Under Assumptions \ref{assump:saturations}--\ref{assump:Bernoulli} and \ref{assump:onesided}--\ref{assump:exclusion}, $(S_g, Z_{ig}, \bar{D}_{ig}) \indep (\mathbf{B}_{ig}, C_{ig}) | (\bar{C}_{ig}, N_g)$.
\end{thm}

\autoref{thm:conditional_indep} implies that conditioning on $(\bar{C}_{ig}, N_g)$ is sufficient to break any dependence between $\mathbf{f}(\bar{D}_{ig})$ and $(\mathbf{B}_{ig},C_{ig})$ that may be present.
The intuition for this result is as follows.
Conditional on $\bar{C}_{ig}$ and $N_g$, we know precisely how many of $(i,g)$'s neighbors are compliers.
Given this information, IOR implies that all remaining variation in $\bar{D}_{ig}$ is arises solely from experimental variation in the saturation $S_g$ assigned to different groups, and the share of compliers offered treatment across groups assigned the same saturation.
So long as $Z_{ig}$ and $S_g$ do not affect $(\mathbf{B}_{ig},C_g)$, \autoref{assump:exclusion}, it follows that $(Z_{ig}, \bar{D}_{ig},S_g)$ are exogenous given $(\bar{C}_{ig}, N_g)$, even when individuals decide whether or not to take up treatment based on knowledge of their potential outcome functions.
In effect, our identification approach is a combination of instrumental variables and control function methods.
First $(\bar{C}_{ig}, N_{g})$ serves as a control function for the endogenous regressor $\bar{D}_{ig}$, similar to \cite{masten2016}.
Second, $Z_{ig}$ serves as an instrument for $D_{ig}$, because this regressor remains endogenous even conditional on $(\bar{C}_{ig}, N_g)$.

\subsection{An Inverse-Weighting Instrumental Variables Approach}

Before stating our identification results, we require some additional notation and one further assumption.
Define the vector $\mathbf{W}_{ig}$ and matrix-valued functions $\mathbf{Q}, \mathbf{Q}_0, \mathbf{Q}_1$ as follows:
\begin{align}
  \label{eq:Qdef}
  \mathbf{Q}(\bar{c}, n) &\equiv \mathbbm{E}\left[ \mathbf{W}_{ig}\mathbf{W}_{ig}'|\bar{C}_{ig} = \bar{c}, N_g = n \right], \quad \mathbf{W}_{ig} \equiv \begin{bmatrix} 1 & Z_{ig} \end{bmatrix}' \otimes \mathbf{f}(\bar{D}_{ig})\\
  \label{eq:Q0def}
  \mathbf{Q}_0(\bar{c}, n) &\equiv \mathbbm{E}\left[ (1 - Z_{ig}) \mathbf{f}(\bar{D}_{ig})\mathbf{f}(\bar{D}_{ig})'|\bar{C}_{ig} = \bar{c}, N_g = n \right]\\
  \label{eq:Q1def}
  \mathbf{Q}_1(\bar{c}, n) &\equiv \mathbbm{E}\left[ Z_{ig} \mathbf{f}(\bar{D}_{ig})\mathbf{f}(\bar{D}_{ig})'|\bar{C}_{ig} = \bar{c}, N_g = n \right].
\end{align}
These functions depend only on the distribution of $\bar{D}_{ig}|(Z_{ig}, \bar{C}_{ig}, N_g)$, which can be calculated from \autoref{lem:DbarDist}, and the distribution of $Z_{ig}|(\bar{C}_{ig},N_g)$, which coincides with its unconditional distribution by \autoref{lem:Zindep}. 
As such, under our assumptions $\mathbf{Q},\mathbf{Q}_0, \mathbf{Q}_1$ are \emph{completely determined} by the design of the randomized saturation experiment.
We can always calculate them by simulating the experimental design.
Depending on the choice of $\mathbf{f}$, analytical expressions may also be available, as shown  in \autoref{subsec:identification_example} for the linear potential outcomes model from \eqref{eq:LinearModel}.

We use $\mathbf{Q}, \mathbf{Q}_0, \mathbf{Q}_1$ to construct valid instrumental variables by inverse-weighting.
Rather than using the randomly assigned saturation $S_g$ as a source of instruments for $\mathbf{f}(\bar{D}_{ig})$ we \emph{transform} the endogenous regressors $\mathbf{W}_{ig}$ into a set of exogenous instruments using $\mathbf{Q}_0(\bar{C}_{ig}, N_g)^{-1}$ and $\mathbf{Q}_{1}(\bar{C}_{ig}, N_g)^{-1}$, in particular
\begin{align}
	\label{eq:Zw_def}
\boldsymbol{\mathcal{Z}}_{ig}^W &\equiv \mathbf{Q}(\bar{C}_{ig}, N_g)^{-1}\mathbf{W}_{ig} \\
\label{eq:Z0_def}
\boldsymbol{\mathcal{Z}}_{ig}^0 &\equiv \mathbf{Q}_0(\bar{C}_{ig}, N_g)^{-1}\mathbf{f}(\bar{D}_{ig}) \\
\label{eq:Z1_def}
\boldsymbol{\mathcal{Z}}_{ig}^1 &\equiv \mathbf{Q}_1(\bar{C}_{ig}, N_g)^{-1}\mathbf{f}(\bar{D}_{ig}).
\end{align}
Constructing these instruments requires us to evaluate  $\mathbf{Q}_0$ and $\mathbf{Q}_1$ at $(\bar{C}_{ig}, N_g)$.\footnote{The function $\mathbf{Q}$ can be constructed from $\mathbf{Q}_0$ and $\mathbf{Q}_1$, as shown in \autoref{eq:QfromQ0Q1}.}
The group size $N_g$ is observed, while the share of compliers $\bar{C}_{ig}$ is not.
In large groups, $\bar{C}_{ig}$ can be precisely estimated by calculating the rate of treatment take-up among the neighbors of $(i,g)$ who are offered treatment. We formally establish the rates of convergence of IV estimators that  plug-in a proxy for $\bar{C}_{ig}$ in Section \ref{sec:estimationinference}.
For the remainder of this section, however, we consider identification \emph{conditional} on knowledge of $\bar{C}_{ig}$. 

To understand the intuition behind $\boldsymbol{\mathcal{Z}}_{ig}^W$, $\boldsymbol{\mathcal{Z}}_{ig}^0$, and  $\boldsymbol{\mathcal{Z}}_{ig}^1$, consider the linear potential outcomes example from \eqref{eq:LinearModel} above.
Here we have $\mathbf{f}(x) = (1, x)'$ and thus
\[
  \mathbf{Q}_z(\bar{C}_{ig}, N_g) = \mathbbm{P}(Z_{ig} = z)\mathbbm{E}\left[ \left.
      \begin{pmatrix}
        1 & \bar{D}_{ig} \\
        \bar{D}_{ig} & \bar{D}_{ig}^2
    \end{pmatrix}\right|\bar{C}_{ig}, N_g, Z_{ig} = z\right], \quad 
    z \in \left\{ 0, 1 \right\}
\]
using the fact that $Z_{ig} \indep (\bar{C}_{ig}, N_g)$ by \autoref{lem:Zindep}.
It follows after a few steps of algebra that 
\[
  \mathbf{Q}_z(\bar{C}_{ig}, N_g)^{-1} \mathbf{f}(\bar{D}_{ig}) = 
  \frac{1}{\mathbbm{P}(Z_{ig} = z)}
  \begin{bmatrix}
    \displaystyle
    \frac{\mathbbm{E}(\bar{D}_{ig}^2|\bar{C}_{ig}, N_g, Z_{ig} = z) - \bar{D}_{ig} \mathbbm{E}(\bar{D}_{ig}|\bar{C}_{ig}, N_g, Z_{ig} = z)}{ \text{Var}(\bar{D}_{ig}|\bar{C}_{ig}, N_g, Z_{ig} = z)}\\ \\
    \displaystyle
  \frac{\bar{D}_{ig} - \mathbbm{E}(\bar{D}_{ig}|\bar{C}_{ig}, N_g, Z_{ig} = z)}{ \text{Var}(\bar{D}_{ig}|\bar{C}_{ig}, N_g, Z_{ig} = z)}
\end{bmatrix}.
\]
While $\bar{D}_{ig}$ is endogenous, the scaled difference between $\bar{D}_{ig}$ and its conditional expectation is a valid instrument under the linear potential outcomes model.
Intuitively, this transformation \emph{adjusts} for the first-stage heterogeneity discussed in \autoref{subsec:compliers}: after controlling for differences in  $(\bar{C}_{ig}, N_g)$, the remaining variation in $\bar{D}_{ig}$ arises only from the experimentally--assigned saturations.
Thus, rather than using $S_g$ as an instrument directly, we use it indirectly to generate variation in $\bar{D}_{ig}$ given $(\bar{C}_{ig}, N_g)$.
The final ingredient that we require is a rank condition.

\begin{assump}[Rank Condition]
  \mbox{}
  \label{assump:rank}
  \begin{enumerate}[(i)]
    \item $0 < \mathbbm{E}(C_{ig}) < 1$
    \item $\mathbf{Q}(\bar{c}, n)$ is invertible at every point $(\bar{c},n)$ in the support of $(\bar{C}_{ig}, N_g)$.
  \end{enumerate}
\end{assump}
Part (i) of \autoref{assump:rank} asserts that the population contains at least some never-takers,  $\mathbbm{E}(C_{ig}) < 1$, and at least some compliers, $\mathbbm{E}(C_{ig}) > 0$.\footnote{Note that this condition applies to the \emph{entire experiment} rather than any particular group. If $\mathbbm{E}(C_{ig})=1$, then there is perfect compliance in every group and no need for our methods. If $\mathbbm{E}(C_{ig}) = 0$, then no one in any group takes the treatment so it is impossible to identify treatment effects.}
Part (ii) requires that the matrix-valued function $\mathbf{Q}$ defined in \eqref{eq:Qdef} is full rank when evaluated at any share of compliers $\bar{c}$ and group size $n$ that occur in the population.
\autoref{assump:rank} does not explicitly restrict $\mathbf{Q}_0$ or $\mathbf{Q}_1$.
By the linearity of conditional expectation, however, 
\begin{equation}
  \mathbf{Q}(\bar{c},n) = \begin{bmatrix}
    \mathbf{Q}_0(\bar{c},n) + \mathbf{Q}_1(\bar{c}, n) & \mathbf{Q}_1(\bar{c}, n) \\
    \mathbf{Q}_1(\bar{c}, n) & \mathbf{Q}_1(\bar{c}, n)
  \end{bmatrix}
  \label{eq:QfromQ0Q1}
\end{equation}
so \autoref{assump:rank}(ii) could equivalently be stated in terms of $\mathbf{Q}_0$ and $\mathbf{Q}_1$.

\begin{lem}
  \label{lem:QfromQ0Q1}
  $\mathbf{Q}(\bar{c}, n)$ is invertible iff $\mathbf{Q}_0(\bar{c}, n)$ and $\mathbf{Q}_1(\bar{c}, n)$ are both invertible, in which case
  \[
    \mathbf{Q}(\bar{c}, n)^{-1} = \begin{bmatrix}
      \mathbf{Q}_0(\bar{c}, n)^{-1} & -\mathbf{Q}_0(\bar{c}, n)^{-1} \\
      -\mathbf{Q}_0(\bar{c}, n)^{-1} & \mathbf{Q}_0(\bar{c}, n)^{-1} + \mathbf{Q}_1(\bar{c}, n)^{-1}
    \end{bmatrix}.
  \]
\end{lem}

We discuss low-level conditions for the invertibility of $(\mathbf{Q}_0, \mathbf{Q}_1)$, and hence $\mathbf{Q}$, below in \autoref{subsec:identification_example} and Appendix \ref{sec:BasisFunctionsRank}.
As a preview: researchers should not include more basis functions $K$ than there are saturations in the experimental design. 
Having assumed the necessary rank condition, we can now state our main identification results.
The following theorem shows how $\mathbf{Q}_0(\bar{C}_{ig}, N_g)$ and $\mathbf{Q}_1(\bar{C}_{ig}, N_g)$ can be used to construct instrumental variables that identify average values of the random coefficients for well-defined groups of individuals.  

\begin{thm}
  \label{thm:expect}
  Let $\boldsymbol{\mathcal{Z}}_{ig}^W$, $\boldsymbol{\mathcal{Z}}_{ig}^0$, and $\boldsymbol{\mathcal{Z}}_{ig}^1$ be as defined in \eqref{eq:Zw_def}--\eqref{eq:Z1_def}.
Then, under Assumptions \ref{assump:randcoef}--\ref{assump:IOR} and \ref{assump:rank} and assuming that $(Z_{ig}, \bar{D}_{ig}) \indep (\mathbf{B}_{ig}, C_{ig})|(\bar{C}_{ig}, N_g)$, we have

  \begin{enumerate}[(i)]

    \item $\begin{bmatrix}
        \mathbbm{E}(\boldsymbol{\theta}_{ig})\\
        \mathbbm{E}(\boldsymbol{\psi}_{ig} - \boldsymbol{\theta}_{ig}| C_{ig} = 1)\\
      \end{bmatrix}= \mathbbm{E}\left[\boldsymbol{\mathcal{Z}}_{ig}^W \mathbf{X}_{ig}'\right]^{-1}  \mathbbm{E}\left[ \boldsymbol{\mathcal{Z}}_{ig}^W Y_{ig} \right]$,

\item $\mathbbm{E}\left[ \boldsymbol{\psi}_{ig}|C_{ig}=1 \right] = \mathbbm{E}\left[ \boldsymbol{\mathcal{Z}}_{ig}^1 \left\{ D_{ig} \mathbf{f}(\bar{D}_{ig}) \right\}' \right]^{-1}\mathbbm{E}\left[ \boldsymbol{\mathcal{Z}}_{ig}^1\left\{ D_{ig}Y_{ig} \right\}\right]$

\item $\mathbbm{E}\left[ \boldsymbol{\theta}_{ig}| C_{ig} = 0 \right] = \mathbbm{E}\left[ \boldsymbol{\mathcal{Z}}_{ig}^1 \left\{ Z_{ig}(1 - D_{ig})\mathbf{f}(\bar{D}_{ig}) \right\}' \right]^{-1}\mathbbm{E}\left[ \boldsymbol{\mathcal{Z}}_{ig}^{1}\left\{ Z_{ig}(1 - D_{ig})Y_{ig} \right\}\right]$, and

\item $\mathbbm{E}\left[ \boldsymbol{\theta}_{ig} \right] = \mathbbm{E}\left[ \boldsymbol{\mathcal{Z}}_{ig}^0 \left\{ (1 - Z_{ig})\mathbf{f}(\bar{D}_{ig}) \right\}' \right]^{-1}\mathbbm{E}\left[ \boldsymbol{\mathcal{Z}}_{ig}^0 \left\{ (1 - Z_{ig})Y_{ig} \right\}\right]$.

  \end{enumerate}
\end{thm}

The first part of \autoref{thm:expect} identifies the average effects that the na\"{i}ve IV approach from \autoref{lem:linear} in general fails to.
Parts (ii) and (iii) use a similar approach to obtain moment equations for the average value of $\boldsymbol{\psi}_{ig}$ for compliers and $\boldsymbol{\theta}_{ig}$ for never-takers. 
Given part (i), part (iv) is technically redundant, but it is convenient to have an expression for $\mathbbm{E}(\boldsymbol{\theta}_{ig})$ in isolation.
As discussed below in Section \ref{subsec:identification_example}, having sufficient variation in the saturations is crucial for part (ii) of \autoref{assump:rank}. 
  
Notice that \autoref{thm:expect} does not explicitly invoke the randomized saturation design, Assumptions \ref{assump:saturations}--\ref{assump:Bernoulli}, or the exclusion restriction, \autoref{assump:exclusion}. 
Using this result for identification, however, requires us to satisfy $(Z_{ig}, \bar{D}_{ig})\indep (\mathbf{B}_{ig}, C_{ig})|(\bar{C}_{ig}, N_g)$.
As shown in \autoref{thm:conditional_indep} above, the randomized saturation design and exclusion restriction are sufficient for this condition to hold under one-sided non-compliance and IOR, Assumptions \ref{assump:onesided} and \ref{assump:IOR}.
The following result catalogues the full set of causal effects that are identified under our assumptions.

\begin{thm}
  Given knowledge of $\bar{C}_{ig}$ the following are identified under Assumptions \ref{assump:saturations}--\ref{assump:rank}: 
  \label{thm:identification}
  \begin{enumerate}[(i)]
    \item $\text{IE}_0(\bar{d},\Delta) \equiv \mathbbm{E}[Y_{ig}(0, \bar{d} + \Delta) - Y_{ig}(0, \bar{d})]$, 
    \item $\text{DE}_1(\bar{d}|D_{ig}=1) \equiv \mathbbm{E}\left[ Y_{ig}(1, \bar{d}) - Y_{ig}(0,\bar{d})|D_{ig} = 1 \right]$,
    \item $\text{IE}_0(\bar{d},\Delta|D_{ig} = 1) \equiv \mathbbm{E}[Y_{ig}(0, \bar{d} + \Delta) - Y_{ig}(0, \bar{d})|D_{ig} = 1]$, 
    \item $\text{IE}_1(\bar{d},\Delta|D_{ig} = 1) \equiv \mathbbm{E}[Y_{ig}(1, \bar{d} + \Delta) - Y_{ig}(1, \bar{d})|D_{ig} = 1]$,
    \item $\text{IE}_0(\bar{d},\Delta|D_{ig} = 0) \equiv \mathbbm{E}[Y_{ig}(0, \bar{d} + \Delta) - Y_{ig}(0, \bar{d})|D_{ig} = 0]$, 
  \end{enumerate}
\end{thm}

Part (i) of \autoref{thm:identification} is a population average indirect treatment effect, as defined in \eqref{eq:IE} above.
It measures the causal impact of increasing the treatment take-up rate among Alice's neighbors from $\bar{d}$ to $(\bar{d} + \Delta)$ when Alice's own treatment is held fixed at zero.
In the \cite{crepon2013} experiment discussed in our empirical example below, this corresponds to the average labor market displacement effect.
Whereas part (i) is an average treatment effect, parts (ii)--(iv) are the effects of treatment-on-the-treated.\footnote{Because this is a setting with one-sided non-compliance, any participant with $D_{ig} = 1$ must be a complier.}
Part (ii) gives the direct effect of treating Alice while holding the treatment take-up rate of her neighbors fixed at $\bar{d}$, while (iii) and (iv) give the indirect effect of increasing her neighbors' treatment take-up from $\bar{d}$ to $\bar{d} + \Delta$ while holding Alice's treatment fixed at either zero, part (iii), or one, part (iv). 
Part (v) is a treatment-on-the-untreated version of \autoref{eq:IE}: it gives the indirect effect for never-takers, holding their treatment fixed at zero.
While we identify the full set of direct and indirect effects for the treated sub-population, we only identify a subset of these effects for other groups.
By definition, never-takers cannot be observed with $D_{ig} = 1$.
As such, we cannot identify direct treatment effects for this group or indirect treatment effects when $D_{ig}$ is held fixed at one.
This in turn implies that we cannot identify the average direct effect for the population as a whole, $\text{DE}(\bar{d})$, or the average indirect effect when $D_{ig}$ is held fixed at one, $\text{IE}_1(\bar{d}, \Delta)$.

The treatment effects identified in \autoref{thm:identification} provide information that should be of interest to policymakers who are concerned about the distributional consequences of policies that may generate spillovers. 
For example, comparing $\text{IE}_0$ to $\text{IE}_1$ allows policymakers to determine whether the treated mainly create spillovers on the untreated, or whether they create spillovers on each other.
Similarly, $\text{DE}_1$ gives the average effect of program participation for those who are willing to participate, allowing policymakers to determine whether the treatment is on net beneficial to those who receive it.
Intent-to-treat effects alone do not provide this information.

\subsection{Identification in Practice}
\label{subsec:identification_example}
Given that $\mathbf{Q}_0$ and $\mathbf{Q}_1$ are completely determined by the experimental design, we can directly check part (ii) of \autoref{assump:rank} for any choice of basis functions $\mathbf{f}$ and probability distribution over saturations.
Consider again the linear potential outcomes model from \eqref{eq:LinearModel}.
In this example $\mathbf{f}\left(x \right) = (1, x)'$ and thus, 
\begin{align}
  \mathbf{Q}_0(\bar{c},n) &= 
  \begin{bmatrix}
  \label{eq:Q0LinearBernoulli}
  \mathbbm{E}\left\{ 1 - S_g \right\} & \bar{c}\, \mathbbm{E}\left\{ S_g(1-S_g) \right\} \\
    \bar{c} \, \mathbbm{E}\left\{ S_g (1 - S_g) \right\} &
    \bar{c}^2 \, \mathbbm{E}\left\{ S_g^2(1 - S_g) \right\} + \frac{\bar{c}}{n - 1} \mathbbm{E}\left\{ S_g(1 - S_g)^2 \right\}
  \end{bmatrix}\\ 
  \label{eq:Q1LinearBernoulli}
  \mathbf{Q}_1(\bar{c},n) &= 
  \begin{bmatrix}
  \mathbbm{E}\left\{ S_g \right\} & \bar{c}\, \mathbbm{E}\left\{ S_g^2\right\} \\
    \bar{c} \, \mathbbm{E}\left\{ S_g^2 \right\} &
    \bar{c}^2 \, \mathbbm{E}\left\{ S_g^3 \right\} + \frac{\bar{c}}{n - 1} \mathbbm{E}\left\{ S_g^2(1 - S_g) \right\}.
  \end{bmatrix}
\end{align}
by Bayes' Theorem, the Law of Total Probability, and Lemmas \ref{lem:DbarDist} and \ref{lem:Zindep}.
Suppose first that there is a single saturation $s$.
Then \eqref{eq:Q0LinearBernoulli} and \eqref{eq:Q1LinearBernoulli} simplify to yield 
\[
  \left|\mathbf{Q}_0(\bar{c},n)\right| = \frac{\bar{c}s(1-s)^3}{n-1}, \quad
  \left|\mathbf{Q}_1(\bar{c},n)\right| = \frac{\bar{c}s^3(1-s)}{n-1}.
\]
so that $\mathbf{Q}_0(\bar{c}, n)$ and $\mathbf{Q}_1(\bar{c}, n)$ are both invertible for any $n$ and all $\bar{c}$ greater than zero provided that $0 < s < 1$.
The identifying power of this ``degenerate'' randomized saturation design, however, is weak: $\mathbf{Q}_0, \mathbf{Q}_1$ are arbitrarily close to being singular for any $\bar{c}$ if $n$ is sufficiently large.
Consider next a so-called ``cluster randomized'' experiment in which there are two saturations, 0 and 1, and $\mathbbm{P}(S_g = 1) = p$.
Calculating the expectations in \eqref{eq:Q0LinearBernoulli} and \eqref{eq:Q1LinearBernoulli}, 
\[
  \mathbf{Q}_0(\bar{c},n) = \begin{bmatrix}
    (1-p) & 0 \\
    0 & 0
  \end{bmatrix}, \quad
  \mathbf{Q}_1(\bar{c},n) = \begin{bmatrix}
    p & \bar{c}p \\
    \bar{c}p & \bar{c}^2 p
  \end{bmatrix}.
\]
In this case neither $\mathbf{Q}_0$ nor $\mathbf{Q}_1$ is invertible for \emph{any} values of $n$ and $\bar{c}$.
Finally, consider a design with two distinct, equally likely saturations $s_L < s_H$.
For this design, straightforward but tedious algebra gives
\begin{align*}
  \left|\mathbf{Q}_0(\bar{c}, n)\right| &= \frac{\bar{c}^2}{4} (1 - s_L) (1 - s_H)(s_H - s_L)^2 + \frac{\bar{c}\left[ (1 - s_L) + (1 - s_H) \right]\left[ s_L(1 - s_L)^2 + s_H(1 - s_H)^2 \right]}{4(n-1)}\\
  \left| \mathbf{Q}_1(\bar{c}, n) \right| &= \frac{\bar{c}^2}{4}s_L s_H (s_H - s_L)^2 + \frac{\bar{c}\left( s_L + s_H \right)\left[ s_L^2(1 - s_L) + s_H^2 (1 - s_H) \right]}{4 (n-1)}.
\end{align*}
So long as neither $s_L$ nor $s_H$ equals zero or one, both terms in each expression are strictly positive for any $\bar{c} > 0$, so that $\mathbf{Q}_0$ and $\mathbf{Q}_1$ are invertible.
Moreover, in contrast to the single saturation design discussed above, this design does not suffer from a weak identification problem.
While the second term in each of the preceding equalities vanishes for large $n$, the first term does not.
Thus, two interior saturations are sufficient to strongly identify the linear potential outcomes model that we use in our empirical example and simulation study below.

As the preceding examples show, two distinct sources of experimental variation determine the rank of $\mathbf{Q}_0(\bar{c},n)$ and $\mathbf{Q}_1(\bar{c},n)$: ``between'' saturation variation, and ``within'' saturation variation.
Our first example lacks ``between'' variation because each group is assigned the same saturation, $S_g = s$.
Yet even with a single saturation, there is still ``within'' variation under \autoref{assump:Bernoulli}, because the number of offers made to a given group is random.
This ``within'' variation, however, is negligible when $n$ is large. 
In our second example, the cluster randomized experiment, the situation is reversed.
Because everyone in a given group is either offered ($S_g = 0$) or unoffered ($S_g = 1)$, this design generates no ``within'' variation.
While a cluster randomized design does generate some ``between'' variation, it is too coarse to identify our effects of interest: under our assumptions $\bar{D}_{ig}$ equals zero when $S_g = 0$ and $\bar{C}_{ig}$ when $S_g = 1$.
Our third example, with two saturations $0 < s_L < s_H < 1$, features sufficient ``between'' variation to identify the effects of interest even when $n$ is so large that ``within'' variation becomes negligible.

In general, sufficient conditions for \autoref{assump:rank}(ii) will depend on the specific choice of basis functions $\mathbf{f}$. 
For large $n$, however, a necessary condition is that the design contains at least as many distinct interior saturations as there are elements in $\mathbf{f}$. 
Appendix \ref{sec:BasisFunctionsRank} provides a detailed explanation of this result.


\section{Estimation and Inference}
\label{sec:estimationinference}

If $\bar{C}_{ig}$ were observed, a handful of just-identified IV regressions would suffice to estimate the causal effects from \autoref{thm:identification}.
While $\bar{C}_{ig}$ is unobserved in practice, fortunately we can estimate it under one-sided non-compliance by comparing treatment take-up to the share of treatment offers, i.e.\
\begin{equation}
  \label{eq:Chat}
  \widehat{C}_{ig} \equiv
  \begin{cases}
    \bar{D}_{ig}/\bar{Z}_{ig}, & \text{if } \bar{Z}_{ig} > 0\\
    0, & \text{otherwise}
  \end{cases}
\end{equation}
where we arbitrarily define $\widehat{C}_{ig} = 0$ if none of $(i,g)$'s neighbors are offered treatment.\footnote{Under \autoref{assump:Bernoulli} it is possible, although unlikely, that $\bar{Z}_{ig}$ could be zero even if $S_g>0$.}
In this section we use \eqref{eq:Chat} to derive feasible, consistent, and asymptotically normal estimators of the direct and indirect causal effects identified in \autoref{sec:identification}.
\autoref{sec:StepByStep} provides full implementation details specialized to the linear outcome model from \eqref{eq:LinearModel}.
For simplicity, we assume throughout that the random saturation $S_g$ is bounded below by $\underline{s} > 0$.
Because we cannot estimate $\bar{C}_{ig}$ when $S_g = 0$, experiments that include a 0\% saturation require a slightly different approach.
We explain these differences in Appendix \ref{sec:0percent}.

In the interest of brevity, we introduce shorthand notation and high-level regularity conditions that apply to all four of our sample analogue estimators.
These take the form
\begin{equation}
  \label{eq:varthetahat}
  \widehat{\boldsymbol{\vartheta}} 
  \equiv \left(\sum_{g=1}^{G} \sum_{i=1}^{N_g} \widehat{\boldsymbol{\mathcal{Z}}}_{ig} \boldsymbol{X}_{ig}'\right)^{-1} \left(\sum_{g=1}^{G} \sum_{i=1}^{N_g} \widehat{\boldsymbol{\mathcal{Z}}}_{ig} Y_{ig}\right), \quad  \widehat{\boldsymbol{\mathcal{Z}}}_{ig} \equiv \mathbf{R}(\widehat{C}_{ig}, N_g)^{+} \boldsymbol{W}_{ig} 
\end{equation}
where $Y_{ig}$ is the outcome variable from \autoref{assump:randcoef}, and  $\mathbf{M}^+$ denotes the Moore-Penrose inverse of a square matrix $\mathbf{M}$.
\autoref{tab:estimators} gives the definitions of 
$\boldsymbol{X}_{ig}, \mathbf{R}$, and  $\boldsymbol{W}_{ig}$ corresponding to each part of \autoref{thm:expect}. 
The ``estimated'' instrument $\widehat{\boldsymbol{\mathcal{Z}}}_{ig}$ is a stand-in for the unobserved ``true'' instrument $\boldsymbol{\mathcal{Z}}_{ig} \equiv \mathbf{R}(\bar{C}_{ig}, N_g)^{-1} \boldsymbol{W}_{ig}$.
While $\mathbf{R}(\bar{C}_{ig}, N_g)$ is invertible under \autoref{assump:rank}, $\mathbf{R}(\widehat{C}_{ig}, N_g)$ may not be so, since $\widehat{C}_{ig}$ could fall outside the support set of $\bar{C}_{ig}$ or even equal zero. 
For this reason we define $\widehat{\boldsymbol{\mathcal{Z}}}_{ig}$ using the Moore-Penrose inverse, which always exists and coincides with the ordinary matrix inverse when $\mathbf{R}(\widehat{C}_{ig}, N_g)$ is indeed invertible. 

\begin{table}
  \centering
  \begin{tabular}{rlll}
    & \multicolumn{1}{c}{$\boldsymbol{X}_{ig}$} & \multicolumn{1}{c}{$\mathbf{R}$} & \multicolumn{1}{c}{$\boldsymbol{W}_{ig}$}\\
    \hline\\
    (i) 
    &$\begin{bmatrix}
      1 \\ D_{ig}
    \end{bmatrix}\otimes \mathbf{f}(\bar{D}_{ig})$& $\mathbf{Q}$ & $\begin{bmatrix}
      1 \\ Z_{ig}
  \end{bmatrix}\otimes \mathbf{f}(\bar{D}_{ig})$   \\ \\
  (ii) 
  & $\mathbf{f}(\bar{D}_{ig})$ & $\mathbf{Q}_1$ & $\mathbf{f}(\bar{D}_{ig})D_{ig}$  \\ \\ 
  (iii) 
  & $\mathbf{f}(\bar{D}_{ig})$ & $\mathbf{Q}_1$ & $\mathbf{f}(\bar{D}_{ig}) Z_{ig}(1 - D_{ig})$  \\ \\
  (iv)
  & $\mathbf{f}(\bar{D}_{ig})$ 
  & $\mathbf{Q}_0$ & $\mathbf{f}(\bar{D}_{ig})(1 - Z_{ig})$
  \end{tabular}
  \caption{This table defines the shorthand from \eqref{eq:varthetahat} for the four sample analogue estimators corresponding to the parts of \autoref{thm:expect}. In each part, 
the vector of regressors is $\boldsymbol{X}_{ig}$, the true instrument vector is $\boldsymbol{\mathcal{Z}}_{ig} \equiv \mathbf{R}(\bar{C}_{ig}, N_g)^{-1} \boldsymbol{W}_{ig}$, and the estimated instrument vector is $\widehat{\boldsymbol{\mathcal{Z}}}_{ig} \equiv \mathbf{R}(\widehat{C}_{ig}, N_g)^{+} \boldsymbol{W}_{ig}$, where  $\mathbf{M}^{+}$ denotes the Moore-Penrose inverse of a square matrix $\mathbf{M}$, and $\widehat{C}_{ig}$ is as defined in \eqref{eq:Chat}. The functions $\mathbf{Q}, \mathbf{Q}_0, \mathbf{Q}_1$ are as defined in \eqref{eq:Qdef}--\eqref{eq:Q1def}.}
  \label{tab:estimators}
\end{table}

As $G$ grows, so does the number of unknown values $\bar{C}_{ig}$ that we must estimate to construct the instrument vectors $\widehat{\boldsymbol{\mathcal{Z}}}_{ig}$.\footnote{While $\bar{C}_{ig}$ can vary across individuals in the same group, it takes on at most two distinct values for fixed $g$. If a group contains $T$ total individuals, of whom $c$ are compliers and $n$ never-takers, then the share of compliers among a given person's neighbors is either $(c-1)/(T-1)$ if she is a complier or $c/(T-1)$ if she is a never-taker. Thus, the number of incidental parameters is $2G$.}
For this reason, we consider an asymptotic sequence in which the minimum group size $\underline{n}$ grows along with the number of groups $G$.
Under appropriate assumptions, this implies that the limit behavior of $\widehat{\boldsymbol{\vartheta}}$, which we refer to as the ``random saturation IV" (RS-IV), coincides with that of the infeasible estimator that uses the true instrument vector $\boldsymbol{\mathcal{Z}}_{ig}$ instead of its estimate $\widehat{\boldsymbol{\mathcal{Z}}}_{ig}$. 

Like \cite{baird2018}, we take an infinite population approach to inference, assuming that the researcher observes a random sample of size $G$ from a population of groups.
Unlike \cite{baird2018}, we allow these groups to differ in size.
Upon drawing a group $g$ from the population, we observe the group-level random variables $(S_g, N_g)$ along with the individual-level random variables $(Y_{ig}, D_{ig}, Z_{ig})$ for each member of the group: $1 \leq i \leq N_g$.
We further assume that observations are identically distributed, but not independent, within groups.\footnote{The assumption that observations are identically distributed within group amounts to stipulating that the indices $1 \leq i \leq N_g$ are assigned at random.} 

Groups are only observed \emph{as a unit}: either everyone from the group appears in the sample or no one does.
For this reason, some care is needed in defining random variables to represent our sampling procedure and expectations to represent the population averages that define our causal effects of interest.
The expectations in Theorems \ref{thm:expect}--\ref{thm:identification} are averages that give equal weight to each individual in the population, or sub-population if we condition on $C_{ig}$.
Analogously, the estimator in \eqref{eq:varthetahat} is an average that gives equal weight to each individual in the sample.
Both of these are precisely what we want, as our goal is to identify and estimate average causal effects for individuals.
Under iid sampling of groups, however, $(Y_{ig}, D_{ig}, Z_{ig}, \bar{D}_{ig})$ represent a single person chosen at random from a randomly-selected group.
If all groups were the same size, this would be equivalent to choosing a person uniformly at random from the population of \emph{individuals}.
When groups vary in size, however, the equivalence no longer holds.\footnote{Consider a population of 100 groups, half of which have 5 members and the rest of which have 15 members so that 250 of the 1000 people in the population belong to a small group and the remaining 750 belong to a large group. Suppose first that we choose a single group at random and then a single person within the selected group. Then someone from a small group has probability 1/500 of being selected while someone from a large group has probability 1/1500 of being selected.}
This creates the possibility for ambiguity when taking the expectation of an individual-level random variable, such as $Y_{ig}$, without conditioning on group size.

Fortunately this is only a question of defining appropriate notation.
Our group sampling procedure unambiguously gives equal weight to each individual in the population because we observe not isolated individuals but whole groups.
While small groups are just as likely to be drawn as large groups, large groups make a greater contribution to the sample averages from \eqref{eq:varthetahat} because they contain more people.\footnote{Continuing from the example in the preceding footnote: suppose we randomly sample $10$ groups and observe \emph{everyone} in them. Then, on average, our sample will contain 5 small groups and 5 large groups. While the total sample size is random, on average we will observe 100 people, of whom 25 come from small groups and the rest from large groups, matching the shares of each kind of person in the population.} 
The question is merely how to represent this mathematically.
Let $\rho_g \equiv N_g / \mathbbm{E}(N_g)$ denote the \emph{relative size} of group $g$.
We write $\mathbbm{E}[Y_{ig}]$ to denote the average that gives equal weight to groups---choosing one person at random from a randomly-chosen group---and $\mathbbm{E}[\rho_g Y_{ig}]$ to denote the average that gives equal weight to individuals---observing an entire group chosen at random.
It is the latter expectation that appears in our results below, as it denotes the population equivalent of the double sums from \eqref{eq:varthetahat}.
While this is a slight abuse of notation, expectations from \autoref{sec:identification} that involve individual-level random variables but do not condition on group size should be interpreted as (implicitly) weighting by relative group size.
Using the notation and sampling scheme defined above, we now state high-level sufficient conditions for the consistency of $\widehat{\boldsymbol{\vartheta}}$ in \eqref{eq:varthetahat}.

\begin{thm}
  \label{thm:consistency}
  Let $\rho_g \equiv N_g / \mathbbm{E}(N_g)$ and suppose that
  \mbox{}
  \begin{enumerate}[(i)]
    \item we observe a random sample of $G$ groups, where observations within a given group are identically distributed although not necessarily independent,
    \item $Y_{ig} = \boldsymbol{X}_{ig}' \boldsymbol{\vartheta} + U_{ig}$ for $1 \leq g \leq G$, $1 \leq i \leq N_g$,
    \item $\mathbbm{E}\left( \rho_g \boldsymbol{\mathcal{Z}}_{ig} U_{ig} \right) = \mathbf{0}$ and $\mathbbm{E}\left( \rho_g \boldsymbol{\mathcal{Z}}_{ig} \boldsymbol{X}_{ig}' \right) = \mathbbm{I}$, 
    \item $\mathbbm{E}\left[ \rho_g^2 \lvert| \boldsymbol{\mathcal{Z}}_{ig} \boldsymbol{X}_{ig}' \rvert|^2 \right] = o(G)$,
    \item $\mathbbm{E}\left[ \rho_g^{2} \lvert| \boldsymbol{\mathcal{Z}}_{ig} U_{ig} \rvert|^{2} \right] = o(G)$,
    \item $\lvert| \sum_{g=1}^G \frac{1}{N_g}\sum_{i=1}^{N_g} \rho_g ( \widehat{\boldsymbol{\mathcal{Z}}}_{ig} - \boldsymbol{\mathcal{Z}}_{ig} ) \boldsymbol{X}_{ig}' \rvert| = o_{\mathbbm{P}}(G)$, and
    \item $\lvert| \sum_{g=1}^G \frac{1}{N_g}\sum_{i=1}^{N_g} \rho_g ( \widehat{\boldsymbol{\mathcal{Z}}}_{ig} - \boldsymbol{\mathcal{Z}}_{ig} ) U_{ig} \rvert| = o_{\mathbbm{P}}(G)$.
  \end{enumerate}
  Then $\widehat{\boldsymbol{\vartheta}}$, defined in \eqref{eq:varthetahat}, is consistent for $\boldsymbol{\vartheta}$ as $G \rightarrow \infty$.
\end{thm}


Condition (i) of \autoref{thm:consistency} simply restates our group sampling assumption.
Conditions (ii) and (iii) hold under the assumptions of \autoref{thm:expect}, as shown in the proof of that result: for each average effect $\boldsymbol{\vartheta}$ from the theorem, we can define an appropriate error term $U_{ig}$, vector of regressors $\boldsymbol{X}_{ig}$, and vector of instruments $\boldsymbol{\mathcal{Z}}_{ig}$ such that $Y_{ig} = \boldsymbol{X}_{ig}' \boldsymbol{\vartheta} + U_{ig}$ where $\boldsymbol{\mathcal{Z}}_{ig}$ is an exogenous and relevant instrument.
Moreover, for each part of \autoref{thm:expect}, $\mathbbm{E}(\rho_g \boldsymbol{\mathcal{Z}}_{ig} \boldsymbol{X}_{ig}')$ equals the identity matrix.\footnote{For effects that condition on $C_{ig}=c$, e.g.\ those from parts (ii) and (iii) of \autoref{thm:expect}, the appropriate definition of $\rho_g$ becomes $N_g \mathbbm{E}[\mathbbm{1}(C_{ig}=c)]/\mathbbm{E}[N_g \mathbbm{1}(C_{ig} = c)]$.}\textsuperscript{,}\footnote{Given that $\mathbbm{E}(\rho_g \boldsymbol{\mathcal{Z}}_{ig} \boldsymbol{X}_{ig}') = \mathbbm{I}$, we could have defined our estimator to be $\frac{1}{N} \sum_{g=1}^G \sum_{i=1}^{N_g} \widehat{\boldsymbol{\mathcal{Z}}}_{ig} Y_{ig}$ rather than $\widehat{\boldsymbol{\vartheta}}$. It is more convenient both for our asymptotic derivations and practical implementation, however, to work with an IV estimator.}
Conditions (iv) and (v) of \autoref{thm:consistency} would be implied by requiring that the second moments of $\rho_g \boldsymbol{\mathcal{Z}}_{ig} \mathbf{X}_{ig}'$ and $\rho_g \boldsymbol{\mathcal{Z}}_{ig} U_{ig}$ exist and are bounded.
We choose to state these conditions in a slightly weaker form because the distribution of $\rho_g$ necessarily changes with $G$ if we consider an asymptotic sequence in which the minimum group size $\underline{n}$ increases with the number of groups, as we will assume below.
Requiring the relevant expectations to be $o(G)$ in principle allows the variance of relative group size $\rho_g$ to grow along with the number of groups, provided that it does not grow too quickly. 
Conditions (i)--(v) together are sufficient for the consistency of 
\begin{equation}
  \label{eq:varthetatilde}
  \widetilde{\boldsymbol{\vartheta}} \equiv \left(\sum_{g=1}^G \sum_{i=1}^{N_g} \boldsymbol{\mathcal{Z}}_{ig} \boldsymbol{X}_{ig}' \right)^{-1}\left(\sum_{g=1}^G \sum_{i=1}^{N_g} \boldsymbol{\mathcal{Z}}_{ig} Y_{ig}\right),
\end{equation}
an infeasible estimator that uses the true instrument vector $\boldsymbol{\mathcal{Z}}_{ig}$ instead of its estimate $\widehat{\boldsymbol{\mathcal{Z}}}_{ig}$.
The final two conditions of \autoref{thm:consistency} assume that $\widehat{\boldsymbol{\mathcal{Z}}}_{ig}$ is a sufficiently accurate estimator of $\boldsymbol{\mathcal{Z}}_{ig}$ to ensure that $\widehat{\boldsymbol{\vartheta}} = \widetilde{\boldsymbol{\vartheta}} + o_{\mathbbm{P}}(1)$. 
In the setting we consider here, this will require a condition on how quickly the minimum group size $\underline{n}$ grows relative to $G$, as we discuss in detail below.
Strengthening conditions (v) and (vii) and adding one further assumption implies that $\widehat{\boldsymbol{\vartheta}}$ is asymptotically normal.


\begin{thm}
  \label{thm:asymptoticnormality}
Suppose that
  \begin{enumerate}[(i)]
    \item $\text{Var}\left( \frac{1}{N_g}\sum_{i=1}^{N_g} \rho_g \boldsymbol{\mathcal{Z}}_{ig} U_{ig} \right) \rightarrow \boldsymbol{\Sigma}$ as $G \rightarrow \infty$, 
    \item $\mathbbm{E}\left[ \rho_g^{2 + \delta} \lvert| \boldsymbol{\mathcal{Z}}_{ig} U_{ig} \rvert|^{2 + \delta} \right] = o(G^{\delta/2})$ for some $\delta > 0$, and
    \item $\lvert| \sum_{g=1}^G \frac{1}{N_g}\sum_{i=1}^{N_g} \rho_g ( \widehat{\boldsymbol{\mathcal{Z}}}_{ig} - \boldsymbol{\mathcal{Z}}_{ig} ) U_{ig} \rvert| = o_{\mathbbm{P}}(G^{1/2})$.
  \end{enumerate}
  Then, under the conditions of \autoref{thm:consistency}, $\sqrt{G}(\widehat{\boldsymbol{\vartheta}} - \boldsymbol{\vartheta}) \rightarrow_d N(\mathbf{0}, \boldsymbol{\Sigma})$.
\end{thm}


Combined with the first four conditions of \autoref{thm:consistency}, (i) and (ii) from \autoref{thm:asymptoticnormality} are sufficient for the asymptotic normality of $\widetilde{\boldsymbol{\vartheta}}$, the infeasible estimator defined in \eqref{eq:varthetatilde}. 
Condition (i) implies that the rate of convergence of $\widetilde{\boldsymbol{\vartheta}}$ is $G^{-1/2}$.
Obtaining a rate of convergence that depends on the total number of \emph{individuals} rather than groups in the sample would require assumptions that are implausible in typical applications of the randomized saturation design.\footnote{Obtaining the faster rate of convergence would require $\text{Var}\left( \frac{1}{N_g} \sum_{i=1}^{N_g} \rho_g \boldsymbol{\mathcal{Z}}_{ig} U_{ig} \right) \rightarrow \mathbf{0}$ as $G \rightarrow \infty$. Because we consider an asymptotic sequence in which the minimum group size grows with $G$, this is technically possible. It would, however, require us to assume that both heterogeneity between groups and dependence within groups vanish in the limit.}
Conditions (ii) and (iii) strengthen (v) and (vii), respectively, from \autoref{thm:consistency}: (ii) is sufficient for the Lindeberg condition, which we use to establish a central limit theorem, while (iii) ensures that the limit distribution of the feasible estimator $\widehat{\boldsymbol{\vartheta}}$ coincides with that of the infeasible estimator $\widetilde{\boldsymbol{\vartheta}}$.

Conditions (vi)--(vii) of \autoref{thm:consistency}, along with condition (iii) of \autoref{thm:asymptoticnormality}, require the difference $(\widehat{\boldsymbol{\mathcal{Z}}}_{ig} - \boldsymbol{\mathcal{Z}}_{ig})$ to be sufficiently small on average that the limiting behavior of $\widehat{\boldsymbol{\vartheta}}$ coincides with that of the infeasible estimator.
We now provide low-level sufficient conditions for this to obtain.
By definition, 
\begin{equation}
  \label{eq:ZtoR}
  \widehat{\boldsymbol{\mathcal{Z}}}_{ig} - \boldsymbol{\mathcal{Z}}_{ig} = \left[\mathbf{R}(\widehat{C}_{ig}, N_g)^{+} -  \mathbf{R}(\bar{C}_{ig}, N_g)^{-1} \right]\boldsymbol{W}_{ig}.
\end{equation}
Accordingly, so long as $\mathbf{R}$ is a sufficiently well-behaved function, $(\widehat{\boldsymbol{\mathcal{Z}}}_{ig} - \boldsymbol{\mathcal{Z}}_{ig})$ will be small if $|\widehat{C}_{ig} - \bar{C}_{ig}|$ is.
As shown in the following lemma, a sufficient condition for this difference to vanish \emph{uniformly} over $(i,g)$ is for the minimum group size $\underline{n}$ to be large relative to $\log G$.

\begin{lem}
  \label{lem:Chat}
  Suppose that $0 < \underline{s} \leq S_g$ and $\underline{n} \leq N_{g}$.
  Under Assumptions \ref{assump:saturations}--\ref{assump:Bernoulli} and \ref{assump:onesided}--\ref{assump:exclusion} 
\[
  \max_{1 \leq g \leq G} \left(\max_{1\leq i \leq N_g} \left| \widehat{C}_{ig} - \bar{C}_{ig}\right|\right) = O_{\mathbbm{P}}\left( \sqrt{\frac{\log G}{\underline{n}}} \right) \text{as } (\underline{n}, G) \rightarrow \infty.
\]
\end{lem}

The following regularity conditions are sufficient for $\mathbf{R}(\widehat{C}_{ig}, N_g)^+ - \mathbf{R}(\bar{C}_{ig}, N_g)^{-1}$ to inherit the asymptotic behavior of $(\widehat{C}_{ig} - \bar{C}_{ig})$.

  \begin{assump}[Regularity Conditions for $\mathbf{R}$]
  \label{assump:R} 
  \mbox{}
  \begin{enumerate}[(i)] 
    \item $\mathbf{R}(\bar{c}, n)$ is well-defined and symmetric for all $\bar{c} \in [\bar{c}_L/2, 1), \, n \geq \underline{n}$ where $0 < \bar{c}_L \leq \bar{C}_{ig}$;
    \item $\displaystyle\inf_{\bar{c} \geq \bar{c}_L/2,\, n \geq \underline{n}} \sigma\left( \mathbf{R}(\bar{c}, n) \right) > \underline{\sigma} > 0$, where $\sigma(\mathbf{M})$ denotes the minimum eigenvalue of $\mathbf{M}$;
    \item $\lvert|\mathbf{R}(\bar{c}_1, n) - \mathbf{R}(\bar{c}_2, n) \rvert| \leq L \left\{ \left| \bar{c}_1 - \bar{c}_2 \right| + O(n^{-1/2}) \right\}$ as $n \rightarrow \infty$ for some $0 < L <\infty$.
  \end{enumerate}
\end{assump}

Parts (i) and (ii) of \autoref{assump:R} require that $\mathbf{R}$ is well-defined and uniformly invertible over a range of values for $\bar{c}$ that includes the support of $\bar{C}_{ig}$ and excludes zero.
Part (iii) is a variant of Lipschitz continuity that holds in the limit as $n$ grows. 
These conditions are mild: they amount to a slight strengthening of the rank condition from \autoref{assump:rank}.
In the linear basis function example from \eqref{eq:Q0LinearBernoulli} and \eqref{eq:Q1LinearBernoulli}, for instance, \autoref{assump:R} holds whenever $\bar{C}_{ig}$ is bounded away from zero and $S_g$ takes on at least two distinct values between zero and one.\footnote{See the discussion in \autoref{sec:identification} immediately following \eqref{eq:Q0LinearBernoulli} for details.} 
More generally, provided that \autoref{assump:rank} holds, whenever $\bar{C}_{ig}$ is bounded away from zero and the basis functions $\mathbf{f}$ are well-behaved, we can always \emph{extend} the definitions of $\mathbf{Q}_0, \mathbf{Q}_1$ from \eqref{eq:Q0def}--\eqref{eq:Q1def} to ensure that \autoref{assump:R} holds.
See \autoref{sec:Qextend} for full details.
Under this assumption, we can derive sufficient conditions on the rates at which $G$ and $\underline{n}$ approach infinity to ensure that the difference between $\widehat{\boldsymbol{\mathcal{Z}}}_{ig}$ and $\boldsymbol{\mathcal{Z}}_{ig}$ is negligible.

\begin{thm}
  \label{thm:R}
  Suppose that $\mathbbm{E}\left[\rho_g^2 \lvert|\boldsymbol{W}_{ig} \boldsymbol{X}_{ig}' \rvert|^2  \right]$ and $\mathbbm{E}\left[\rho_g^2 \lvert|\boldsymbol{W}_{ig} U_{ig} \rvert|^2  \right]$ are both $o(G)$. 
  Then, under condition (i) of \autoref{thm:consistency} and the conditions of \autoref{lem:Chat}, 
  \begin{enumerate}[(i)]
    \item $\log G / \underline{n} \rightarrow 0$ is sufficient for conditions (vi)--(vii) of \autoref{thm:consistency}. 
    \item $G \log G/\underline{n} \rightarrow 0$ is sufficient for condition (iii) of \autoref{thm:asymptoticnormality}.
  \end{enumerate}
\end{thm}

Taken together, Theorems \ref{thm:consistency}--\ref{thm:R} establish that $\widehat{\boldsymbol{\vartheta}}$ from \eqref{eq:varthetahat} is consistent, and asymptotically normal in the limit as $G$ and $\underline{n}$ grow at an appropriate rate. 
In practical terms, our estimators are appropriate for settings with many large groups such as the experiment of \cite{crepon2013}.
To implement them in practice, all that is required is to calculate the estimated instrument $\widehat{\boldsymbol{\mathcal{Z}}}_{ig}$ and then run the appropriate just-identified IV regression from \autoref{tab:estimators} with standard errors clustered by group.
\autoref{sec:StepByStep} provides full implementation details for the linear outcome model from \eqref{eq:LinearModel}.

\section{Application: Job Placement Program in the French Labor Market}
\label{sec:application}

\noindent In this section we illustrate our methods using data from \cite{crepon2013}, who implemented a large-scale randomized saturation experiment across French cities, offering job placement program services to young workers seeking employment.
In doing so, we uncover patterns of spillovers that could prove relevant for the design of similar labor market programs.
The intervention included 235 cities (labor markets), covering a sample of \num[group-separator={,}]{21431} workers of whom \num[group-separator={,}]{11806} were unemployed at the time of randomization.\footnote{The formal criteria for eligibility included ``aged below 30, with at least a two-year college degree, and having spent either 12 out of the last 18 months or 6 months continuously unemployed or underemployed'' \cite[p. 545]{crepon2013}.} 
Two questions of interest arise in this setting.
First, the presence of direct effects: whether receiving job placement services impacts subsequent labor market outcomes of participants, in particular the likelihood of being employed.
Second, the presence of indirect (spillover) effects: whether the receipt of job placement services by others in the same labor market impacts subsequent labor market outcomes of participants.
For example, in such a large-scale experiment one may worry that increasing some workers' likelihood of obtaining a job may hurt the labor market prospects of other workers.

Cities were initially randomly assigned to five saturation bins $\mathcal{S} = \{0, 0.25, 0.5, 0.75, 1 \}$. 
For reasons outside of the experiment, 43 of the 47 cities initially assigned to the $25\%$ saturation bin in fact received a $50\%$ saturation, and 12 of the 47 cities initially assigned to the $75\%$ saturation bin received a $100\%$ saturation.\footnote{The reassignment of cities across bins is not a problem for the analysis in \cite{crepon2013}, because the main results in that study make only a binary comparison between the cities assigned to the $0\%$ saturation bin and the pooled group of cities assigned to positive saturation bins.}
For this reason all of the results we present below restrict attention to the subset of cities that received their initially assigned saturation.\footnote{Naturally, the validity of this restriction relies on the assumption that the reassignment of cities across saturation bins was unrelated to their underlying characteristics.}
Thus, our estimation sample consists of 47 cities in the $0\%$ saturation bin, 4 cities in the $25\%$ saturation bin, 47 cities in the $50\%$ saturation bin, 35 cities in the $75\%$ saturation bin, and 47 cities in the $100\%$ saturation bin.

Eligible workers in each city then received offers with a probability equal to the saturation assigned to their city.
As mentioned in the introduction, the overall take-up rate of job placement services was 35\%.
Only workers who were assigned to treatment could receive it, so \autoref{assump:onesided} (one-sided non-compliance) holds. 
In addition, \autoref{assump:IOR} (IOR) is reasonable in this setting: using a simple regression-based test, \autoref{sec:IOR} shows that an individual's probability of treatment take-up is statistically unrelated to her group's randomly assigned saturation.\footnote{As far as we are aware, subjects in the experiment of \cite{crepon2013} were not informed of their groups' saturations, making IOR \emph{a priori} plausible as well.}
Researchers collected data on labor market outcomes in a follow-up 8 months after treatment receipt.
Here we present results for two outcome variables: long-term employment (indefinite contract or fixed-term contract longer than 6 months) and any employment.
We estimate a linear outcome model, $\mathbf{f}(\bar{d}) = (1, \bar{d})$, so that
\begin{equation}
\label{eq:Y}
Y_{ig} = \alpha_{ig} + \beta_{ig} D_{ig} + \gamma_{ig} \bar{D}_{ig} + \delta_{ig} D_{ig} \bar{D}_{ig} \text{.}
\end{equation}
The linear specification is simple to implement and easy to interpret.
For full implementation details, see Appendices \ref{sec:StepByStep} and \ref{sec:0percent}.
As discussed in Appendix \ref{sec:BasisFunctionsRank}, the number of basis functions that can be included in practice is limited by the number of saturations.
Given the re-assignment of 25\% saturations in this experiment, we ``effectively'' have only three interior saturations, one more than the minimum needed to identify a linear outcome model.
(See \autoref{subsec:identification_example}.)
Given the limits imposed by the design, we limit attention to the linear model throughout this section. 
Our simulation study, discussed in \autoref{sec:simulations} suggests that the sample size of \cite{crepon2013} is sufficient to permit reasonably precise estimation of the linear specification. 

Recall that our RS-IV estimator recovers average coefficients for compliers $(\alpha^c,\beta^c,\gamma^c,\delta^c)$, for never-takers $(\alpha^n,\gamma^n)$, and for the whole population $(\alpha,\gamma)$.\footnote{In the more general notation in section  \autoref{sec:notationAssumptions}, $\boldsymbol{\theta}_{ig}=(\alpha_{ig},\gamma_{ig})$ and $(\boldsymbol{\psi}_{ig} - \boldsymbol{\theta}_{ig}) = (\beta_{ig},\delta_{ig})$.}
Using these, we can reconstruct the average potential outcome functions for treated and untreated compliers, for untreated never-takers, and for the whole population.\footnote{Because non-compliance is one-sided, compliers are synonymous with ``the treated'' and never-takers with ``the untreated.''}

\begin{table}
\centering
\begin{tabular}{lcccccccc}
  \hline
  & $\alpha$ & $\gamma$ & $\alpha^n$ & $\gamma^n$ & $\alpha^c$ & $\gamma^c$ & $\beta^c$ & $\delta^c$ \\ 
  \hline
 &  &  &  &  &  &  &  &  \\[-0.2cm]
  \textit{Outcome: long-term employment} &  &  &  &  &  &  &  &  \\ 
  \ Estimate & 0.47 & -0.09 & 0.47 & 0.14 & 0.48 & -0.51 & -0.09 & 0.62 \\ 
  \ Std. error & 0.01 & 0.07 & 0.02 & 0.09 & 0.04 & 0.24 & 0.05 & 0.25 \\ 
   &  &  &  &  &  &  &  &  \\[-0.2cm]
  \textit{Outcome: any employment} &  &  &  &  &  &  &  &  \\ 
  \ Estimate & 0.60 & -0.11 & 0.57 & 0.14 & 0.66 & -0.56 & -0.10 & 0.62 \\ 
  \ Std. error & 0.01 & 0.06 & 0.02 & 0.09 & 0.04 & 0.24 & 0.05 & 0.25 \\ 
   \cmidrule(lr){2-3}\cmidrule(lr){4-5}\cmidrule(lr){6-9}
  Observations & \multicolumn{2}{c}{7,440} & \multicolumn{2}{c}{5,814}  & \multicolumn{4}{c}{3,104}  \\ 
   \hline
\end{tabular}
\caption{Estimated coefficients for long-term employment and any employment. Standard errors are clustered at the city level. See \autoref{eq:Y} for the coefficient definitions. \label{tab:creponresults}}
\end{table}

\autoref{tab:creponresults} presents estimates and standard errors (clustered at the city level) of the average effects for the whole population, for never-takers, and for compliers using long-term employment and any employment as outcome variables.\footnote{We include observations from the 0\% saturation cities as described in \autoref{sec:0percent}.}
We estimate large negative spillovers ($\gamma^c = -0.51$) for untreated compliers, and effectively no spillovers ($\gamma^c + \delta^c = 0.62 - 0.51 = 0.11$) for treated compliers.
For the average untreated complier, increasing the treated share among his neighbors from 10 percent to 50 percent would decrease his likelihood of employment by 20 percentage points.
This is a considerable negative indirect effect of the policy intervention. 
However, this negative spillover effect is nullified --and possibly reversed-- when compliers are assigned to, and therefore receive, the treatment.

\begin{figure}
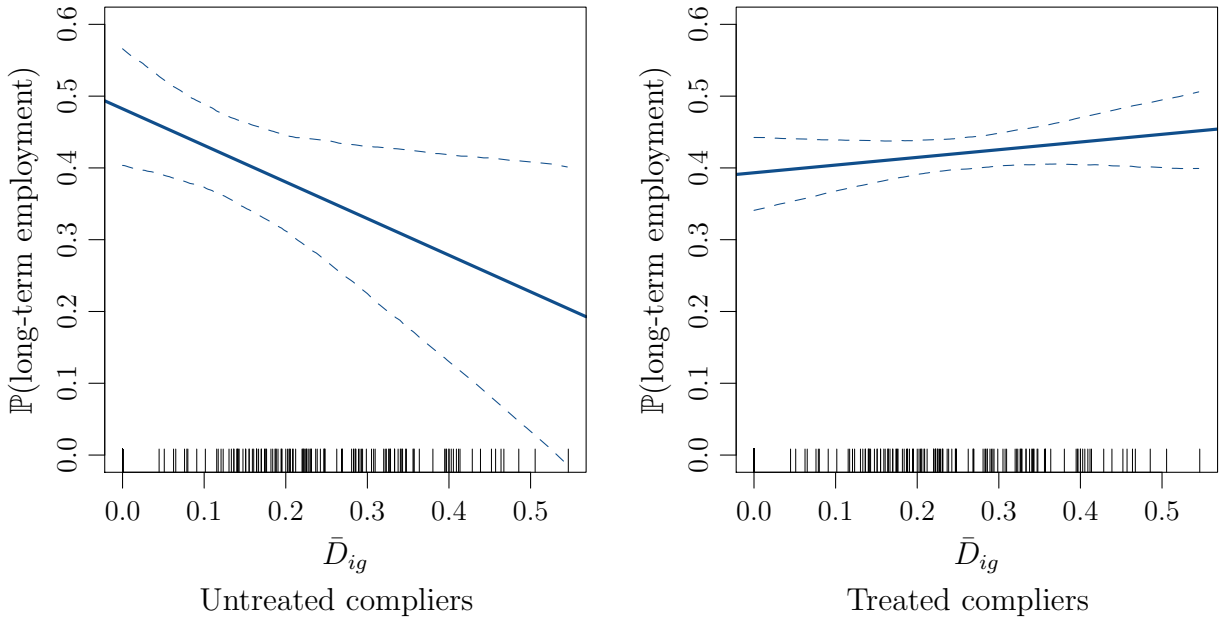

	\vspace{-1.3cm}
	\hspace{-0.4cm}
	\begin{subfigure}[t]{0.5\textwidth}
		{\input{./R/figures/crepon/FigPart1_y3_def2_CI95.tex}}
		\vspace{-0.7cm} \caption{\normalsize \hspace{1.4cm} Untreated compliers} 
		\label{fig:pot_outcomes1a}
	\end{subfigure}
	\begin{subfigure}[t]{0.5\textwidth}
		{\input{./R/figures/crepon/FigPart2_y3_def2_CI95.tex}}
		\vspace{-0.7cm} \caption{\normalsize \hspace{1.4cm} Treated compliers} 
		\label{fig:pot_outcomes1b}
	\end{subfigure}
	\caption{Potential outcomes as a function of $\bar{d}_{ig}$ using the probability of long-term employment as outcome. The left-hand side panel illustrates the average potential outcome function for untreated compliers: $\alpha^c + \gamma^c \bar{d}_{ig}$. The right-hand side panel illustrates the average potential outcome function for treated compliers: $(\alpha^c+\beta^c)+(\gamma^c+\delta^c)\bar{d}_{ig}$. The dashed curves represent 95\% confidence intervals. Each tick in the rug plot on the horizontal axis represents a realized value of $\bar{D}_{g}$ in a city in the experiment. \label{fig:pot_outcomes1}}
\end{figure}

\begin{figure}
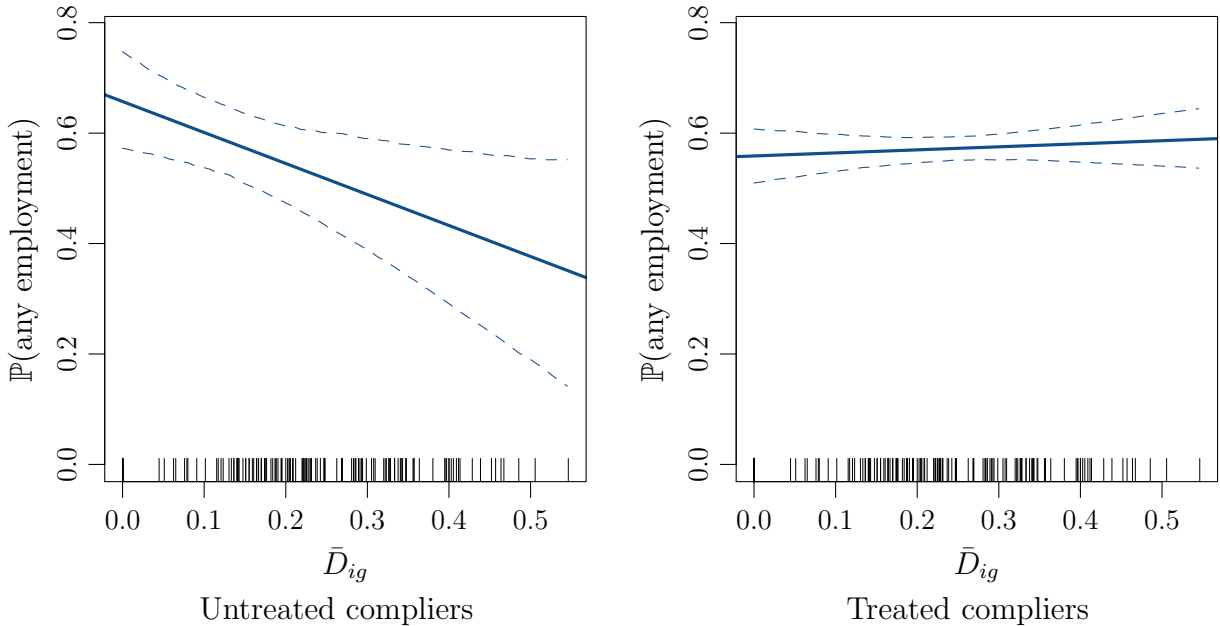

	\vspace{-0.5cm}
	\hspace{-0.4cm}
	\begin{subfigure}[b]{0.5\textwidth}
		{\input{./R/figures/crepon/FigPart1_y1_def2_CI95.tex}}
		\vspace{-0.7cm} \caption{\normalsize \hspace{1.4cm} Untreated compliers} 
		\label{fig:pot_outcomes2a}
	\end{subfigure}
	\begin{subfigure}[b]{0.5\textwidth}
		{\input{./R/figures/crepon/FigPart2_y1_def2_CI95.tex}}
		\vspace{-0.7cm} \caption{\normalsize \hspace{1.4cm} Treated compliers} 
		\label{fig:pot_outcomes2b}
	\end{subfigure}
	\caption{Potential outcomes as a function of $\bar{d}_{ig}$ using the probability of any employment as outcome. The left-hand side panel illustrates the average potential outcome function for untreated compliers: $\alpha^c + \gamma^c \bar{d}_{ig}$. The right-hand side panel illustrates the average potential outcome function for treated compliers: $(\alpha^c+\beta^c)+(\gamma^c+\delta^c)\bar{d}_{ig}$. The dashed curves represent 95\% confidence intervals. Each tick in the rug plot on the horizontal axis represents a realized value of $\bar{D}_{g}$ in a city in the experiment. \label{fig:pot_outcomes2}}
\end{figure}

For completeness, \autoref{fig:pot_outcomes1} depicts the implied average potential outcome functions for untreated and treated compliers, using long-term employment as the outcome variable.
\autoref{fig:pot_outcomes2} depicts the corresponding functions using any employment as the outcome variable instead.
We report average functions as bold lines, and corresponding (pointwise) 95\% confidence intervals as dashed curves.
The downward sloping functions on the left of both figures illustrate the negative estimated spillover for untreated compliers: employment prospects for those who would have taken up treatment if offered worsen rapidly as more job seekers in their city take up the job placement program.
The flat curves on the right, in contrast, reveal that employment prospects for those who take up treatment are unaffected by the average city-level treatment take up.
These patterns are consistent with the idea that compliers who did not receive job placement assistance are hurt by competition in the labor market, while job placement assistance shields those who take it up from these negative spillovers.

\begin{figure}
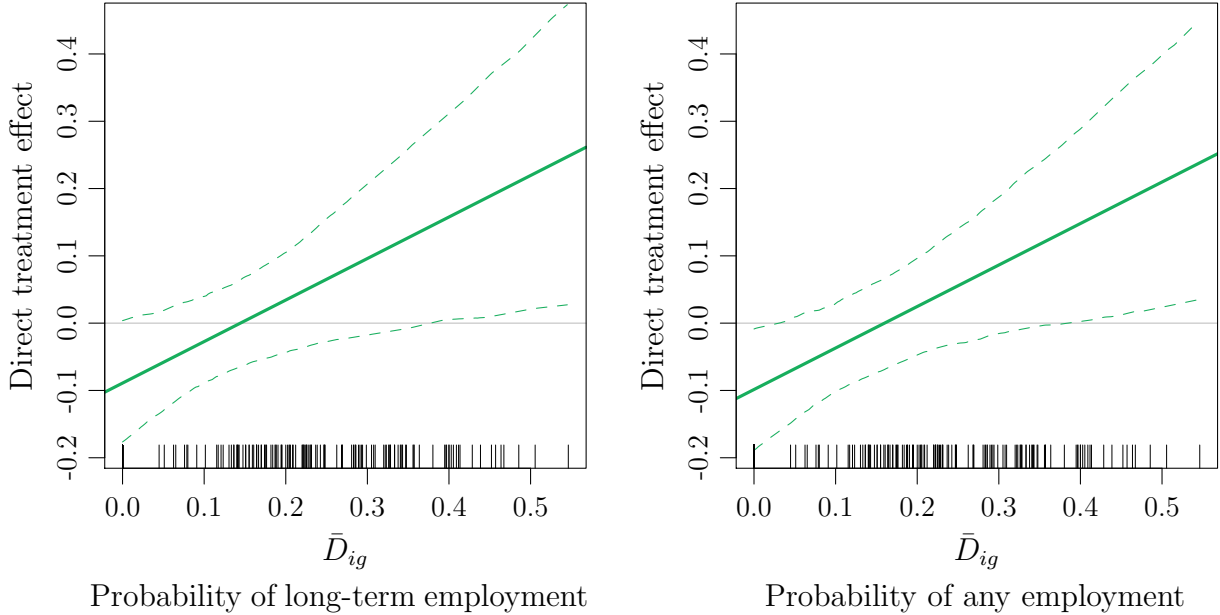

	\vspace{-0.5cm}
	\hspace{-0.4cm}
	\begin{subfigure}[b]{0.5\textwidth}
		{\input{./R/figures/crepon/Fig_TE_y3_def2_CI95.tex}}
		\vspace{-0.7cm} \caption{\normalsize \hspace{1.5cm} Probability of long-term employment} 
		\label{fig:creponTEsa}
	\end{subfigure}
	\begin{subfigure}[b]{0.5\textwidth}
		{\input{./R/figures/crepon/Fig_TE_y1_def2_CI95.tex}}
		\vspace{-0.7cm} \caption{\normalsize \hspace{1.6cm} Probability of any employment} 
		\label{fig:creponTEsb}
	\end{subfigure}
	\caption{Direct treatment effect as a function of $\bar{d}_{ig}$ for compliers: $\beta^c + \delta^c \bar{d}_{ig}$. The left-hand side figure uses long-term employment as outcome variable. The right-hand side figure uses any employment as outcome. The dashed curves represent 95\% confidence intervals. Each tick in the rug plot on the horizontal axis represents a realized value of $\bar{D}_{g}$ in a city in the experiment. \label{fig:creponTEs}}
\end{figure}

Thus, among those willing to receive job placement services, more widespread take-up of the program, possibly via increased labor market competition, has a differential impact across those who do receive and those who do not receive treatment.
This difference is driven by the direct treatment effects on compliers, which we plot in \autoref{fig:creponTEs}. 
The estimated direct effect increases with $\bar{D}_{ig}$ and is positive for most values of $\bar{D}_{ig}$ observed in the data, although the 95\% confidence interval contains an effect size of zero for most observations.
Finally, although we cannot recover full treatment effects for never-takers or for the population as a whole, \autoref{tab:creponresults} also illustrates that the average spillover $\gamma^n$ for never takers is positive albeit statistically insignificant.
The resulting average spillover for the population as a whole, $\gamma$, although much smaller in magnitude compared to the one for compliers, is negative and marginally significant for any employment ($\gamma = -0.11$).

In settings with potential non-compliance such as this one, participants' take-up decisions may be driven by the expected gains from participation.
Our findings are consistent with such behavior: those who decline participation may do so precisely if they expect they will not suffer negative spillovers from others receiving the program.
In turn, compliance may in part be driven by the knowledge that, in the absence of treatment, program receipt by others hurts own labor market prospects.\footnote{Notice that these forms of `selection on gains' are compatible with the IOR assumption holding.}
Indeed, in \autoref{tab:lpm_compliers} we report results from a regression of compliance indicators on pre-treatment characteristics for the sub-sample of offered individuals.
Compared to never-takers, compliers appear to be a more vulnerable sub-population: at baseline they are less likely to cohabit, less educated, less likely to be employed or to have a stable labor contract, and are more likely to receive unemployment insurance.\footnote{Compliers are also less likely to have young children, which may indicate that never takers are less able to participate in the program and possibly in the labor market.}
Knowledge of this pattern of effects may prove valuable for the design of other similar large-scale labor market programs.

\section{Simulation study}
\label{sec:simulations}
We now present the results of a simulation study to demonstrate the performance of our estimator in a setting similar to that of our empirical example.
As in \autoref{sec:application}, we work with the linear outcome model $\mathbf{f}(x)' = (1, x)'$ from (\ref{eq:LinearModel}); for full implementation details, see \autoref{sec:StepByStep}.
We compare the results of our estimator to those of a `na\"{i}ve' IV regression of $Y_{ig}$ on  $\mathbf{X}_{ig} \equiv (1, D_{ig}, \bar{D}_{ig}, D_{ig} \bar{D}_{ig})'$ with instruments $\boldsymbol{\mathcal{Z}}_{ig} \equiv (1, Z_{ig}, S_g, Z_{ig}S_g)'$. 
As detailed in \autoref{lem:linear}, this estimator yields consistent estimates of $\alpha$ and $\beta^c$, but inconsistent estimates of $\gamma$ and $\delta^c$ when the random coefficients are correlated with the share of compliers.

Our simulation design broadly follows the sampling and experimental design of \cite{crepon2013}, employing a simple data generating process that allows for correlation between the random coefficients and the share of compliers in a city, $\bar{C}_{ig}$. 
We present results from three simulation studies with different numbers of groups, $G$. 
Our main simulations set $G=235$ to match the experimental design in \cite{crepon2013}; comparison exercises 150 and 500 groups. 
For simplicity we consider groups of equal size, 116 individuals each, to match the average group size from \cite{crepon2013}.
We randomly assign exactly $1/5$ of groups to each of five saturations, $S_g \in \{0, 0.25, 0.5, 0.75, 1\}$, then draw individual Bernoulli offers at the assigned saturation.

\begin{table}[htbp!]
\begin{center}
\begin{tabular}{lcccccccc}
  \hline
  & $\alpha$ & $\gamma$ & $\alpha^n$ & $\gamma^n$ & $\alpha^c$ & $\gamma^c$ & $\beta^c$ & $\delta^c$ \\
  \hline
  &  &  &  &  &  &  &  &  \\[-0.3cm]
  True values & 0.50 & -0.70 & 0.50 & -0.73 & 0.50 & -0.63 & 0.20 & 0.94 \\[0.2cm]
  \hline
  &  &  &  &  &  &  &  &  \\[-0.3cm]
  \textbf{150 groups} &  &  &  &  &  &  &  &  \\[0.1cm] 
  \textit{RS-IV} &  &  &  &  &  &  &  &  \\ 
  \ Average coefficient & 0.50 & -0.69 & 0.50 & -0.73 & 0.50 & -0.59 & 0.21 & 0.89 \\ 
  \ Std. dev. & 0.00 & 0.08 & 0.01 & 0.10 & 0.04 & 0.36 & 0.07 & 0.44 \\ 
  \ Coverage & 0.97 & 0.95 & 0.91 & 0.91 & 0.98 & 0.97 & 0.96 & 0.96 \\ 
   &  &  &  &  &  &  &  &  \\ 
  \textit{Na\"{i}ve IV} &  &  &  &  &  &  &  &  \\ 
  \ Average coefficient & 0.50 & -0.63 &  &  &  &  & 0.21 & 1.02 \\ 
  \ Std. dev. & 0.00 & 0.05 &  &  &  &  & 0.06 & 0.29 \\ 
  \ Coverage & 0.97 & 0.65 &  &  &  &  & 0.95 & 0.91 \\ 
   &  &  &  &  &  &  &  &  \\[-0.3cm] 
   \hline
   &  &  &  &  &  &  &  &  \\[-0.3cm]
   \textbf{235 groups} &  &  &  &  &  &  &  &  \\[0.1cm] 
   \textit{RS-IV} &  &  &  &  &  &  &  &  \\ 
  \ Average coefficient & 0.50 & -0.69 & 0.50 & -0.73 & 0.50 & -0.60 & 0.20 & 0.91 \\ 
  \ Std. dev. & 0.00 & 0.06 & 0.01 & 0.08 & 0.03 & 0.28 & 0.05 & 0.34 \\ 
  \ Coverage & 0.97 & 0.94 & 0.91 & 0.92 & 0.98 & 0.97 & 0.96 & 0.96 \\ 
   &  &  &  &  &  &  &  &  \\ 
   \textit{Na\"{i}ve IV} &  &  &  &  &  &  &  &  \\ 
  \ Average coefficient & 0.50 & -0.63 &  &  &  &  & 0.20 & 1.03 \\ 
  \ Std. dev. & 0.00 & 0.04 &  &  &  &  & 0.05 & 0.22 \\ 
  \ Coverage & 0.97 & 0.50 &  &  &  &  & 0.95 & 0.90 \\ 
   &  &  &  &  &  &  &  &  \\[-0.3cm]
   \hline
   &  &  &  &  &  &  &  &  \\[-0.3cm]
   \textbf{500 groups} &  &  &  &  &  &  &  &  \\[0.1cm] 
   \textit{RS-IV} &  &  &  &  &  &  &  &  \\ 
  \ Average coefficient & 0.50 & -0.69 & 0.50 & -0.73 & 0.50 & -0.60 & 0.20 & 0.91 \\ 
  \ Std. dev. & 0.00 & 0.04 & 0.01 & 0.05 & 0.02 & 0.19 & 0.04 & 0.23 \\ 
  \ Coverage & 0.97 & 0.95 & 0.91 & 0.91 & 0.98 & 0.97 & 0.96 & 0.95 \\ 
   &  &  &  &  &  &  &  &  \\ 
   \textit{Na\"{i}ve IV} &  &  &  &  &  &  &  &  \\ 
  \ Average coefficient & 0.50 & -0.63 &  &  &  &  & 0.20 & 1.04 \\ 
  \ Std. dev. & 0.00 & 0.02 &  &  &  &  & 0.03 & 0.15 \\ 
  \ Coverage & 0.97 & 0.20 &  &  &  &  & 0.95 & 0.87 \\ 
   &  &  &  &  &  &  &  &  \\[-0.3cm] 
   \hline
\end{tabular}
\end{center}
\vspace{-0.3cm}
\caption{Comparison of our RS-IV and the `na\"{i}ve' IV in simulations with 150, 235 or 500 groups. We show the mean, standard deviation, and coverage for estimates using the RS-IV and `na\"{i}ve' IV over 5000 simulations. \label{tab:simresults}}
\end{table}


\begin{figure}[!htbp]
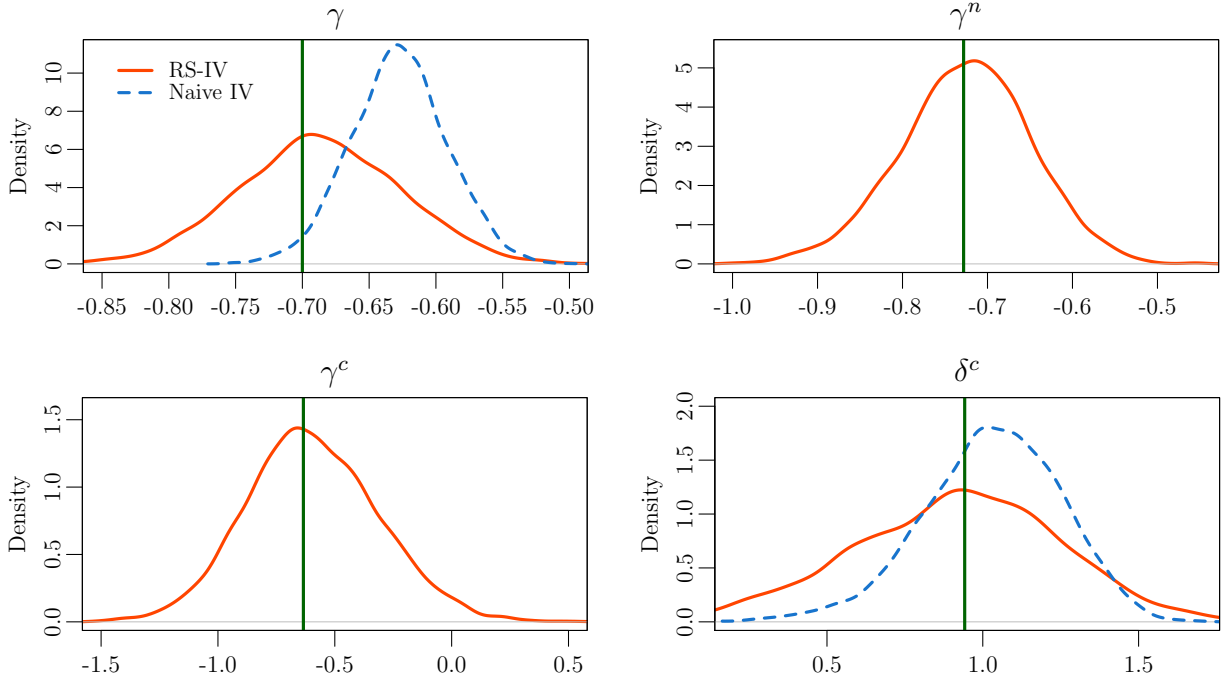

	\vspace{-0.2cm}
	{\centering
		\input{./R/figures/simulations/theta2_e_Pars10.tex}
		\input{./R/figures/simulations/theta2_n_Pars10.tex}
		\\[-0.35cm]
		\input{./R/figures/simulations/gamma_c_Pars10.tex}
		\input{./R/figures/simulations/delta_c_Pars10.tex}
		\label{fig:FrankDGP_ParSet10}
	}
	\vspace{-1cm}
	\caption{Distribution of the estimates of the spillover terms, $(\gamma,\gamma^n,\gamma^c,\delta^c)$, for our IV and the `na\"{i}ve' IV (where available) for simulations with 235 groups, over 5000 simulations. The true parameter values are given by green vertical lines. Analogous figures for simulations with 150 and 500 groups appear in \autoref{sec:additional}. \label{fig:sims235}}
\end{figure}

We randomly assign to each group $g$ a share of compliers $\bar{C}_g \in \{0.1,0.2,0.3,0.4,0.5\}$ with equal probability.
Individuals are assigned a compliance status in the corresponding proportion. 
To generate the random coefficients, we first set values for the unconditional average parameters, $(\alpha,\beta,\gamma,\delta) = (0.5,0.2,-0.7,0.8)$. 
For $\theta \in \{\alpha, \beta, \gamma, \delta\}$, and $\theta_{ig} \in \{\alpha_{ig}, \beta_{ig}, \gamma_{ig}, \delta_{ig}\}$, we then draw the individual random coefficients according to
\[
  \theta_{ig} = \theta + \left[\frac{\bar{C}_{ig} - \mathbb{E}[\bar{C}_{ig}]}{\text{SD}(\bar{C}_{ig})} \frac{\kappa_\theta}{\sqrt{\kappa_\theta^2+1}} + \frac{u_{ig}}{\sqrt{\kappa_\theta^2+1}}\right] \sigma_\theta, \quad u_{ig} \overset{iid}{\sim} \mathcal{N}(0,1)
\]
where $\kappa_\theta$ controls the strength of correlation between $\bar{C}_{ig}$ and a given random coefficient so that $\text{Corr}(\theta_{ig},\bar{C}_{ig}) = \kappa_\theta / \sqrt{\kappa_\theta^2 + 1}$.
We normalize the random coefficients so that their means are given by the unconditional parameters, $(\alpha,\beta,\gamma,\delta)$, and their standard deviations are given by $\boldsymbol{\sigma} = (\sigma_\alpha,\sigma_\beta,\sigma_\gamma,\sigma_\delta)$. 
In the simulations below, we set $\boldsymbol{\kappa}=(0,0,1.2,1.5)$ and $\boldsymbol{\sigma}= (0.3,0.3,0.2,0.4)$, which gives $\text{Corr}(\gamma_{ig},\bar{C}_{ig})\approx0.77$ and $\text{Corr}(\delta_{ig},\bar{C}_{ig})\approx0.83$.

\autoref{tab:simresults} presents means and standard deviations of estimated coefficients along with the actual coverage of the associated nominal 95\% confidence intervals for both our estimator and the ``na\"{i}ve'' IV estimator, based on 5000 simulation replications.\footnote{In principle, one could estimate $(\alpha^n,\gamma^n)$ by estimating a na\"{i}ve IV regression of $Y_{ig}$ on a constant and $\bar{D}_{ig}$ on a subset of the data with $(Z_{ig}=1,D_{ig}=0)$, using $S_g$ as an instrument for $\bar{D}_{ig}$. Similarly, one could estimate $(\alpha^c+\beta^c,\gamma^c+\delta^c)$ by estimating the same regression on a subset of the data with $(Z_{ig}=1,D_{ig}=1)$. However, both sets of estimated parameters would be biased if $\gamma$ is correlated with $\bar{C}_{ig}$, and the second set of estimates would be biased if $\delta$ is correlated with $\bar{C}_{ig}$.} 
The second panel presents the results for a sample size similar to the experimental design in \cite{crepon2013}: 235 groups. 
Our estimator performs well at this sample size---the average coefficients are very close to the true values and the coverage is close to its nominal level for all eight parameter values---and its performance improves in larger samples, as expected. 
In contrast, the na\"{i}ve IV estimates of $\gamma$ and $\delta^c$ are substantially biased, as predicted by \autoref{lem:linear}. 
The performance of the na\"{i}ve IV estimator does not improve as we increase the sample size---the average coefficients remain unchanged and the coverage worsens as the standard errors shrink. 
\autoref{fig:sims235} shows the empirical distribution of the point estimates for our estimator in the simulations with 235 groups and compares this to the na\"{i}ve IV for $\gamma$ and $\delta^c$. 
Again, our estimator performs well and the bias of the na\"{i}ve IV is clearly visible, as is the mean-variance tradeoff between the two estimators. Appendix \ref{sec:additional} presents similar figures for simulations with 150 and 500 groups.


\section{Conclusion}
\label{sec:conclusion}

In this paper we have proposed methods to identify and estimate direct and indirect causal effects under one-sided non-compliance, using data from a randomized saturation experiment.
Under appropriate assumptions, we show that the key source of unobserved heterogeneity is the share of compliers within a given group.
In a setting with many large groups, this quantity can be estimated and yields a simple IV estimator that is consistent and asymptotically normal in the limit as group size and the number of groups grow.
We have also illustrated the applicability of our methods using data from a large-scale job-placement program randomized saturation experiment.
In this setting, we find negative spillover effects on the sub-population willing to take up the program.
The direct effects, however, shield those who take up treatment from these negative indirect effects.
These findings illustrate how our methods allow researchers to go beyond intent-to-treat effects and reveal important information that may be relevant for the design of real-world policies.

A possible extension of the methods described above would be to consider settings with two-sided non-compliance.
In this case our identification approach would condition on the share of always-takers in addition to the share of compliers. 
A related idea would be to relax the assumption of anonymous interactions by allowing individuals' potential outcome functions to depend on the take-up rates of different \emph{sub-groups} within their group. 
For example, male students may experience stronger spillovers from their male classmates, and female students from their female classmates.
In this case, one would need to condition on the compliance rate in each sub-group.
A more challenging extension would consider relaxing IOR.
Without this assumption, the estimands identified in this paper lack a straightforward causal interpretation.
However it may be possible to identify, or at least partially identify, well-defined causal effects under somewhat weaker restrictions on treatment take-up behavior.
While a full analysis of policy relevant treatment effects in the presence of spillovers is beyond the scope or this paper, we think there is ample scope for future research in this direction.
It could be interesting, for example, to consider applying the marginal treatment effects approach to settings with spillovers and non-compliance.
We leave this possibility for future research.


\small
\singlespacing
\bibliographystyle{elsarticle-harv}
\bibliography{spillovers-noncompliance}

\begin{thebibliography}{39}
\expandafter\ifx\csname natexlab\endcsname\relax\def\natexlab#1{#1}\fi
\providecommand{\url}[1]{\texttt{#1}}
\providecommand{\href}[2]{#2}
\providecommand{\path}[1]{#1}
\providecommand{\DOIprefix}{doi:}
\providecommand{\ArXivprefix}{arXiv:}
\providecommand{\URLprefix}{URL: }
\providecommand{\Pubmedprefix}{pmid:}
\providecommand{\doi}[1]{\href{http://dx.doi.org/#1}{\path{#1}}}
\providecommand{\Pubmed}[1]{\href{pmid:#1}{\path{#1}}}
\providecommand{\bibinfo}[2]{#2}
\ifx\xfnm\relax \def\xfnm[#1]{\unskip,\space#1}\fi
\bibitem[{Abebe et~al.(2021)Abebe, Caria, Fafchamps, Falco, Franklin and
  Quinn}]{abebe2021anonymity}
\bibinfo{author}{Abebe, G.}, \bibinfo{author}{Caria, A.S.},
  \bibinfo{author}{Fafchamps, M.}, \bibinfo{author}{Falco, P.},
  \bibinfo{author}{Franklin, S.}, \bibinfo{author}{Quinn, S.},
  \bibinfo{year}{2021}.
\newblock \bibinfo{title}{Anonymity or distance? {J}ob search and labour market
  exclusion in a growing african city}.
\newblock \bibinfo{journal}{The Review of Economic Studies}
  \bibinfo{volume}{88}, \bibinfo{pages}{1279--1310}.
\bibitem[{Akram et~al.(2018)Akram, Chowdhury and Mobarak}]{Akram2018}
\bibinfo{author}{Akram, A.A.}, \bibinfo{author}{Chowdhury, S.},
  \bibinfo{author}{Mobarak, A.M.}, \bibinfo{year}{2018}.
\newblock \bibinfo{title}{Effects of emigration on rural labor markets}
  \URLprefix
  \url{http://faculty.som.yale.edu/mushfiqmobarak/papers/migrationge.pdf}.
\bibitem[{Altonji and Matzkin(2005)}]{altonji2005cross}
\bibinfo{author}{Altonji, J.G.}, \bibinfo{author}{Matzkin, R.L.},
  \bibinfo{year}{2005}.
\newblock \bibinfo{title}{Cross section and panel data estimators for
  nonseparable models with endogenous regressors}.
\newblock \bibinfo{journal}{Econometrica} \bibinfo{volume}{73},
  \bibinfo{pages}{1053--1102}.
\bibitem[{Anderson et~al.(2014)Anderson, Huttenlocher, Kleinberg and
  Leskovec}]{anderson2014}
\bibinfo{author}{Anderson, A.}, \bibinfo{author}{Huttenlocher, D.},
  \bibinfo{author}{Kleinberg, J.}, \bibinfo{author}{Leskovec, J.},
  \bibinfo{year}{2014}.
\newblock \bibinfo{title}{Engaging with massive online courses}, in:
  \bibinfo{booktitle}{Proceedings of the 23rd international conference on World
  wide web}, \bibinfo{organization}{ACM}. pp. \bibinfo{pages}{687--698}.
\bibitem[{Angelucci and De~Giorgi(2009)}]{angelucci2009}
\bibinfo{author}{Angelucci, M.}, \bibinfo{author}{De~Giorgi, G.},
  \bibinfo{year}{2009}.
\newblock \bibinfo{title}{Indirect effects of an aid program: how do cash
  transfers affect ineligibles' consumption?}
\newblock \bibinfo{journal}{American Economic Review} \bibinfo{volume}{99},
  \bibinfo{pages}{486--508}.
\bibitem[{Baird et~al.(2018)Baird, Bohren, McIntosh and {\"O}zler}]{baird2018}
\bibinfo{author}{Baird, S.}, \bibinfo{author}{Bohren, J.A.},
  \bibinfo{author}{McIntosh, C.}, \bibinfo{author}{{\"O}zler, B.},
  \bibinfo{year}{2018}.
\newblock \bibinfo{title}{Optimal design of experiments in the presence of
  interference}.
\newblock \bibinfo{journal}{Review of Economics and Statistics}
  \bibinfo{volume}{100}, \bibinfo{pages}{844--860}.
\bibitem[{Baird et~al.(2011)Baird, McIntosh and {\"O}zler}]{baird2011}
\bibinfo{author}{Baird, S.}, \bibinfo{author}{McIntosh, C.},
  \bibinfo{author}{{\"O}zler, B.}, \bibinfo{year}{2011}.
\newblock \bibinfo{title}{Cash or condition? {E}vidence from a cash transfer
  experiment}.
\newblock \bibinfo{journal}{The Quarterly Journal of Economics}
  \bibinfo{volume}{126}, \bibinfo{pages}{1709--1753}.
\bibitem[{Banerjee et~al.(2012)Banerjee, Chattopadhyay, Duflo, Keniston and
  Singh}]{banerjee2012}
\bibinfo{author}{Banerjee, A.V.}, \bibinfo{author}{Chattopadhyay, R.},
  \bibinfo{author}{Duflo, E.}, \bibinfo{author}{Keniston, D.},
  \bibinfo{author}{Singh, N.}, \bibinfo{year}{2012}.
\newblock \bibinfo{title}{Can institutions be reformed from within? evidence
  from a randomized experiment with the {R}ajasthan police} .
\bibitem[{Barrera-Osorio et~al.(2011)Barrera-Osorio, Bertrand, Linden and
  Perez-Calle}]{barrera2011}
\bibinfo{author}{Barrera-Osorio, F.}, \bibinfo{author}{Bertrand, M.},
  \bibinfo{author}{Linden, L.L.}, \bibinfo{author}{Perez-Calle, F.},
  \bibinfo{year}{2011}.
\newblock \bibinfo{title}{Improving the design of conditional transfer
  programs: Evidence from a randomized education experiment in {Colombia}}.
\newblock \bibinfo{journal}{American Economic Journal: Applied Economics}
  \bibinfo{volume}{3}, \bibinfo{pages}{167--95}.
\bibitem[{Bhattacharya et~al.(2021)Bhattacharya, Dupas and
  Kanaya}]{bhattacharya2021demand}
\bibinfo{author}{Bhattacharya, D.}, \bibinfo{author}{Dupas, P.},
  \bibinfo{author}{Kanaya, S.}, \bibinfo{year}{2021}.
\newblock \bibinfo{title}{Demand and welfare analysis in discrete choice models
  with social interactions}.
\newblock \bibinfo{type}{Technical Report}.
\bibitem[{Bobba and Gignoux(2014)}]{bobba2014}
\bibinfo{author}{Bobba, M.}, \bibinfo{author}{Gignoux, J.},
  \bibinfo{year}{2014}.
\newblock \bibinfo{title}{Neighborhood effects and take-up of transfers in
  integrated social policies: Evidence from {P}rogresa}.
\newblock \bibinfo{type}{Technical Report}.
\bibitem[{Bobonis and Finan(2009)}]{bobonis2009}
\bibinfo{author}{Bobonis, G.J.}, \bibinfo{author}{Finan, F.},
  \bibinfo{year}{2009}.
\newblock \bibinfo{title}{Neighborhood peer effects in secondary school
  enrollment decisions}.
\newblock \bibinfo{journal}{The Review of Economics and Statistics}
  \bibinfo{volume}{91}, \bibinfo{pages}{695--716}.
\bibitem[{Bond et~al.(2012)Bond, Fariss, Jones, Kramer, Marlow, Settle and
  Fowler}]{bond2012}
\bibinfo{author}{Bond, R.M.}, \bibinfo{author}{Fariss, C.J.},
  \bibinfo{author}{Jones, J.J.}, \bibinfo{author}{Kramer, A.D.},
  \bibinfo{author}{Marlow, C.}, \bibinfo{author}{Settle, J.E.},
  \bibinfo{author}{Fowler, J.H.}, \bibinfo{year}{2012}.
\newblock \bibinfo{title}{A 61-million-person experiment in social influence
  and political mobilization}.
\newblock \bibinfo{journal}{Nature} \bibinfo{volume}{489},
  \bibinfo{pages}{295}.
\bibitem[{Bursztyn et~al.(2021)Bursztyn, Cantoni, Yang, Yuchtman and
  Zhang}]{bursztyn2021}
\bibinfo{author}{Bursztyn, L.}, \bibinfo{author}{Cantoni, D.},
  \bibinfo{author}{Yang, D.Y.}, \bibinfo{author}{Yuchtman, N.},
  \bibinfo{author}{Zhang, Y.J.}, \bibinfo{year}{2021}.
\newblock \bibinfo{title}{Persistent political engagement: Social interactions
  and the dynamics of protest movements}.
\newblock \bibinfo{journal}{American Economic Review: Insights}
  \bibinfo{volume}{3}, \bibinfo{pages}{233--50}.
\bibitem[{Callen et~al.(2019)Callen, De~Mel, McIntosh and
  Woodruff}]{callen2019headwaters}
\bibinfo{author}{Callen, M.}, \bibinfo{author}{De~Mel, S.},
  \bibinfo{author}{McIntosh, C.}, \bibinfo{author}{Woodruff, C.},
  \bibinfo{year}{2019}.
\newblock \bibinfo{title}{What are the headwaters of formal savings?
  experimental evidence from {S}ri {L}anka}.
\newblock \bibinfo{journal}{The Review of Economic Studies}
  \bibinfo{volume}{86}, \bibinfo{pages}{2491--2529}.
\bibitem[{Constantinou and Dawid(2017)}]{constantinou2017}
\bibinfo{author}{Constantinou, P.}, \bibinfo{author}{Dawid, A.P.},
  \bibinfo{year}{2017}.
\newblock \bibinfo{title}{Extended conditional independence and applications in
  causal inference}.
\newblock \bibinfo{journal}{The Annals of Statistics} \bibinfo{volume}{45},
  \bibinfo{pages}{2618--2653}.
\bibitem[{Cr{\'e}pon et~al.(2013)Cr{\'e}pon, Duflo, Gurgand, Rathelot and
  Zamora}]{crepon2013}
\bibinfo{author}{Cr{\'e}pon, B.}, \bibinfo{author}{Duflo, E.},
  \bibinfo{author}{Gurgand, M.}, \bibinfo{author}{Rathelot, R.},
  \bibinfo{author}{Zamora, P.}, \bibinfo{year}{2013}.
\newblock \bibinfo{title}{Do labor market policies have displacement effects?
  {E}vidence from a clustered randomized experiment}.
\newblock \bibinfo{journal}{The Quarterly Journal of Economics}
  \bibinfo{volume}{128}, \bibinfo{pages}{531--580}.
\bibitem[{Dawid(1979)}]{Dawid1979}
\bibinfo{author}{Dawid, A.P.}, \bibinfo{year}{1979}.
\newblock \bibinfo{title}{Conditional independence in statistical theory}.
\newblock \bibinfo{journal}{Journal of the Royal Statistical Society: Series B
  (Methodological)} \bibinfo{volume}{41}, \bibinfo{pages}{1--15}.
\bibitem[{Duflo and Saez(2003)}]{duflo2003}
\bibinfo{author}{Duflo, E.}, \bibinfo{author}{Saez, E.}, \bibinfo{year}{2003}.
\newblock \bibinfo{title}{The role of information and social interactions in
  retirement plan decisions: Evidence from a randomized experiment}.
\newblock \bibinfo{journal}{{The Quarterly Journal of Economics}}
  \bibinfo{volume}{118}, \bibinfo{pages}{815--842}.
\bibitem[{Eckles et~al.(2016)Eckles, Kizilcec and Bakshy}]{eckles2016}
\bibinfo{author}{Eckles, D.}, \bibinfo{author}{Kizilcec, R.F.},
  \bibinfo{author}{Bakshy, E.}, \bibinfo{year}{2016}.
\newblock \bibinfo{title}{Estimating peer effects in networks with peer
  encouragement designs}.
\newblock \bibinfo{journal}{Proceedings of the National Academy of Sciences}
  \bibinfo{volume}{113}, \bibinfo{pages}{7316--7322}.
\bibitem[{Graham and de~Xavier~Pinto(2022)}]{graham2022semiparametrically}
\bibinfo{author}{Graham, B.S.}, \bibinfo{author}{de~Xavier~Pinto, C.C.},
  \bibinfo{year}{2022}.
\newblock \bibinfo{title}{Semiparametrically efficient estimation of the
  average linear regression function}.
\newblock \bibinfo{journal}{Journal of Econometrics} \bibinfo{volume}{226},
  \bibinfo{pages}{115--138}.
\bibitem[{Haushofer and Shapiro(2016)}]{haushofer2016}
\bibinfo{author}{Haushofer, J.}, \bibinfo{author}{Shapiro, J.},
  \bibinfo{year}{2016}.
\newblock \bibinfo{title}{The short-term impact of unconditional cash transfers
  to the poor: experimental evidence from {K}enya}.
\newblock \bibinfo{journal}{{The Quarterly Journal of Economics}}
  \bibinfo{volume}{131}, \bibinfo{pages}{1973--2042}.
\bibitem[{Heckman and Vytlacil(1998)}]{heckman1998}
\bibinfo{author}{Heckman, J.}, \bibinfo{author}{Vytlacil, E.},
  \bibinfo{year}{1998}.
\newblock \bibinfo{title}{Instrumental variables methods for the correlated
  random coefficient model: Estimating the average rate of return to schooling
  when the return is correlated with schooling}.
\newblock \bibinfo{journal}{Journal of Human Resources} ,
  \bibinfo{pages}{974--987}.
\bibitem[{Hoeffding(1963)}]{Hoeffding1963}
\bibinfo{author}{Hoeffding, W.}, \bibinfo{year}{1963}.
\newblock \bibinfo{title}{Probability inequalities for sums of bounded random
  variables}.
\newblock \bibinfo{journal}{Journal of the American Statistical Association}
  \bibinfo{volume}{58}, \bibinfo{pages}{13--30}.
\bibitem[{Horn and Johnson(2013)}]{horn2013matrix}
\bibinfo{author}{Horn, R.A.}, \bibinfo{author}{Johnson, C.R.},
  \bibinfo{year}{2013}.
\newblock \bibinfo{title}{Matrix analysis}.
\bibitem[{Hudgens and Halloran(2008)}]{hudgens2008}
\bibinfo{author}{Hudgens, M.G.}, \bibinfo{author}{Halloran, M.E.},
  \bibinfo{year}{2008}.
\newblock \bibinfo{title}{{Toward causal inference with interference}}.
\newblock \bibinfo{journal}{Journal of the American Statistical Association}
  \bibinfo{volume}{103}, \bibinfo{pages}{832--842}.
\newblock \DOIprefix\doi{10.1198/016214508000000292}.
\bibitem[{Imai et~al.(2020)Imai, Jiang and Malani}]{Imai2018}
\bibinfo{author}{Imai, K.}, \bibinfo{author}{Jiang, Z.},
  \bibinfo{author}{Malani, A.}, \bibinfo{year}{2020}.
\newblock \bibinfo{title}{Causal inference with interference and noncompliance
  in two-stage randomized experiments}.
\newblock \bibinfo{journal}{Journal of the American Statistical Association} ,
  \bibinfo{pages}{1--13}.
\bibitem[{Imbens and Newey(2009)}]{imbens2009identification}
\bibinfo{author}{Imbens, G.W.}, \bibinfo{author}{Newey, W.K.},
  \bibinfo{year}{2009}.
\newblock \bibinfo{title}{Identification and estimation of triangular
  simultaneous equations models without additivity}.
\newblock \bibinfo{journal}{Econometrica} \bibinfo{volume}{77},
  \bibinfo{pages}{1481--1512}.
\bibitem[{Kang and Imbens(2016)}]{Kang2016}
\bibinfo{author}{Kang, H.}, \bibinfo{author}{Imbens, G.}, \bibinfo{year}{2016}.
\newblock \bibinfo{title}{{Peer Encouragement Designs in Causal Inference with
  Partial Interference and Identification of Local Average Network Effects}} ,
  \bibinfo{pages}{1--39}\URLprefix \url{http://arxiv.org/abs/1609.04464},
  \href{http://arxiv.org/abs/1609.04464}{{\tt arXiv:1609.04464}}.
\bibitem[{Manski(2013)}]{Manski2013}
\bibinfo{author}{Manski, C.F.}, \bibinfo{year}{2013}.
\newblock \bibinfo{title}{{Identification of treatment response with social
  interactions}}.
\newblock \bibinfo{journal}{Econometrics Journal} \bibinfo{volume}{16},
  \bibinfo{pages}{1--23}.
\newblock \DOIprefix\doi{10.1111/j.1368-423X.2012.00368.x}.
\bibitem[{Masten and Torgovitsky(2016)}]{masten2016}
\bibinfo{author}{Masten, M.A.}, \bibinfo{author}{Torgovitsky, A.},
  \bibinfo{year}{2016}.
\newblock \bibinfo{title}{Identification of instrumental variable correlated
  random coefficients models}.
\newblock \bibinfo{journal}{Review of Economics and Statistics}
  \bibinfo{volume}{98}, \bibinfo{pages}{1001--1005}.
\bibitem[{Miguel and Kremer(2004)}]{miguel2004}
\bibinfo{author}{Miguel, E.}, \bibinfo{author}{Kremer, M.},
  \bibinfo{year}{2004}.
\newblock \bibinfo{title}{Worms: identifying impacts on education and health in
  the presence of treatment externalities}.
\newblock \bibinfo{journal}{Econometrica} \bibinfo{volume}{72},
  \bibinfo{pages}{159--217}.
\bibitem[{Pearl(1988)}]{Pearl1988}
\bibinfo{author}{Pearl, J.}, \bibinfo{year}{1988}.
\newblock \bibinfo{title}{Probabilistic reasoning in intelligent systems:
  Networks of plausible inference}.
\bibitem[{Vazquez-Bare(2021)}]{VazquezBare}
\bibinfo{author}{Vazquez-Bare, G.}, \bibinfo{year}{2021}.
\newblock \bibinfo{title}{Causal spillover effects using instrumental
  variables}.
\newblock \bibinfo{journal}{Journal of the American Statistical Association} ,
  \bibinfo{pages}{1--35}.
\bibitem[{Wooldridge(1997)}]{Wooldridge1997}
\bibinfo{author}{Wooldridge, J.M.}, \bibinfo{year}{1997}.
\newblock \bibinfo{title}{{On two stage least squares estimation of the average
  treatment effect in a random coefficient model}}.
\newblock \bibinfo{journal}{Economics Letters} \bibinfo{volume}{56},
  \bibinfo{pages}{129--133}.
\newblock \DOIprefix\doi{10.1016/s0165-1765(97)81890-3}.
\bibitem[{Wooldridge(2003)}]{Wooldridge2003}
\bibinfo{author}{Wooldridge, J.M.}, \bibinfo{year}{2003}.
\newblock \bibinfo{title}{{Further results on instrumental variables estimation
  of average treatment effects in the correlated random coefficient model}}.
\newblock \bibinfo{journal}{Economics Letters} \bibinfo{volume}{79},
  \bibinfo{pages}{185--191}.
\newblock \DOIprefix\doi{10.1016/S0165-1765(02)00318-X}.
\bibitem[{Wooldridge(2004)}]{wooldridge2004}
\bibinfo{author}{Wooldridge, J.M.}, \bibinfo{year}{2004}.
\newblock \bibinfo{title}{Estimating average partial effects under conditional
  moment independence assumptions}.
\newblock \bibinfo{type}{Technical Report}. cemmap working paper.
\bibitem[{Wooldridge(2016)}]{Wooldridge2016}
\bibinfo{author}{Wooldridge, J.M.}, \bibinfo{year}{2016}.
\newblock \bibinfo{title}{{Instrumental variables estimation of the average
  treatment effect in the correlated random coefficient model}}.
\newblock \bibinfo{journal}{Advances in Econometrics} \bibinfo{volume}{21},
  \bibinfo{pages}{93--116}.
\newblock \DOIprefix\doi{10.1016/S0731-9053(07)00004-7}.
\bibitem[{Yi et~al.(2015)Yi, Song, Liu, Huang, Zhang, Bai, Ren, Shi, Loyalka,
  Chu et~al.}]{yi2015}
\bibinfo{author}{Yi, H.}, \bibinfo{author}{Song, Y.}, \bibinfo{author}{Liu,
  C.}, \bibinfo{author}{Huang, X.}, \bibinfo{author}{Zhang, L.},
  \bibinfo{author}{Bai, Y.}, \bibinfo{author}{Ren, B.}, \bibinfo{author}{Shi,
  Y.}, \bibinfo{author}{Loyalka, P.}, \bibinfo{author}{Chu, J.}, et~al.,
  \bibinfo{year}{2015}.
\newblock \bibinfo{title}{Giving kids a head start: The impact and mechanisms
  of early commitment of financial aid on poor students in rural {China}}.
\newblock \bibinfo{journal}{Journal of Development Economics}
  \bibinfo{volume}{113}, \bibinfo{pages}{1--15}.

\end{thebibliography}

\appendix
\numberwithin{equation}{section}
\numberwithin{table}{section}
\numberwithin{figure}{section}

\section{Proofs}

The following lemma, taken from \cite{constantinou2017}, summarizes several useful properties of conditional independence that we use in our proofs below.
The names attached to properties (i) and (iii)--(v) originate with \cite{Pearl1988}.
For the purposes of this document, we call the second property ``redundancy.'' 

\begin{lem}[Axioms of Conditional Independence]
  \label{lem:axioms}
  Let $X,Y,Z,W$ be random vectors defined on a common probability space, and let $h$ be a measurable function.
  Then:
  \begin{enumerate}[(i)]
    \item (Symmetry): $X \indep Y|Z \implies Y \indep X|Z$.
    \item (Redundancy): $X \indep Y|Y$.
    \item (Decomposition): $X \indep Y|Z$ and $W = h(Y) \implies X \indep W|Z$.
    \item (Weak Union): $X\indep Y|Z$ and $W = h(Y) \implies X \indep Y|(W,Z)$.
    \item (Contraction): $X\indep Y|Z$ and $X\indep W|(Y,Z) \implies X\indep (Y,W)|Z$.
  \end{enumerate}
\end{lem}

For simplicity, our proofs below freely use the ``Symmetry'' property without comment, although we reference the other properties when used.
We also rely on the following corollary of \autoref{lem:axioms}.
\begin{cor}
  \label{cor:example}
  $X\indep Y|Z$ implies $(X,Z)\indep Y|Z$.
\end{cor}


\begin{proof}[Proof of \autoref{lem:DbarDist}]
  Applying \autoref{cor:example} and the  Decomposition property to \autoref{assump:exclusion}(ii) yields $\boldsymbol{Z}_g \indep (\boldsymbol{C}_g, \bar{C}_{ig})|(N_g,S_g)$.
  By the definition of conditional independence, it follows that the distribution of $\boldsymbol{Z}_g|(N_g, S_g, \boldsymbol{C}_g,\bar{C}_{ig})$ is the same as that of $\boldsymbol{Z}_g|(N_g, S_g)$:
\begin{equation}
\label{eq:design_indep2}
  \mathbbm{P}(\boldsymbol{Z}_g=\boldsymbol{z}|N_g = n, S_g = s, \boldsymbol{C}_g, \bar{C}_{ig}) = \mathbbm{P}(\boldsymbol{Z}_g = \boldsymbol{z}|N_g = n, S_g = s).
\end{equation}
Now, define the shorthand $A \equiv \left\{ N_g = n, S_g = s, \boldsymbol{C}_g = \boldsymbol{c}, \bar{C}_{ig} = \bar{c}\right\}$ and let $\mathcal{C}(i)$ be the indices of all non-zero components of $\boldsymbol{c}$, \emph{excluding} the $i$th component, i.e.\ $\mathcal{C}(i) \equiv \left\{ j\neq i\colon c_j = 1 \right\}$.
By the definition of $\bar{D}_{ig}$, the event $\left\{ \bar{D}_{ig} = d  \right\}$ is equivalent to $\left\{ \sum_{j\neq i} C_{jg}Z_{jg} = d (N_g-1) \right\}$.
Consequently, 
\[
  \mathbbm{P}(\bar{D}_{ig} = d|A, Z_{ig}) = \mathbbm{P}\left(\left.\left[\sum_{j\neq i} C_{jg} Z_{jg}\right] = d(n - 1)\right|A,Z_{ig}\right) = \mathbbm{P}\left(\left. \left[\sum_{j\in \mathcal{C}(i)} Z_{jg}\right] = d(n-1) \right|A,Z_{ig}\right)
\]
where the first equality uses the fact that $A$ implies $N_g = n$, and the second uses the fact that $A$ implies $\boldsymbol{C}_g = \boldsymbol{c}$, so we know precisely which of the indicators $C_{jg}$ equal zero and which equal one. 
Under \autoref{assump:Bernoulli}, \eqref{eq:design_indep2} implies that $\boldsymbol{Z}_g|A \sim \mbox{iid Bernoulli}(s)$.
By our definition of $\mathcal{C}(i)$ it follows that, conditional on $A$, the subvector of $\boldsymbol{Z}_g$ that corresponds to $\mathcal{C}(i)$ constitutes an iid sequence of $\bar{c}(n-1)$ Bernoulli$(s)$ random variables, each of which is \emph{independent of} $Z_{ig}$.
Hence, conditional on $(A,Z_{ig})$, we see that $\sum_{j \in \mathcal{C}(i)} Z_{jg} \sim \text{Binomial}\big(\bar{c}(n-1), s \big)$.
\end{proof}


\begin{proof}[Proof of \autoref{lem:linear}]
  Under \eqref{eq:LinearModel}, $Y_{ig} = \mathbf{X}_{ig}' \mathbf{B}_{ig}$ where $\mathbf{B}_{ig} = (\alpha_{ig}, \beta_{ig}, \gamma_{ig}, \delta_{ig})'$.
  Now, let $\mathcal{R}_{ig} \equiv \left\{ S_g, Z_{ig}, N_g, \bar{C}_{ig}, C_{ig}, \mathbf{B}_{ig} \right\}$ and $\boldsymbol{\Lambda}_{ig} \equiv \text{diag}\left\{ 1, C_{ig}, \bar{C}_{ig}, C_{ig} \bar{C}_{ig} \right\}$.
  From \autoref{lem:DbarDist} we see that $\mathbbm{E}[\bar{D}_{ig}|\mathcal{R}] = \bar{C}_{ig} S_g$.  
  Since $D_{ig} = C_{ig} Z_{ig}$ under one-sided non-compliance and IOR, it follows that $\mathbbm{E}[\mathbf{X}_{ig}'|\mathcal{R}_{ig}] = \boldsymbol{\mathcal{Z}}_{ig}' \boldsymbol{\Lambda}_{ig}$.
  Hence, 
  \begin{align*}
    \mathbbm{E}\left[ \boldsymbol{\mathcal{Z}}_{ig} Y_{ig} \right]
    &=  \mathbbm{E}\left[ \boldsymbol{\mathcal{Z}}_{ig}\mathbbm{E}( \mathbf{X}_{ig}' |\mathcal{R}_{ig}) \mathbf{B}_{ig}\right]
    =  \mathbbm{E}\left[ \left(\boldsymbol{\mathcal{Z}}_{ig}\boldsymbol{\mathcal{Z}}_{ig}'\right) \left( \boldsymbol{\Lambda}_{ig}\mathbf{B}_{ig}\right)\right]\\
    \mathbbm{E}\left[ \boldsymbol{\mathcal{Z}}_{ig} \mathbf{X}_{ig}' \right] &= \mathbbm{E}\left[ \boldsymbol{\mathcal{Z}}_{ig} \mathbbm{E}\left( \mathbf{X}_{ig}'|\mathcal{R}_{ig} \right) \right] =  \mathbbm{E}\left[ \left( \boldsymbol{\mathcal{Z}}_{ig} \boldsymbol{\mathcal{Z}}_{ig}' \right) \boldsymbol{\Lambda}_{ig} \right]
  \end{align*}
  since $\boldsymbol{\mathcal{Z}}_{ig}$ and $\mathbf{B}_{ig}$ are $\mathcal{R}_{ig}$--measurable.
  Now, applying Decomposition and \autoref{cor:example} to part (ii) of \autoref{assump:exclusion} gives $Z_{ig} \indep (C_{ig}, \bar{C}_{ig}, \mathbf{B}_{ig})|(S_g, N_g)$. 
  Under \autoref{assump:Bernoulli}, however, the conditional distribution of $Z_{ig}|(S_g=s,N_g=n)$ does not involve $n$, so we obtain 
  \begin{equation}
    (C_{ig}, \bar{C}_{ig}, \mathbf{B}_{ig})\indep Z_{ig} | S_g.
    \label{eq:BernoulliIndep}
  \end{equation}
  Similarly, applying Decomposition to part (ii) of \autoref{cor:example}, we see that $(C_{ig}, \bar{C}_{ig}, \mathbf{B}_{ig})\indep S_g$.
  Combining this with \eqref{eq:BernoulliIndep}, the Contraction axiom yields $(C_{ig}, \bar{C}_{ig}, \mathbf{B}_{ig}) \indep (Z_{ig}, S_g)$, implying that $(\boldsymbol{\mathcal{Z}}_{ig} \boldsymbol{\mathcal{Z}}_{ig}')$ is independent of both $\boldsymbol{\Lambda}_{ig}$ and $(\boldsymbol{\Lambda}_{ig} \mathbf{B}_{ig})$.
  Accordingly,  
  \begin{align*}
    \boldsymbol{\vartheta}_{\text{IV}} 
    &=  \left\{\mathbbm{E}\left[ \left( \boldsymbol{\mathcal{Z}}_{ig} \boldsymbol{\mathcal{Z}}_{ig}' \right) \boldsymbol{\Lambda}_{ig} \right]\right\}^{-1} \mathbbm{E}\left[ \left(\boldsymbol{\mathcal{Z}}_{ig}\boldsymbol{\mathcal{Z}}_{ig}'\right) \left( \boldsymbol{\Lambda}_{ig}\mathbf{B}_{ig}\right)\right]
    = \mathbbm{E}\left[  \boldsymbol{\Lambda}_{ig}\right]^{-1}\mathbbm{E}\left[  \boldsymbol{\Lambda}_{ig}\mathbf{B}_{ig}\right].
  \end{align*}
  By the definitions of $\boldsymbol{\vartheta}_{\text{IV}}$, $\boldsymbol{\Lambda}_{ig}$ and $\mathbf{B}_{ig}$ it follows that
  \[
    \alpha_{\text{IV}} = \mathbbm{E}\left[ \alpha_{ig} \right], \quad
    \beta_{\text{IV}} = \frac{\mathbbm{E}\left[ C_{ig} \beta_{ig} \right]}{\mathbbm{E}\left[ C_{ig} \right]}, \quad
    \gamma_{\text{IV}} = \frac{\mathbbm{E}\left[ \bar{C}_{ig} \gamma_{ig} \right]}{\mathbbm{E}\left[ \bar{C}_{ig} \right]}, \quad
    \delta_{\text{IV}} = \frac{\mathbbm{E}\left[ C_{ig}\bar{C}_{ig} \delta_{ig} \right]}{\mathbbm{E}\left[ C_{ig} \bar{C}_{ig} \right]}.
  \]
  By iterated expectations over $C_{ig}$, we obtain $\beta_{\text{IV}} = \mathbbm{E}\left[ \beta_{ig}|C_{ig}=1 \right]$ while
  \[
    \gamma_{\text{IV}} = \frac{\mathbbm{E}\left[ \bar{C}_{ig} \gamma_{ig} \right]}{\mathbbm{E}\left[ \bar{C}_{ig} \right]} = 
    \frac{\text{Cov}(\bar{C}_{ig}, \gamma_{ig}) + \mathbbm{E}(\bar{C}_{ig})\mathbbm{E}(\gamma_{ig})}{\mathbbm{E}(\bar{C}_{ig})} = \mathbbm{E}[\gamma_{ig}] + \frac{\text{Cov}(\bar{C}_{ig}, \gamma_{ig})}{\mathbbm{E}(\bar{C}_{ig})}.
  \]
  Similarly, again taking iterated expectations over $C_{ig}$, 
\[
  \delta_{\text{IV}} = \frac{\mathbbm{E}\left[ \bar{C}_{ig} \delta_{ig}|C_{ig} = 1 \right]}{\mathbbm{E}\left[\bar{C}_{ig}|C_{ig} = 1 \right]} = \mathbbm{E}\left[ \delta_{ig}|C_{ig} = 1 \right] + \frac{\text{Cov}(\bar{C}_{ig}, \delta_{ig}|C_{ig}=1)}{\mathbbm{E}(\bar{C}_{ig}|C_{ig} = 1)}.
\]
\end{proof}

\begin{proof}[Proof of \autoref{thm:conditional_indep}]
  \autoref{assump:exclusion}(i) implies $(\boldsymbol{C}_g, \mathbf{B}_g)\indep S_g|N_g$ by Weak Union and Decomposition.
Combining this with \autoref{assump:exclusion}(ii) gives
\begin{equation}
  (\mathbf{Z}_g, S_g)\indep (\mathbf{B}_g, \boldsymbol{C}_g)|N_g
  \label{eq:OldHighLevel}
\end{equation}
by Contraction.
Now let $\boldsymbol{C}_{-ig}$ denote the subvector of $\boldsymbol{C}_g$ that excludes element $i$.
Applying Decomposition, \autoref{cor:example}, and Weak Union to \eqref{eq:OldHighLevel},
\begin{equation}
  \label{eq:s_z}
  (S_g, \boldsymbol{Z}_g) \indep (B_{ig}, C_{ig}, \boldsymbol{C}_{-ig},N_g)|(N_g,\bar{C}_{ig}). 
\end{equation}
because $\bar{C}_{ig}$ is a function of $(\boldsymbol{C}_{g},N_g)$.
By \autoref{lem:DbarDist},
\begin{equation}
  \bar{D}_{ig} \indep \boldsymbol{C}_{-ig} | (N_g, \bar{C}_{ig}, S_g, Z_{ig}).
\label{eq:DbarCindep}
\end{equation}
Applying Decomposition to \eqref{eq:s_z} gives  
$\boldsymbol{C}_{-ig} \indep (S_g, Z_{ig}) |(N_g, \bar{C}_{ig})$.
Combining this with \eqref{eq:DbarCindep},  
\begin{equation}
  (S_g, Z_{ig},\bar{D}_{ig}) \indep\boldsymbol{C}_{-ig} |(N_g, \bar{C}_{ig})
  \label{eq:contraction1}
\end{equation}
by Contraction.
Now, applying Weak Union and Decomposition to \eqref{eq:s_z}, 
\begin{equation}
(S_g, \boldsymbol{Z}_g) \indep (B_{ig}, C_{ig})|(\boldsymbol{C}_{-ig},\bar{C}_{ig}, N_g).
\label{eq:contraction2}
\end{equation}
Applying \autoref{cor:example}, we can move $(\boldsymbol{C}_{-ig},N_g)$ from the conditioning set onto the left side of the conditional independence relation, yielding
\begin{equation}
(S_g, Z_{ig},\boldsymbol{C}_{-ig},N_g) \indep (B_{ig}, C_{ig})|(\boldsymbol{C}_{-ig},\bar{C}_{ig}, N_g).
\label{eq:contraction3}
\end{equation}
Since $(S_g,Z_{ig},\bar{D}_{ig})$ is a function of $(S_g,\boldsymbol{Z}_g, \boldsymbol{C}_{-ig}, N_g)$, applying Decomposition to \eqref{eq:contraction3}, gives 
\begin{equation}
(S_g, Z_{ig},\bar{D}_{ig}) \indep (B_{ig}, C_{ig})|(\boldsymbol{C}_{-ig},\bar{C}_{ig}, N_g).
\label{eq:contraction4}
\end{equation}
Finally, applying Contraction to \eqref{eq:contraction1} and \eqref{eq:contraction4}, 
\[
(S_g, Z_{ig}, \bar{D}_{ig}) \indep (\boldsymbol{C}_{-ig},B_{ig}, C_{ig})|(\bar{C}_{ig}, N_g)
\]
and the result follows by a final application of Decomposition.
\end{proof}

\begin{proof}[Proof of \autoref{lem:QfromQ0Q1}]
  Define the shorthand $U \equiv \mathbf{Q}(\bar{c}, n), A \equiv \mathbf{Q}_0(\bar{c},n)$, and $B = \mathbf{Q}_1(\bar{c},n)$ so that
\[
  U = \begin{bmatrix}
    A + B & B \\
    B & B
  \end{bmatrix}.
\]
Using this notation, we are asked to show that $U$ is invertible if and only if $A$ and $B$ are both invertible, in which case $U^{-1} = V$ where 
\[
  V \equiv \begin{bmatrix}
    A^{-1} & -A^{-1}\\
    -A^{-1} & A^{-1} + B^{-1}
  \end{bmatrix}.
\]
  The ``if'' direction follows by direct calculation: $VU = UV= \mathbbm{I}$.
  For the ``only if'' direction, suppose that $U$ is invertible.
  Partitioning $U^{-1}$ into blocks $(C,D,E,F)$ conformably with the partition of $U$, we have 
  \[
    U U^{-1} = 
    \begin{bmatrix}
      A + B & B \\ B & B
    \end{bmatrix}
    \begin{bmatrix}
      C & D \\ E & F
    \end{bmatrix} = 
    \begin{bmatrix}
      \mathbbm{I} & 0 \\ 0 & \mathbbm{I}
    \end{bmatrix} = 
    \begin{bmatrix}
      C & D \\ E & F
    \end{bmatrix}  
    \begin{bmatrix}
      A + B & B \\ B & B
    \end{bmatrix} = U^{-1} U.
  \]

We begin by showing that $A$ is invertible.
Consider the product $U U^{-1}$.
Multiplying the first row of $U$ by the first column of $U^{-1}$ gives the equation $AC + B(C+E) = \mathbbm{I}$; multiplying the second row of $U$ by the first column of $U^{-1}$ gives $B(C + E) = 0$.
Combining these, $AC = \mathbbm{I}$.
Now consider the product $U^{-1}U$.
Multiplying the first row of $U^{-1}$ by the first column of $U$ gives $CA + (C + D)B = \mathbbm{I}$; multiplying the first row of $U^{-1}$ by the second column of $U$ gives $(C + D)B = 0$.
Combining these, $CA = \mathbbm{I}$.
Since $AC = CA = \mathbbm{I}$, we have shown that $A$ is invertible with $A^{-1} = C$.

We next show that $D = E = -C$.
Consider again the product $UU^{-1}$.
Multiplying the first row of $U$ by the second column of $U^{-1}$ gives $AD + B(D + F) = 0$; multiplying the second row of $U$ by the second column of $U^{-1}$ gives $B(D + F) = \mathbbm{I}$.
Combining these, $AD = -\mathbbm{I}$ and because $A^{-1}=C$ we can solve this equation to yield $D = -C$.
Now consider $U^{-1}U$.
Multiplying the second row of $U^{-1}$ by the first column of $U$ gives $EA + (E + F)B = 0$; multiplying the second row of $U^{-1}$ by the second column of $U$ gives $(E + F)B = \mathbbm{I}$.
Combining these, $EA = -\mathbbm{I}$ and solving for $E$, we have $E = -C$ since $A^{-1}=C$.

Finally we show that $B$ is invertible.
Multiplying the second row of $U$ by the second column of $U^{-1}$ gives $B(D + F) = \mathbbm{I}$, but since $D = -C$ this becomes $B(F - C) = \mathbbm{I}$
Multiplying the second row of $U^{-1}$ by the first column of $U$ gives $(E + F) B + EA = 0$ but because $E = -C = -A^{-1}$ this becomes $(F - C) B = \mathbbm{I}$.
Thus, $B (F - C) = (F - C) B = \mathbbm{I}$ so we have shown that $B$ is invertible with $B^{-1} = F - C$. 
\end{proof}


\begin{proof}[Proof of \autoref{thm:expect}]

  For each part, it suffices to find an appropriate outcome variable $\widetilde{Y}_{ig}$, regressor vector $\widetilde{\mathbf{X}}_{ig}$, and instrument set $\widetilde{\boldsymbol{\mathcal{Z}}}_{ig}$ such that we can write 
  $\widetilde{Y}_{ig} = \widetilde{\mathbf{X}}_{ig}' \boldsymbol{\vartheta} + U_{ig}$ where $\boldsymbol{\vartheta}$ is the parameter of interest, $\mathbbm{E}[\boldsymbol{\widetilde{\mathcal{Z}}}_{ig} U_{ig}] = \boldsymbol{0}$, and $\mathbbm{E}[\widetilde{\boldsymbol{\mathcal{Z}}}_{ig} \widetilde{\mathbf{X}}_{ig}']$ is invertible.
  Note that $(\widetilde{\mathbf{X}}_{ig}, \widetilde{Y}_{ig}, \widetilde{\boldsymbol{\mathcal{Z}}}_{ig})$ are placeholders for quantities that differ in each part of the proof: for part (i) they represent $(\mathbf{X}_{ig}, Y_{ig}, \boldsymbol{\mathcal{Z}}_{ig}^W)$ while for part (ii) they stand for $\left(D_{ig} \mathbf{f}(\bar{D}_{ig}),D_{ig} Y_{ig}, \boldsymbol{\mathcal{Z}}^1_{ig}\right)$, for example.
  The definitions of $U_{ig}$ and $\boldsymbol{\vartheta}$ are also specific to each part of the proof.

\paragraph{Part (i)}

By \eqref{eq:XBdef} we can write $\widetilde{Y}_{ig} = \widetilde{\mathbf{X}}_{ig}' \boldsymbol{\vartheta} + U_{ig}$ where 
$\boldsymbol{\vartheta}' \equiv \begin{bmatrix} 
 \mathbbm{E}(\boldsymbol{\theta}_{ig}') &  \mathbbm{E}(\boldsymbol{\psi}_{ig}' - \boldsymbol{\vartheta}_{ig}'|C_{ig} = 1) 
  \end{bmatrix}$, 
$\widetilde{Y}_{ig} \equiv Y_{ig}$, $\widetilde{\mathbf{X}}_{ig} \equiv \mathbf{X}_{ig}$, and  
$U_{ig} \equiv \mathbf{X}_{ig}'(\mathbf{B}_{ig} - \boldsymbol{\vartheta})$.
Under IOR $D_{ig} = C_{ig} Z_{ig}$.
Hence, defining $\mathbf{M}_{ig} \equiv \text{diag}\left\{ 1, C_{ig} \right\} \otimes \mathbbm{I}_K$, 
  \begin{align*}
    \mathbf{X}_{ig} 
    &= \left(\begin{bmatrix}
      1 & 0 \\ 
      0 & C_{ig}
    \end{bmatrix} \begin{bmatrix}
      1 \\ Z_{ig}
    \end{bmatrix}\right) \otimes \left[\mathbbm{I}_K \mathbf{f}(\bar{D}_{ig})\right]
    = \left( \begin{bmatrix}
      1 & 0 \\ 0 & C_{ig}
    \end{bmatrix} \otimes \mathbbm{I}_K\right)\left( \begin{bmatrix}
      1 \\ Z_{ig}
  \end{bmatrix}\otimes \mathbf{f}(\bar{D}_{ig})\right) = \mathbf{M}_{ig} \mathbf{W}_{ig}.
  \end{align*}
  Since $\mathbf{M}_{ig}$ is symmetric, $U_{ig} = \mathbf{W}_{ig}'\left[ \mathbf{M}_{ig} \left(\mathbf{B}_{ig} - \boldsymbol{\vartheta}\right) \right]$.
  Thus, taking $\widetilde{\boldsymbol{\mathcal{Z}}}_{ig} \equiv \boldsymbol{\mathcal{Z}}_{ig}^W$, we have
\begin{align*}
  \mathbbm{E}[\widetilde{\boldsymbol{\mathcal{Z}}}_{ig} U_{ig}] &= \mathbbm{E}\left\{ \mathbbm{E}\left[\left. \widetilde{\boldsymbol{\mathcal{Z}}}_{ig} U_{ig} \right| \bar{C}_{ig}, N_g \right] \right\} = \mathbbm{E}\left\{\mathbf{Q}(\bar{C}_{ig}, N_g)^{-1} \mathbbm{E}\left[\left. \mathbf{W}_{ig}\mathbf{W}_{ig}' \mathbf{M}_{ig} \left(\mathbf{B}_{ig} - \boldsymbol{\vartheta}\right)\right|\bar{C}_{ig}, N_g  \right]\right\}
\end{align*}
by iterated expectations.
By assumption $(Z_{ig}, \bar{D}_{ig}) \indep (C_{ig}, \mathbf{B}_{ig})|(\bar{C}_{ig}, N_g)$. Hence, 
\[
  \mathbbm{E}\left[ \left.\mathbf{W}_{ig}\mathbf{W}_{ig}' \mathbf{M}_{ig}(\mathbf{B}_{ig} - \boldsymbol{\vartheta})\right|\bar{C}_{ig}, N_g\right] = 
    \mathbbm{E}\left[ \left.\mathbf{W}_{ig}\mathbf{W}_{ig}' \right|\bar{C}_{ig}, N_g \right] \mathbbm{E}\left[\left.\mathbf{M}_{ig}(\mathbf{B}_{ig} - \boldsymbol{\vartheta})\right|\bar{C}_{ig}, N_g\right] 
\]
by Decomposition, since  $\mathbf{W}_{ig}\mathbf{W}_{ig}'$ is a measurable function of $(Z_{ig}, \bar{D}_{ig})$ and $\mathbf{M}_{ig}(\mathbf{B}_{ig} - \boldsymbol{\vartheta})$ is a measurable function of $(C_{ig}, \mathbf{B}_{ig})$.
Substituting into the expression for $\mathbbm{E}[ \widetilde{\boldsymbol{\mathcal{Z}}}_{ig} U_{ig}]$,
\[
  \mathbbm{E}\left[ \widetilde{\boldsymbol{\mathcal{Z}}}_{ig} U_{ig} \right] = \mathbbm{E}\left\{  \mathbbm{E}\left[\left.\mathbf{M}_{ig}(\mathbf{B}_{ig} - \boldsymbol{\vartheta})\right|\bar{C}_{ig}, N_g\right] \right\} = \mathbbm{E}\left[ \mathbf{M}_{ig} \left( \mathbf{B}_{ig} - \boldsymbol{\vartheta} \right) \right]
\]
by iterated expectations, since $\mathbf{Q}(\bar{C}_{ig}, N_g)^{-1} = \mathbbm{E}[ \mathbf{W}_{ig}\mathbf{W}_{ig}'|\bar{C}_{ig}, N_g]^{-1}$.
  Now, substituting the definitions of $\mathbf{M}_{ig}$, $\mathbf{B}_{ig}$, and $\boldsymbol{\vartheta}$, 
\[
  \mathbbm{E}\left[ \mathbf{M}_{ig} \left( \mathbf{B}_{ig} - \boldsymbol{\vartheta} \right) \right] = \mathbbm{E} 
\begin{bmatrix}
  \left(\boldsymbol{\theta}_{ig} - \mathbbm{E}\left\{ \boldsymbol{\theta}_{ig} \right\} \right)\\
  C_{ig} \left(\left\{ \boldsymbol{\psi}_{ig} - \boldsymbol{\theta}_{ig}  \right\} - \mathbbm{E}\left\{\left.\boldsymbol{\psi}_{ig} - \boldsymbol{\theta}_{ig}\right|C_{ig} = 1\right\} \right) 
    \end{bmatrix} = \mathbf{0}
\]
since $\mathbbm{E}\left[ C_{ig} \left( \boldsymbol{\psi}_{ig} - \boldsymbol{\theta}_{ig} \right) \right] = \mathbbm{E}(C_{ig}) \mathbbm{E}\left(\left. \boldsymbol{\psi}_{ig} - \boldsymbol{\theta}_{ig}\right| C_{ig} = 1 \right)$.
  Therefore $\mathbbm{E}[\widetilde{\boldsymbol{\mathcal{Z}}}_{ig} U_{ig}] = \mathbf{0}$. 
  Similarly,
\begin{align*}
  \mathbbm{E}\left[ \widetilde{\boldsymbol{\mathcal{Z}}}_{ig} \widetilde{\mathbf{X}}_{ig}' \right] 
  &= \mathbbm{E}\left\{ \mathbf{Q}(\bar{C}_{ig}, N_g)^{-1} \mathbbm{E}\left[ \mathbf{W}_{ig} \mathbf{W}_{ig}' \mathbf{M}_{ig}|\bar{C}_{ig}, N_g \right]\right\}\\
  &= \mathbbm{E}\left\{ \mathbf{Q}(\bar{C}_{ig}, N_g)^{-1} \mathbbm{E}\left[ \mathbf{W}_{ig} \mathbf{W}_{ig}'|\bar{C}_{ig}, N_g\right] \mathbbm{E}\left[ \mathbf{M}_{ig}|\bar{C}_{ig}, N_g \right]\right\} = \mathbbm{E}\left[ \mathbf{M}_{ig} \right].
\end{align*}
Since $\mathbbm[\mathbf{M}_{ig}]$ is invertible if and only if $\mathbbm{E}(C_{ig}) \neq 0$, it follows that $\mathbbm{E}[\widetilde{\boldsymbol{\mathcal{Z}}}_{ig} \widetilde{\mathbf{X}}_{ig}']$ is invertible by \autoref{assump:rank}.  

\paragraph{Part (ii)}
Since $D_{ig}^2 = D_{ig}$ and $D_{ig}(1 - D_{ig}) = 0$, multiplying both sides of \eqref{eq:XBdef} by $D_{ig}$ and simplifying gives $D_{ig} Y_{ig} = D_{ig} \mathbf{f}(\bar{D}_{ig}) \boldsymbol{\psi}_{ig}$.
Thus $\widetilde{Y}_{ig} = \widetilde{\mathbf{X}}_{ig}' \boldsymbol{\vartheta} + U_{ig}$ where $\boldsymbol{\vartheta} \equiv \mathbbm{E}(\boldsymbol{\psi}_{ig}|C_{ig} = 1)$, $\widetilde{Y} \equiv D_{ig} Y_{ig}$, $\widetilde{\mathbf{X}}_{ig} \equiv  D_{ig} \mathbf{f}(\bar{D}_{ig})$, and $U_{ig} \equiv \left[D_{ig} \mathbf{f}(\bar{D}_{ig})\right]' (\boldsymbol{\psi}_{ig} - \boldsymbol{\vartheta})$.
The remainder of the argument is similar to that of part (i).
Taking $\widetilde{\boldsymbol{\mathcal{Z}}}_{ig} \equiv \boldsymbol{\mathcal{Z}}_{ig}^1$ and substituting $D_{ig} = Z_{ig}C_{ig}$ gives 
\begin{align*}
  \mathbbm{E}[\widetilde{\boldsymbol{\mathcal{Z}}}_{ig} U_{ig}]  
&= \mathbbm{E}\left\{  \mathbf{Q}_1(\bar{C}_{ig}, N_g)^{-1} \mathbbm{E}\left[ \left. \mathbf{f}(\bar{D}_{ig}) \mathbf{f}(\bar{D}_{ig})' Z_{ig}|\bar{C}_{ig}, N_g\right] \mathbbm{E}\left[ C_{ig} (\boldsymbol{\psi}_{ig} - \boldsymbol{\vartheta}) \right| \bar{C}_{ig}, N_g \right] \right\}\\
  &= \mathbbm{E}\left\{\mathbbm{E}\left[ C_{ig} (\boldsymbol{\psi}_{ig} - \boldsymbol{\vartheta})|\bar{C}_{ig}, N_g \right]\right\} = \mathbbm{E}\left[ C_{ig} (\boldsymbol{\psi}_{ig} - \boldsymbol{\vartheta})\right].
\end{align*}
Since $\mathbbm{E}[C_{ig} \boldsymbol{\psi}_{ig}] = \mathbbm{E}(C_{ig})\mathbbm{E}(\boldsymbol{\psi}_{ig}|C_{ig} = 1) = \mathbbm{E}(C_{ig} \boldsymbol{\vartheta})$, we obtain $\mathbbm{E}(\widetilde{\boldsymbol{\mathcal{Z}}}_{ig} U_{ig}) = \boldsymbol{0}$.
Similarly,
\begin{align*}
  \mathbbm{E}\left[ \widetilde{\boldsymbol{\mathcal{Z}}}_{ig} \widetilde{\mathbf{X}}_{ig}' \right] &= \mathbbm{E}\left\{ \mathbf{Q}_1(\bar{C}_{ig}, N_g)^{-1} \mathbbm{E}\left[  \mathbf{f}(\bar{D}_{ig}) \mathbf{f}(\bar{D}_{ig})' Z_{ig} C_{ig}|\bar{C}_{ig}, N_g \right] \right\}\\
  &= \mathbbm{E}\left\{ \mathbf{Q}_1(\bar{C}_{ig}, N_g)^{-1} \mathbbm{E}\left[  \mathbf{f}(\bar{D}_{ig}) \mathbf{f}(\bar{D}_{ig})' Z_{ig} | \bar{C}_{ig}, N_g\right] \mathbbm{E} \left[C_{ig}|\bar{C}_{ig}, N_g \right] \right\} = \mathbbm{E}(C_{ig}) \mathbbm{I}_K.
\end{align*}
Hence, $\mathbbm{E}[\widetilde{\boldsymbol{\mathcal{Z}}}_{ig} \widetilde{\mathbf{X}}_{ig}]'$ is invertible by \autoref{assump:rank}.


\paragraph{Part (iii)}
Since $(1 - D_{ig})^2 = (1 - D_{ig})$ and $D_{ig}(1 - D_{ig}) = 0$, multiplying both sides of \eqref{eq:XBdef} by $Z_{ig}(1 - D_{ig})$ and simplifying gives $Z_{ig}(1 - D_{ig}) Y_{ig} = Z_{ig}(1 - D_{ig}) \mathbf{f}(\bar{D}_{ig}) \boldsymbol{\theta}_{ig}$.
Thus we have $\widetilde{Y}_{ig} = \widetilde{\mathbf{X}}_{ig}' \boldsymbol{\vartheta} + U_{ig}$ where $\boldsymbol{\vartheta} \equiv \mathbbm{E}(\boldsymbol{\theta}_{ig}|C_{ig} = 0)$, $\widetilde{Y}_{ig} \equiv Z_{ig}(1 - D_{ig})Y_{ig}$, $\widetilde{\mathbf{X}}_{ig} \equiv Z_{ig}(1 - D_{ig}) \mathbf{f}(\bar{D}_{ig})$, and $U_{ig} \equiv \left[Z_{ig}(1 - D_{ig}) \mathbf{f}(\bar{D}_{ig})\right]' (\boldsymbol{\theta}_{ig} - \boldsymbol{\vartheta})$.
The remainder of the argument is similar to that of part (i).
Taking
$\widetilde{\boldsymbol{\mathcal{Z}}}_{ig} \equiv \boldsymbol{\mathcal{Z}}_{ig}^1$ and substituting $Z_{ig} (1 - D_{ig}) = Z_{ig} (1 - C_{ig})$ gives 
\begin{align*}
  \mathbbm{E}[\boldsymbol{\widetilde{\mathcal{Z}}_{ig}} U_{ig}]  
    &= \mathbbm{E}\left\{  \mathbf{Q}_1(\bar{C}_{ig}, N_g)^{-1} \mathbbm{E}\left[ \left. \mathbf{f}(\bar{D}_{ig}) \mathbf{f}(\bar{D}_{ig})' Z_{ig}|\bar{C}_{ig}, N_g\right] \mathbbm{E}\left[ (1 - C_{ig}) (\boldsymbol{\theta}_{ig} - \boldsymbol{\vartheta}) \right| \bar{C}_{ig}, N_g \right] \right\}\\
      &= \mathbbm{E}\left\{\mathbbm{E}\left[ (1 - C_{ig}) (\boldsymbol{\theta}_{ig} - \boldsymbol{\vartheta})|\bar{C}_{ig}, N_g \right]\right\} = \mathbbm{E}\left[ (1 - C_{ig}) (\boldsymbol{\theta}_{ig} - \boldsymbol{\vartheta})\right].
\end{align*}
Since $\mathbbm{E}[(1 - C_{ig}) \boldsymbol{\theta}_{ig}] = \mathbbm{E}(1 - C_{ig})\mathbbm{E}(\boldsymbol{\theta}_{ig}|C_{ig} = 1) = \mathbbm{E}[(1 - C_{ig}) \boldsymbol{\vartheta}]$, we obtain $\mathbbm{E}(\widetilde{\boldsymbol{\mathcal{Z}}}_{ig} U_{ig}) = \boldsymbol{0}$.
Similarly,
\begin{align*}
  \mathbbm{E}\left[\widetilde{\boldsymbol{\mathcal{Z}}}_{ig} \widetilde{\mathbf{X}}_{ig}' \right] &= \mathbbm{E}\left\{ \mathbf{Q}_1(\bar{C}_{ig}, N_g)^{-1} \mathbbm{E}\left[  \mathbf{f}(\bar{D}_{ig}) \mathbf{f}(\bar{D}_{ig})' Z_{ig} (1 - C_{ig})|\bar{C}_{ig}, N_g \right] \right\}\\
  &= \mathbbm{E}\left\{ \mathbf{Q}_1(\bar{C}_{ig}, N_g)^{-1} \mathbbm{E}\left[  \mathbf{f}(\bar{D}_{ig}) \mathbf{f}(\bar{D}_{ig})' Z_{ig} | \bar{C}_{ig}, N_g\right] \mathbbm{E} \left[(1 - C_{ig})|\bar{C}_{ig}, N_g \right] \right\} = \mathbbm{E}(1 - C_{ig} )\mathbbm{I}_K.
\end{align*}
It follows that $\mathbbm{E}[\widetilde{\boldsymbol{\mathcal{Z}}}_{ig} \widetilde{\mathbf{X}}_{ig}']$ is invertible by \autoref{assump:rank}.


\paragraph{Part (iv)}
Under one-sided non-compliance and IOR, $(1 - Z_{ig})(1 - D_{ig}) = (1 - Z_{ig})$.
Hence, multiplying both sides of \eqref{eq:XBdef} by $(1 - Z_{ig})$, we obtain $(1 - Z_{ig})Y_{ig} = (1 - Z_{ig}) \mathbf{f}(\bar{D}_{ig})' \boldsymbol{\theta}_{ig}$, using the fact that $Z_{ig}(1 - Z_{ig}) = 0$.  
Thus we can write $\widetilde{Y}_{ig} = \widetilde{\mathbf{X}}_{ig}' \boldsymbol{\vartheta} + U_{ig}$ where $\boldsymbol{\vartheta} \equiv \mathbbm{E}(\boldsymbol{\theta}_{ig})$, $\widetilde{Y}_{ig} \equiv (1 - Z_{ig})Y_{ig}$, $\widetilde{\mathbf{X}}_{ig} \equiv (1 - Z_{ig}) \mathbf{f}(\bar{D}_{ig})$, and $U_{ig} \equiv (1 - Z_{ig})\mathbf{f}(\bar{D}_{ig})'(\boldsymbol{\theta}_{ig} - \boldsymbol{\vartheta})$.
The remainder of the argument is similar to that of part (i).
Taking $\widetilde{\boldsymbol{\mathcal{Z}}}_{ig} \equiv \boldsymbol{\mathcal{Z}}_{ig}^0$, we obtain
\begin{align*}
  \mathbbm{E}[\widetilde{\boldsymbol{\mathcal{Z}}}_{ig} U_{ig}]  
    &= \mathbbm{E}\left\{  \mathbf{Q}_0(\bar{C}_{ig}, N_g)^{-1} \mathbbm{E}\left[ \left. \mathbf{f}(\bar{D}_{ig}) \mathbf{f}(\bar{D}_{ig})' (1 - Z_{ig})|\bar{C}_{ig}, N_g\right] \mathbbm{E}\left[\boldsymbol{\theta}_{ig} - \boldsymbol{\vartheta} \right| \bar{C}_{ig}, N_g \right] \right\}\\
      &= \mathbbm{E}\left\{\mathbbm{E}\left[ \boldsymbol{\theta}_{ig} - \mathbbm{E}(\boldsymbol{\theta}_{ig})|\bar{C}_{ig}, N_g \right]\right\} = \boldsymbol{0}
\end{align*}
and 
$\mathbbm{E}\left[ \widetilde{\boldsymbol{\mathcal{Z}}}_{ig} \widetilde{\mathbf{X}}_{ig}' \right] = \mathbbm{E}\left\{ \mathbf{Q}_0(\bar{C}_{ig}, N_g)^{-1} \mathbbm{E}\left[  \mathbf{f}(\bar{D}_{ig}) \mathbf{f}(\bar{D}_{ig})' (1 - Z_{ig})|\bar{C}_{ig}, N_g \right] \right\} = \mathbbm{I}_K$.
\end{proof}


\begin{lem}
  \label{lem:Zindep}
  Under Assumptions \ref{assump:Bernoulli} and \ref{assump:exclusion}, $(S_g, Z_{ig}) \indep (C_{ig}, \bar{C}_{ig}, N_g, \mathbf{B}_{ig})$.
\end{lem}


\begin{proof}[Proof of \autoref{lem:Zindep}]
  By \autoref{assump:Bernoulli} $Z_{ig} \indep N_g |S_g$ and by \autoref{assump:exclusion} (ii) and Decomposition $Z_{ig} \indep (C_{ig}, \mathbf{B}_{ig}) | (S_g, N_g)$.
  Combining these by Contraction yields
  \begin{equation}
    Z_{ig} \indep (\boldsymbol{C}_{g}, \mathbf{B}_{ig}, N_g)|S_g.
    \label{Zindep1}
  \end{equation}
  Now, by \autoref{assump:exclusion} (i) we have $S_g \indep (\boldsymbol{C}_{g}, \mathbf{B}_{ig}, N_g)$.
  Combining this with \eqref{Zindep1} by a second application of Contraction gives $(Z_{ig}, S_g) \indep (\boldsymbol{C}_{g}, \mathbf{B}_{ig}, N_g)$. 
The result follows by a final application of Decomposition.
\end{proof}


\begin{proof}[Proof of \autoref{thm:identification}]
  Assumptions \ref{assump:saturations}--\ref{assump:exclusion} imply that $(Z_{ig}, \bar{D}_{ig})\indep (\mathbf{B}_{ig}, C_{ig})|(\bar{C}_{ig}, N_g)$ by \autoref{thm:conditional_indep}.
  Hence Assumptions \ref{assump:saturations}--\ref{assump:rank} are sufficient for the conclusions of \autoref{thm:expect} to hold.
  Now, by \autoref{lem:DbarDist}, Assumptions \ref{assump:saturations}--\ref{assump:Bernoulli} and \ref{assump:onesided}--\ref{assump:exclusion} imply that the conditional distribution of $\bar{D}_{ig}|(\bar{C}_{ig}, N_g, Z_{ig})$ is known. 
  Moreover, by \autoref{lem:Zindep},  $Z_{ig} \indep (\bar{C}_{ig}, N_g)$ so the distribution of $\boldsymbol{\mathcal{Z}}_{ig}|(\bar{C}_{ig}, N_g)$ is likewise known.
  It follows that $\mathbf{Q}, \mathbf{Q}_0$ and $\mathbf{Q}_1$ are \emph{known} functions of $(\bar{C}_{ig}, N_g)$.
  Since $N_g$ is observed, knowledge of $\bar{C}_{ig}$ is thus sufficient to identify the quantities
  \[
  \mathbbm{E}(\boldsymbol{\theta}_{ig}), \quad 
  \mathbbm{E}(\boldsymbol{\psi}_{ig} - \boldsymbol{\theta}_{ig}|C_{ig} = 1), \quad
  \mathbbm{E}(\boldsymbol{\psi}_{ig}|C_{ig}=1), \quad
  \mathbbm{E}(\boldsymbol{\theta}_{ig}|C_{ig} = 0)
\]
by the relevant parts of \autoref{thm:expect}.
Now, by iterated expectations,  
\[
  \mathbbm{E}(\boldsymbol{\theta}_{ig}|C_{ig} = 1) = \mathbbm{E}(\boldsymbol{\theta}_{ig}|C_{ig} = 0) + \frac{1}{\mathbbm{E}(C_{ig})} \left[ \mathbbm{E}(\boldsymbol{\theta}_{ig}) - \mathbbm{E}(\boldsymbol{\theta}_{ig}|C_{ig} = 0) \right].
\]
Since $\mathbbm{E}(C_{ig}) = \mathbbm{E}(D_{ig}|Z_{ig}=1)$, it follows that $\mathbbm{E}(\boldsymbol{\theta}_{ig}|C_{ig} = 1)$ is identified.
Under IOR and one-sided non-compliance $\left\{ D_{ig} = 1\right\} = \left\{ C_{ig} = 1, Z_{ig} = 1 \right\}$, and applying Weak Union and Decomposition to \autoref{lem:Zindep}, we see that $Z_{ig} \indep \mathbf{B}_{ig}|C_{ig}$.
Thus,
\[
  \mathbbm{E}(\mathbf{B}_{ig}|D_{ig}=1) = \mathbbm{E}(\mathbf{B}_{ig}|C_{ig} = 1,Z_{ig}=1) = \mathbbm{E}(\mathbf{B}_{ig}|C_{ig}=1) .
\]
The result follows since $Y_{ig}(d,\bar{d}) = \mathbf{f}(\bar{d})' \boldsymbol{\theta}_{ig} + d \mathbf{f}(\bar{d})' (\boldsymbol{\psi}_{ig} - \boldsymbol{\theta}_{ig} )$ under \autoref{assump:randcoef}.
\end{proof}

\begin{proof}[Proof of \autoref{thm:consistency}]
  Substituting the model into the definition of $\widehat{\boldsymbol{\vartheta}}$ and $\rho_g \equiv N_g / \mathbbm{E}(N_g)$,
\begin{align*}
  \widehat{\boldsymbol{\vartheta}} - \boldsymbol{\vartheta} 
  &=  \left(\sum_{g=1}^{G} \sum_{i=1}^{N_g} \widehat{\boldsymbol{\mathcal{Z}}}_{ig} \boldsymbol{X}_{ig}'\right)^{-1} \left(\sum_{g=1}^{G} \sum_{i=1}^{N_g} \widehat{\boldsymbol{\mathcal{Z}}}_{ig} U_{ig}\right)\\
    &= \left( \frac{1}{G} \sum_{g=1}^{G} \mathbf{A}_g + \frac{1}{G} \sum_{g=1}^{G} \mathbf{R}^{(1)}_g  \right)^{-1} \left( \frac{1}{G} \sum_{g=1}^{G} \mathbf{P}_g + \frac{1}{G} \sum_{g=1}^{G} \mathbf{R}^{(2)}_g  \right)
  \end{align*}
where we define
\begin{align*}
 \mathbf{A}_g &\equiv \frac{1}{N_g}\sum_{i=1}^{N_g} \rho_g \widehat{\boldsymbol{\mathcal{Z}}}_{ig} \boldsymbol{X}_{ig}' 
  &  \mathbf{R}_{g}^{(1)} &\equiv \frac{1}{N_g}\sum_{i=1}^{N_g} \rho_g (\widehat{\boldsymbol{\mathcal{Z}}}_{ig} - \boldsymbol{\mathcal{Z}}_{ig}) \boldsymbol{X}_{ig}'\\
 \mathbf{P}_g &\equiv \frac{1}{N_g}\sum_{i=1}^{N_g} \rho_g \widehat{\boldsymbol{\mathcal{Z}}}_{ig} U_{ig}' 
 &  \mathbf{R}_{g}^{(2)} &\equiv \frac{1}{N_g}\sum_{i=1}^{N_g} \rho_g (\widehat{\boldsymbol{\mathcal{Z}}}_{ig} - \boldsymbol{\mathcal{Z}}_{ig}) U_{ig}.
\end{align*}
By assumption, both $\lvert| \sum_{g = 1}^{G} \mathbf{R}^{(1)}_g \rvert|$ and $\lvert| \sum_{g = 1}^{G} \mathbf{R}^{(2)}_g \rvert|$ are $o_{\mathbbm{P}}(G)$ and thus 
\[
  \widehat{\boldsymbol{\vartheta}} - \boldsymbol{\vartheta} 
  = \left( \frac{1}{G} \sum_{g=1}^{G} \mathbf{A}_g + o_{\mathbbm{P}}(1)   \right)^{-1} \left( \frac{1}{G} \sum_{g=1}^{G} \mathbf{P}_g + o_{\mathbbm{P}}(1)\right)
  \]
Now, since we observe a random sample of groups and $\mathbf{A}_g$ is a group-level random variable
\begin{align*}
  \mathbbm{E}\left[ \frac{1}{G}\sum_{g=1}^G \mathbf{A}_g\right] = \mathbbm{E}(\mathbf{A}_g) = \mathbbm{E}\left[ \frac{1}{N_g} \sum_{i=1}^{N_g} \mathbbm{E}\left( \rho_g \boldsymbol{\mathcal{Z}}_{ig} \mathbf{X}_{ig}'|N_g \right) \right] = \mathbbm{E}\left[ \mathbbm{E}\left( \rho_g \boldsymbol{\mathcal{Z}}_{ig} \mathbf{X}_{ig}'|N_g \right) \right] = \mathbbm{E}(\rho_g \boldsymbol{\mathcal{Z}}_{ig} \mathbf{X}_{ig}') 
\end{align*}
where the second equality uses iterated expectations and linearity, the third uses the assumption of identical distribution within groups, and the the fourth uses iterated expectations a second time.
Now consider an arbitrary entry $A^{(j,k)}_g$ of the matrix $\mathbf{A}_g$ and let $\lVert \cdot \rVert_F$ denote the Frobenius norm.
By the triangle and Cauchy-Schwarz inequalities, and using the assumption of identical distribution with group, we have
\begin{align*}
  \text{Var}\left(\frac{1}{G} \sum_{g=1}^G A^{(j,k)}_g   \right) &= \frac{1}{G} \text{Var}\left(A^{(j,k)}_g\right) \leq \frac{1}{G} \mathbbm{E}\left[ \lvert| \mathbf{A}_g \rvert|_F^2\right] 
  =  \frac{1}{G} \mathbbm{E}\left( \frac{1}{N_g^2}\left\Vert  \sum_{i=1}^{N_g} \rho_g \boldsymbol{\mathcal{Z}}_{ig} \mathbf{X}_{ig}' \right\Vert_F^2 \right)\\  
  &\leq  \frac{1}{G} \mathbbm{E}\left[ \frac{1}{N_g^2} \left(\sum_{i=1}^{N_g}\left\Vert  \rho_g \boldsymbol{\mathcal{Z}}_{ig} \mathbf{X}_{ig}' \right\Vert_F\right)^2 \right]\\  
  &=  \frac{1}{G} \mathbbm{E}\left[ \frac{1}{N_g^2} \mathbbm{E}\left(\left.\sum_{i,j \leq N_g}\left\Vert  \rho_g \boldsymbol{\mathcal{Z}}_{ig} \mathbf{X}_{ig}' \right\Vert_F\left\Vert  \rho_g \boldsymbol{\mathcal{Z}}_{jg} \mathbf{X}_{jg}' \right\Vert_F \right| N_g\right) \right]\\  
  &\leq  \frac{1}{G} \mathbbm{E}\left[ \frac{1}{N_g^2} \mathbbm{E}\left(\left.\sum_{i,j \leq N_g}\left\Vert  \rho_g \boldsymbol{\mathcal{Z}}_{ig} \mathbf{X}_{ig}' \right\Vert_F^2 \right| N_g\right) \right]\\  
  &=  \frac{1}{G} \mathbbm{E}\left[ \mathbbm{E}\left(\left. \left\Vert  \rho_g \boldsymbol{\mathcal{Z}}_{ig} \mathbf{X}_{ig}' \right\Vert_F^2 \right| N_g\right) \right] = \frac{1}{G} \mathbbm{E}\left[\rho_g^2 \left\Vert \boldsymbol{\mathcal{Z}}_{ig} \mathbf{X}_{ig}' \right\Vert_F^2  \right] \rightarrow 0
\end{align*}
since all finite-dimensional norms are equivalent and $\mathbbm{E}\left[\rho_g^2 \left\Vert \boldsymbol{\mathcal{Z}}_{ig} \mathbf{X}_{ig}' \right\Vert_F^2  \right] = o(G)$.
Hence, by the $L^2$ weak law of large numbers $G^{-1}\sum_{g=1}^G \mathbf{A}_g \rightarrow_p \mathbbm{E}(\rho_g \boldsymbol{\mathcal{Z}}_{ig} \mathbf{X}_{ig}') = \mathbbm{I}$.
An analogous argument shows that $G^{-1}\sum_{g=1}^G \mathbf{P}_g \rightarrow_p \mathbbm{E}(\rho_g \boldsymbol{\mathcal{Z}}_{ig} U_{ig}) = \mathbf{0}$.
The result follows by the continuous mapping theorem.
\end{proof}

\begin{proof}[Proof of \autoref{thm:asymptoticnormality}]
  Continuing the argument from the proof of \autoref{thm:consistency}, we have
\[
  \sqrt{G}(\widehat{\boldsymbol{\vartheta}} - \boldsymbol{\vartheta}) 
  = \left[ \mathbbm{I} + o_{\mathbbm{P}}(1) \right]^{-1} \left( \frac{1}{\sqrt{G}} \sum_{g=1}^{G} \mathbf{P}_g + \frac{1}{\sqrt{G}} \sum_{g=1}^{G} \mathbf{R}^{(2)}_g  \right).
\]
By assumption, $\lvert| \sum_{g = 1}^{G} \mathbf{R}^{(2)}_g \rvert| = o_{\mathbbm{P}}(G^{1/2})$, and hence $\sqrt{G}(\widehat{\boldsymbol{\vartheta}} - \boldsymbol{\vartheta}) 
= \frac{1}{\sqrt{G}} \sum_{g=1}^{G} \mathbf{P}_g + o_{\mathbbm{P}}(1)$.
Thus, it suffices to apply the Lindeberg-Feller central limit theorem to $\mathbf{P}_g/\sqrt{G}$.
Because we observe a random sample of groups, $\text{Var}(\sum_{g=1}^G \mathbf{P}_g/\sqrt{G}) = \text{Var}(\mathbbm{P}_g)$ which by assumption converges to $\Sigma$.
All that remains is to verify the Lindeberg condition, namely 
\[
  \mathbbm{E}\left[ \lvert| \mathbf{P}_g\rvert|^2 \mathbbm{1}\left\{ \lvert|\mathbf{P}_g \rvert| > \varepsilon \sqrt{G} \right\} \right] \rightarrow 0
\]
for any $\varepsilon > 0$.
A sufficient condition for this to hold is $G^{-\delta/2} \mathbbm{E}\left[ \lvert| \mathbf{P}_g \rvert|^{2 + \delta} \right] \rightarrow 0$ for some $\delta > 0$.
By an argument similar to that used to establish $\mathbbm{E}\left[ \lvert| \mathbf{A}_g \rvert|_F^2 \right] \leq \mathbbm{E}\left[ \rho_g^2 \lvert| \boldsymbol{\mathcal{Z}}_{ig} \mathbf{X}_{ig}'\rvert|_F^2 \right]$ in the proof of \autoref{thm:consistency}, we likewise have 
\[
  G^{-\delta/2} \mathbbm{E}\left[ \lvert| \mathbf{P}_g \rvert|^{2 + \delta} \right] \leq G^{-\delta/2} \mathbbm{E}\left[ \rho_g^{2 + \delta} \lvert| \boldsymbol{\mathcal{Z}}_{ig} \mathbf{X}_{ig}'\rvert|^{2+\delta} \right] = o(1)
\]
so the result follows.
\end{proof}

\begin{lem}
  \label{lem:appendChat1}
  Let $\bar{Z}_g \equiv \sum_{j=1}^{N_g} Z_{jg}/N_g$.
  Under the conditions of \autoref{lem:Chat}, 
  \[
    \mathbbm{P}(\bar{Z}_g < \underline{s}/2) \leq \exp\left\{ -\underline{n} \underline{s}^2/2  \right\}.
\] 
\end{lem}


\begin{proof}[Proof of \autoref{lem:appendChat1}]
Conditional on $(N_g = n, S_g = s)$, the treatment offers $(Z_1, \dots, Z_{N_g})$ are a collection of $n$ iid Bernoulli$(s)$ random variables by \autoref{assump:Bernoulli}.
Hence, by Hoeffding's inequality
  \[
    \mathbbm{P}\left( \bar{Z}_g < \underline{s}/2|N_g = n, S_g = s \right) \leq \exp\left\{ -2n(s - \underline{s}/2)^2 \right\} \leq \exp\left\{ - \underline{n}\underline{s}^2/2 \right\}
  \]
  where the second inequality follows since $\underline{s} \leq s$.
Thus,
\[
  \mathbbm{P}(\bar{Z}_g < \underline{s}/2) = \sum_{n,s} \mathbbm{P}(\bar{Z}_g \leq  \underline{s}/2|N_g = n, S_g = s) \mathbbm{P}(N_g = n, S_g = s) \leq \exp\left\{ -2 \underline{n}\underline{s}^2/4 \right\}
\]
by the law of total probability.
The result follows since $\mathbbm{P}(\bar{Z}_g < \underline{s}/2) \leq \mathbbm{P}(Z_g \leq \underline{s}/2)$.
\end{proof}

\begin{lem}
  \label{lem:appendChat2}
  Let $\bar{C}_g = \sum_{j=1}^{N_g}C_{jg}/N_g$ and $\widehat{C}_g \equiv \sum_{j=1}^{N_g} D_{jg}/(N_g \bar{Z}_g)$, where $\bar{Z}_g$ is as defined in \autoref{lem:appendChat1}.
  Under the conditions of \autoref{lem:Chat} and for any $t > 0$,
  \[
    \mathbbm{P}\left(\left. \left| \widehat{C}_g - \bar{C}_g\right| \geq t \right| \bar{Z}_g \geq \underline{s}/2 \right) 
      \leq 2 \exp\left\{ -\underline{n}\underline{s}^2 t^2 / 2 \right\}.
  \]
\end{lem}


\begin{proof}[Proof of \autoref{lem:appendChat2}]
  Let $\mathcal{A} \equiv \left\{ \mathbf{C}_g = \mathbf{c}, N_g = n, \bar{C}_g = \bar{c}, N_g \bar{Z}_g = m, S_g = s \right\}$ where $m > 0$.
  Suppose first that $\bar{c} \neq 0$.
  In this case
  \[
    \mathbbm{P}\left(\left.\left|\widehat{C}_g - \bar{C}_g\right| > t \right|\mathcal{A}\right)
      = \mathbbm{P}\left(\left.\left| \sum_{j=1}^n \frac{c_j Z_{jg}}{m} - \bar{c} \right| > t \right| \mathcal{A} \right) 
        = \mathbbm{P}\left(\left.\left| \frac{1}{n\bar{c}}\sum_{j \in \mathcal{C}} Z_{jg}^* - \bar{c} \right| > t \right| \mathcal{A} \right) 
  \]
  where $\mathcal{C} \equiv \left\{ j\colon c_j = 1 \right\}$ and $Z_{jg}^* \equiv n\bar{c} Z_{jg}/m$.
  Given $\mathcal{A}$, the $\left\{ Z_{jg} \right\}_{j \in \mathcal{C}}$ are a sequence of $n \bar{c}$ draws made without replacement from a population of $m$ ones and $(n-m)$ zeros.
  Thus
\[
  \mathbbm{E}(Z_{jg}^*) = \frac{n \bar{c}}{m} \mathbbm{P}(Z_{jg} = 1|\mathcal{A}) = \frac{n\bar{c}}{m} \cdot \frac{m}{n} = \bar{c}.
\]
Moreover, since $Z_{jg} \in \left\{ 0, 1 \right\}$, each of the $Z_{jg}^*$ is bounded between $0$ and $n\bar{c}/m$. 
While these random variables are identically distributed, they are not independent---like the $Z_{jg}$ from which they are constructed, $\left\{ Z_{jg}^* \right\}_{j \in \mathcal{C}}$ are draws made without replacement from a finite population.
Under this form of dependence, however, Hoeffding's Inequality continues to apply \citep[p. 28]{Hoeffding1963} and hence
\[
  \mathbbm{P}\left(\left.\left|\widehat{C}_g - \bar{C}_g\right| > t \right|\mathcal{A}\right) \leq 2 \exp\left\{ \frac{-2 t^2 m^2}{n \bar{c}} \right\} \leq 2 \exp\left\{ -2n \left( \frac{m}{n} \right)^2 t^2 \right\}
\]
where the second inequality follows because $0 < \bar{c} \leq 1$.
If $\bar{c} = 0$, we have
  \[
    \mathbbm{P}\left(\left.\left|\widehat{C}_g - \bar{C}_g\right| > t \right|\mathcal{A}\right) 
      = \mathbbm{P}(|0 - 0| > t |\mathcal{A}) = 0 \leq 2 \exp\left\{ -2n \left( \frac{m}{n} \right)^2 t^2 \right\}
  \]
so this inequality holds for any $\bar{c}$.
Applying the law of total probability as in the proof of \autoref{lem:appendChat1}, we see that
\[
  \mathbbm{P}\left(\left.\left|\widehat{C}_g - \bar{C}_g\right| > t \right|N_g = n, N_g \bar{Z}_g = m \right) \leq 2 \exp\left\{ -2n \left( \frac{m}{n} \right)^2 t^2 \right\}
\]
and thus
\begin{align*}
    \mathbbm{P}\left(\left. \left| \widehat{C}_g - \bar{C}_g\right| \geq t \right| \bar{Z}_g \geq \underline{s}/2 \right) 
      &=  \sum_{\left\{(m,n)\colon \frac{m}{n} \geq \underline{s}/2 \right\}} \mathbbm{P}\left(\left.\left|\widehat{C}_g - \bar{C}_g\right| > t \right|N_g = n, N_g \bar{Z}_g = m \right) \\
        &\quad \times \mathbbm{P}(N_g \bar{Z}_g = m, N_g = n| \bar{Z}_g \geq \underline{s}/2)\\
        &\leq \sum_{\left\{(m,n)\colon \frac{m}{n} \geq \underline{s}/2 \right\}}
        2 \exp\left\{ -2n \left( \frac{m}{n} \right)^2 t^2 \right\} \mathbbm{P}(N_g \bar{Z}_g = m, N_g = n| \bar{Z}_g \geq \underline{s}/2)\\
        &\leq \sum_{\left\{(m,n)\colon \frac{m}{n} \geq \underline{s}/2 \right\}} 2 \exp\left\{ -\underline{n} \underline{s}^2 t^2/2 \right\} \mathbbm{P}(N_g \bar{Z}_g = m, N_g = n| \bar{Z}_g \geq \underline{s}/2)\\
        & = \exp\left\{ -\underline{n} \underline{s}^2 t^2/2 \right\}
\end{align*}
by a second application of the law of total probability, since $\underline{n} \leq N_g$.
\end{proof}


\begin{lem}
  \label{lem:appendChat3}
  Suppose that $\underline{s}\underline{n} > 2$.
  Then, under the conditions of \autoref{lem:Chat},
\[
  \mathbbm{P}\left( \left. \max_{1 \leq i \leq N_g} \left| \widehat{C}_{ig} - \bar{C}_{ig} \right| > t \right| \bar{Z}_g \geq \underline{s}/2 \right)
    \leq 2 \exp\left\{ -\underline{n} \underline{s}^2 h(\underline{s}\underline{n},t)^2/2 \right\}
\]
where we define
  \[
    h(x, t) \equiv \left( \frac{x - 2}{x} \right)^2 t - \left[ 1 - \left( \frac{x - 2}{x} \right)^2 \right] \frac{4}{x - 2}.
  \]
\end{lem}


\begin{proof}[Proof of \autoref{lem:appendChat3}]
  If $\bar{Z}_g > \underline{s}/2 > 1 / \underline{n}$, then $N_g \bar{Z}_g - Z_{ig} > 0$ and $N_g \bar{Z}_g > 0$.
  Hence, 
  \[
    \widehat{C}_{ig} \equiv \frac{\bar{D}_{ig}}{\bar{Z}_{ig}} = \frac{N_g \bar{D}_g - D_{ig}}{N_g \bar{Z}_g - Z_{ig}} = \frac{N_g \bar{Z}_g \widehat{C}_g - D_{ig}}{N_g \bar{Z}_g - Z_{ig}} 
    = \left( \frac{N_g \bar{Z}_g}{N_g \bar{Z}_g - Z_{ig}} \right) \widehat{C}_g - \frac{D_{ig}}{N_g \bar{Z}_g - Z_{ig}}.
  \]
Similar manipulations give
\[
  \bar{C}_{ig} = \left( \frac{N_g}{N_g - 1} \right) \bar{C}_g - \frac{C_{ig}}{N_g - 1}
\]
from which it follows that
\[
  \left| \widehat{C}_{ig} - \bar{C}_{ig} \right| \leq
  \left| 
  \left( \frac{N_g \bar{Z}_g}{N_g \bar{Z}_g - Z_{ig}} \right) \widehat{C}_g
  - \left( \frac{N_g}{N_g - 1} \right) \bar{C}_g \right|
  + \left| \frac{C_{ig}}{N_g - 1}
  - \frac{D_{ig}}{N_g \bar{Z}_g - Z_{ig}}
  \right|
\]
by the triangle inequality.
Using the fact that $Z_{ig}, D_{ig}$, and $C_{ig}$ are binary along with $\underline{n} \leq N_g$ and $\bar{Z}_g > \underline{s}/2 > 1/\underline{n}$, tedious but straightforward algebra allows us to bound the right-hand side of the preceding inequality from above, yielding 
\[
  \left| \widehat{C}_{ig} - \bar{C}_{ig} \right| \leq \left( \frac{\underline{s}\underline{n}}{\underline{s}\underline{n} - 2} \right)^2 \left| \widehat{C}_g - \bar{C}_g \right| + \left[ \left( \frac{\underline{s}\underline{n}}{\underline{s}\underline{n} - 2} \right)^2 + 1 \right] \frac{4}{ \underline{s} \underline{n} - 2}.
\]
Since this upper bound for $|\widehat{C}_{ig} - \bar{C}_{ig}|$ does not depend on $i$, it follows that
\[
  \max_{1 \leq i \leq N_g} \left| \widehat{C}_{ig} - \bar{C}_{ig} \right| \leq \left( \frac{\underline{s}\underline{n}}{\underline{s}\underline{n} - 2} \right)^2 \left| \widehat{C}_g - \bar{C}_g \right| + \left[ \left( \frac{\underline{s}\underline{n}}{\underline{s}\underline{n} - 2} \right)^2 + 1 \right] \frac{4}{ \underline{s} \underline{n} - 2}
\]
provided that $\bar{Z}_g > \underline{s}/2 > 1/ \underline{n}$.
In other words, so long as $\underline{s} \underline{n} > 2$ we have
\[
  \left\{ \bar{Z}_g \geq \underline{s}/2 \right\} \cap \left\{ \max_{1 \leq i \leq N_g}\left| \widehat{C}_{ig} - \bar{C}_{ig} \right| > t \right\} \subseteq \left\{ \bar{Z}_g > \underline{s} / 2 \right\} \cap \left\{ \left|\widehat{C}_{g} - \bar{C}_g\right| > h(\underline{s}\underline{n},t) \right\}.
\]
Therefore, by the monotonicity of probability
\[
  \mathbbm{P}\left( \left. \max_{1 \leq i \leq N_g} \left| \widehat{C}_{ig} - \bar{C}_{ig} \right| > t \right| \bar{Z}_g \geq \underline{s}/2 \right) \leq \mathbbm{P}\left( \left. \left| \widehat{C}_g - \bar{C}_g \right| > h(\underline{s}\underline{n},t) \right| \bar{Z}_g \geq \underline{s}/2 \right)
\]
and the result follows by \autoref{lem:appendChat2}.

\end{proof}

\begin{proof}[Proof of \autoref{lem:Chat}]
	By the law of total probability, \autoref{lem:appendChat2}, and \autoref{lem:appendChat3}
	\begin{align*}
		\mathbbm{P}\left( \max_{1 \leq i \leq N_g } \left| \widehat{C}_{ig} - \bar{C}_{ig} \right| > t \right) &\leq 
		\mathbbm{P}\left(\left. \max_{1 \leq i \leq N_g } \left| \widehat{C}_{ig} - \bar{C}_{ig} \right| > t \right| \bar{Z}_g \geq \underline{s}/2\right) + \mathbbm{P}(\bar{Z}_g < \underline{s}/2)\\
		&\leq 2 \exp\left\{ -\underline{n} \underline{s}^2 h(\underline{s}\underline{n},t)^2/2 \right\} + \exp\left\{ -\underline{n} \underline{s}^2/2 \right\}
	\end{align*}
	where $h(\cdot, \cdot)$ is as defined in \autoref{lem:appendChat3}.
	Expanding and simplifying, we see that
	\[
	h(\underline{s}\underline{n}, t)^2 \geq \left( \frac{\underline{s}\underline{n} - 2}{\underline{s}\underline{n}} \right)^4 t^2 - \frac{16 t}{\underline{s}\underline{n} - 2} \equiv h^*(\underline{s}\underline{n}, t).
	\]
	Now, for any $t \geq 1$ we have $\mathbbm{P}\left( \max_{1 \leq i \leq N_g } \left| \widehat{C}_{ig} - \bar{C}_{ig} \right| > t \right)=0$ since both $\widehat{C}_{ig}$ and $\bar{C}_{ig}$ are between zero and one.
	Since $h^*(\underline{s}\underline{n},t) < 1$ for any $t < 1$, it follows that
	\begin{align*}
		\mathbbm{P}\left( \max_{1 \leq i \leq N_g } \left| \widehat{C}_{ig} - \bar{C}_{ig} \right| > t \right) 
		&\leq 2 \exp\left\{ -\underline{n} \underline{s}^2 h(\underline{s}\underline{n},t)^2/2 \right\} + \exp\left\{ -\underline{n} \underline{s}^2/2 \right\}\\
		&\leq 2 \exp\left\{ -\underline{n} \underline{s}^2 h^*(\underline{s}\underline{n},t)/2 \right\} + \exp\left\{ -\underline{n} \underline{s}^2/2 \right\}\\
		&\leq 3 \exp\left\{ -\underline{n} \underline{s}^2 h^*(\underline{s}\underline{n},t)/2 \right\} 
	\end{align*}
	Applying the union bound we obtain
	\begin{align*}
		\mathbbm{P}\left(\max_{1 \leq g \leq G} \max_{1\leq i \leq N_g} \left| \widehat{C}_{ig} - \bar{C}_{ig}\right| > t \right)
		&= \mathbbm{P}\left(\bigcup_{g=1}^G \left\{ \max_{1 \leq i \leq N_g}\left| \widehat{C}_{ig} - \bar{C}_{ig} \right| > t\right\}   \right)\\
		&\leq \sum_{g=1}^G \mathbbm{P}\left( \max_{1 \leq i \leq N_g} \left| \widehat{C}_{ig} - \bar{C}_{ig} \right| > t \right)\\
		&\leq \sum_{g=1}^G 3\exp\left\{ -\underline{n} \underline{s}^2 h^*(\underline{s}\underline{n}, t)/2 \right\}\\
		&= 3 G \exp\left\{ -\underline{n} \underline{s}^2 h^*(\underline{s}\underline{n}, t)/2 \right\}
	\end{align*}
	and accordingly we have
	\begin{align*}
		&p(\underline{n},G,M) \equiv \mathbbm{P}\left(\max_{1 \leq g \leq G} \max_{1\leq i \leq N_g} \left| \widehat{C}_{ig} - \bar{C}_{ig}\right| > M \sqrt{\frac{\log G}{\underline{n}}}\right)  \\
		&\leq 3 G \exp\left\{ \frac{-\underline{n}\underline{s}^2}{2}\left[ \left( \frac{\underline{s}\underline{n}-2}{\underline{s}\underline{n}} \right)^4 \frac{\log G}{\underline{n}}M^2 - \frac{16}{\underline{s}\underline{n} - 2} \sqrt{\frac{\log G}{\underline{n}}}M \right] \right\} \\
		&= 3\exp\left\{  \log G - \frac{-\underline{n}\underline{s}^2}{2}\left[ \left( \frac{\underline{s}\underline{n}-2}{\underline{s}\underline{n}} \right)^4 \frac{\log G}{\underline{n}}M^2 - \frac{16}{\underline{s}\underline{n} - 2} \sqrt{\frac{\log G}{\underline{n}}}M \right]  \right\} \\
		&= 3\exp\left\{  \log G\left( 1-  \frac{-\underline{n}\underline{s}^2}{2}\left[ \left( \frac{\underline{s}\underline{n}-2}{\underline{s}\underline{n}} \right)^4 \frac{1}{\underline{n}}M^2 - \frac{16}{\underline{s}\underline{n} - 2} \sqrt{\frac{1}{\underline{n}\log G}}M \right] \right)  \right\} \\
    &= 3 \exp\left\{ \log G\left[ 1 - \frac{\underline{s}^2}{2} \left( \frac{\underline{s}\underline{n}-2}{\underline{s}\underline{n}} \right)^4 M^2 + 8 \underline{s}\left(\frac{\underline{s}\underline{n}}{\underline{s}\underline{n} - 2}\right) \sqrt{\frac{1}{\underline{n}\log G}}M  \right] \right\} .
	\end{align*}
To complete the proof we need to show that for any $\delta > 0$ we can choose $M$, $n^*$ and $G^*$ such that $p(\underline{n},G,M) \leq \delta$ for all $\underline{n} \geq n^*$ and $G \geq G^*$.
Since we are free to choose $n^*$, set $n^* \geq 4/\underline{s}$. 
Then, for any $\underline{n} \geq n^*$ we have $(\underline{s}\underline{n}-2)/(\underline{s}\underline{n}) \geq 1/2$, $\underline{s}\underline{n}/(\underline{s}\underline{n} - 2) \leq 2$, and $\sqrt{\underline{n}}\geq 2$.
Hence,
\[
  p(\underline{n},G,M) \le 3 \exp\left\{ \log G\left( 1- \frac{\underline{s}^2}{32}M^2 + 8\underline{s} \sqrt{\frac{1}{\log G}}M   \right)  \right\}
\]
for any $\underline{n} \geq n^* \geq 4/\underline{s}$.
Since we are free to choose $G^*$, set $G^* \geq 3$ so that $\log G^* > 1$.
Then we have
\[
  p(\underline{n},G,M) \leq
  3 \exp\left\{ \log G\left( 1- \frac{\underline{s}^2}{32}M^2 + 8\underline{s} M   \right)  \right\}
  = 3 \exp\left\{ -\log G\left( \frac{\underline{s}^2}{32}M^2 - 8\underline{s} M - 1 \right)  \right\}
\]
for any $\underline{n} \geq n^* \geq 4/\underline{s}$, $G \geq G^* \geq 3$.
This upper bound takes the form $3\exp\left\{ -\log G \times \kappa(M) \right\}$ where $\kappa(M)$ is a convex quadratic function of $M$ with roots $4(32 \pm \sqrt{1022})/\underline{s}$.
Thus $\kappa(M) > 0$ for any $M > 256/\underline{s}$.
If $\kappa(M) > 0$ we have $\exp\{-\log G \times \kappa(M)\} < \exp\{-\kappa(M)\}$, and hence 
\begin{equation}
  p(\underline{n}, G, M) \leq 3 \exp\left\{ -\left( \frac{\underline{s}^2}{32}M^2 - 8\underline{s} M - 1 \right)   \right\}
  \label{eq:Mbound}
\end{equation}
for any $\underline{n} \geq n^* \geq 4/\underline{s}$, $G \geq G^* \geq 3$, $M > 256/\underline{s}$.
The RHS of \eqref{eq:Mbound} can be made arbitrarily small by choosing a sufficiently large value of $M$. 
Since \eqref{eq:Mbound} holds for all $\underline{n}>n^*$ and $G>G^*$, the result follows.
\end{proof}

\begin{proof}[Proof of \autoref{thm:R}]
  We provide the argument for condition (vii) of \autoref{thm:consistency} and (iii) of \autoref{thm:asymptoticnormality} only. 
  For (vi) from \autoref{thm:consistency}, simply replace $U_{ig}$ with $\boldsymbol{X}_{ig}$ in the following derivations.
By \eqref{eq:ZtoR} and the triangle inequality
\begin{equation}
  \label{eq:triangleIneq}
  \left \Vert \sum_{g=1}^G \frac{1}{N_g}\sum_{i=1}^{N_g} \rho_g ( \widehat{\boldsymbol{\mathcal{Z}}}_{ig} - \boldsymbol{\mathcal{Z}}_{ig} ) U_{ig} \right \Vert 
\leq \Delta_G \left(\sum_{g=1}^G \frac{1}{N_g} \sum_{i=1}^{N_g} \lvert| \rho_g \boldsymbol{W}_{ig}U_{ig} \rvert|\right)
\end{equation}
where we define the shorthand
\[
  \Delta_{G} \equiv \max_{1 \leq g \leq G} \left(\max_{1 \leq i \leq N_g}\left\Vert \mathbf{R}(\widehat{C}_{ig}, N_g)^{+} - \mathbf{R}(\bar{C}_{ig}, N_g)^{-1} \right \Vert \right).
\]
Consider the second factor on the RHS of \eqref{eq:triangleIneq}. 
By an argument similar to that used in the proof of \autoref{thm:consistency},
\[
  \frac{1}{G}\sum_{g=1}^G \left(\frac{1}{N_g} \sum_{i=1}^{N_g} \lvert| \rho_g \boldsymbol{W}_{ig}U_{ig} \rvert|\right) \rightarrow_p \mathbbm{E}\left[ \lvert|\rho \boldsymbol{W}_{ig} U_{ig} \rvert| \right] < \infty
\]
so that $\sum_{g=1}^G \frac{1}{N_g} \sum_{i=1}^{N_g} \lvert| \rho_g \boldsymbol{W}_{ig}U_{ig} \rvert| = O_{\mathbbm{P}}(G)$.
Now, define the event $\widehat{1}_G$ as
\[
  \widehat{1}_G \equiv \mathbbm{1}\left\{\displaystyle \min_{1 \leq g \leq G} \left(\min_{1 \leq i \leq N_g} \widehat{C}_{ig} \right) \geq \frac{\bar{c}_L}{2} \right\}.
\]
By assumption $\mathbf{R}(\bar{C}_{ig},N_{ig})$ is invertible, and conditional on 
$\widehat{C}_{ig} \geq \bar{c}_L/2$ it follows that $\mathbf{R}(\widehat{C}_{ig}, N_g)$ is likewise invertible. 
Hence, if $\widehat{1}_G = 1$ we can write
\begin{align*}
  \left\Vert \mathbf{R}(\widehat{C}_{ig}, N_g)^{-1} - \mathbf{R}(\bar{C}_{ig}, N_g)^{-1} \right \Vert 
  &= \left\Vert \mathbf{R}(\widehat{C}_{ig}, N_g)^{-1}\left[ \mathbf{R}(\widehat{C}_{ig}, N_g) - \mathbf{R}(\bar{C}_{ig}, N_g) \right] \mathbf{R}(\bar{C}_{ig}, N_g)^{-1}\right \Vert\\
  &\leq \left\Vert \mathbf{R}(\widehat{C}_{ig}, N_g)^{-1}\right\Vert \left\Vert \mathbf{R}(\widehat{C}_{ig}, N_g) - \mathbf{R}(\bar{C}_{ig}, N_g) \right\Vert \left\Vert \mathbf{R}(\bar{C}_{ig}, N_g)^{-1}\right \Vert.
\end{align*}
Let $\lvert| \mathbf{M} \rvert|_2$ denote the spectral norm of a matrix $\mathbf{M}$, i.e.\ its largest singular value.
Since $\mathbf{R}(\bar{C}_{ig}, N_g)$ is square, symmetric, and positive definite we have $\lvert| \mathbf{R}(\bar{C}_{ig}, N_g)^{-1} \rvert|_2 \leq 1/ \underline{\sigma} < \infty$.
Similarly, if $\widehat{1}_G = 1$, then $\lvert| \mathbf{R}(\widehat{C}_{ig}, N_g)^{-1} \rvert|_2 \leq 1/\underline{\sigma} < \infty$.
Because all finite-dimensional norms are equivalent, it follows that
\begin{align*}
  \widehat{1}_G \Delta_G
  &\leq  K \max_{1 \leq g \leq G} \left( \max_{1 \leq i \leq N_g}\left\Vert \mathbf{R}(\widehat{C}_{ig}, N_g) - \mathbf{R}(\bar{C}_{ig}, N_g) \right\Vert \right) 
  \leq  K \left\{\max_{1 \leq g \leq G}\left( \max_{1 \leq i \leq N_g}\left|\widehat{C}_{ig} - \bar{C}_{ig}\right| \right) + O(\underline{n}^{-1/2}) \right\} 
\end{align*}
where $0 < K < \infty$ denotes a generic, unspecified constant.
Applying \autoref{lem:Chat} we see that $\widehat{1}_G \Delta_G = O_{\mathbbm{P}}\left( \sqrt{\log G/\underline{n}} \right)$ as $(\underline{n}, G)\rightarrow \infty$.
Thus, by \eqref{eq:triangleIneq},
\begin{equation}
  \label{eq:rateOnehat}
  \widehat{1}_G \left \Vert \sum_{g=1}^G \frac{1}{N_g}\sum_{i=1}^{N_g} \rho_g ( \widehat{\boldsymbol{\mathcal{Z}}}_{ig} - \boldsymbol{\mathcal{Z}}_{ig} ) U_{ig} \right \Vert = O_{\mathbbm{P}}\left(\sqrt{\frac{\log G}{\underline{n}}}\right) O_{\mathbbm{P}}(G). 
\end{equation}
If $\log G / \underline{n} \rightarrow 0$ as $(\underline{n}, G)\rightarrow \infty$, then the rate on the RHS of \eqref{eq:rateOnehat} becomes $o_{\mathbbm{P}}(G)$.
If $G \log G / \underline{n} \rightarrow 0$, it becomes $o_{\mathbbm{P}}(G^{1/2})$.
Finally, since $\bar{c}_L \leq \bar{C}_{ig}$, it follows that
\[
  \mathbbm{P}\left( \widehat{1}_G \neq 1 \right) 
  \leq \mathbbm{P}\left[ \max_{1 \leq g \leq G}\left( \max_{1 \leq i \leq N_g} \left| \widehat{C}_{ig} - \bar{C}_{ig} \right| \geq \frac{\bar{c}_L}{2}\right)  \right] 
\]
Hence, applying \autoref{lem:Chat}, $\log G/\underline{n} \rightarrow 0$ implies $\widehat{1}_G \rightarrow_p 1$.
The result follows.
\end{proof}

\pagebreak
\normalsize
\section{Additional Tables and Figures \label{sec:additional}}

\begin{table}[htbp!]
	\begin{center}
\begin{tabular}{lc} \hline
\vspace{0pt} & \begin{footnotesize}\end{footnotesize} \\
Age & 0.004 \\
\vspace{4pt} & \begin{footnotesize}(0.002)\end{footnotesize} \\
Cohabits & -0.02 \\
\vspace{4pt} & \begin{footnotesize}(0.010)\end{footnotesize} \\
Has at least one child & -0.13 \\
\vspace{4pt} & \begin{footnotesize}(0.032)\end{footnotesize} \\
Youngest child: 12+ months & 0.12 \\
\vspace{4pt} & \begin{footnotesize}(0.027)\end{footnotesize} \\
Education: less than Bac+2 years & -0.03 \\
\vspace{4pt} & \begin{footnotesize}(0.012)\end{footnotesize} \\
Employed at baseline & -0.09 \\
\vspace{4pt} & \begin{footnotesize}(0.019)\end{footnotesize} \\
Not employed at baseline & 0.03 \\
\vspace{4pt} & \begin{footnotesize}(0.015)\end{footnotesize} \\
Permanent contract at baseline & -0.14 \\
\vspace{4pt} & \begin{footnotesize}(0.017)\end{footnotesize} \\
Fixed term contract at baseline & -0.06 \\
\vspace{4pt} & \begin{footnotesize}(0.015)\end{footnotesize} \\
Duration of contract at baseline: 7-12 months & -0.04 \\
\vspace{4pt} & \begin{footnotesize}(0.018)\end{footnotesize} \\
Duration of contract at baseline: 13+ months & -0.12 \\
\vspace{4pt} & \begin{footnotesize}(0.028)\end{footnotesize} \\
Receives unemployment insurance at baseline & 0.04 \\
 & \begin{footnotesize}(0.009)\end{footnotesize} \\
\vspace{0pt} & \begin{footnotesize}\end{footnotesize} \\
Mean compliance & 0.35 \\
Observations & 11,976 \\
 $R^2$ & 0.055 \\ \hline
\end{tabular}
\end{center}

	\vspace{-0.3cm}
	\caption{\textbf{Predictors of compliance: linear probability model.} OLS estimates of compliance indicators on baseline covariates, estimated on the subsample of participants assigned to treatment. Standard errors clustered at the city level. The following variables are included in the regression but are not reported and are not statistically significant: sex; number of children; youngest child 0-4 months, 4-8 months, 8-12 months; unemployment duration at baseline; did not provide employment status at baseline; unemployment duration in the last 18 months; temporary contract at baseline; 1-3 month contract at baseline; 3-6 month contract at baseline; average city unemployment rate. \label{tab:lpm_compliers}}
\end{table}

\begin{figure}[!htbp]
	\vspace{-0.2cm}
	{\centering
		\input{./R/figures/simulations/theta2_e_Pars9.tex}
		\input{./R/figures/simulations/theta2_n_Pars9.tex}
		\\[-0.35cm]
		\input{./R/figures/simulations/gamma_c_Pars9.tex}
		\input{./R/figures/simulations/delta_c_Pars9.tex}
		\label{fig:FrankDGP_ParSet9}
	}
	\vspace{-1cm}
	\caption{Distribution of the estimates of the spillover terms, $(\gamma,\gamma^n,\gamma^c,\delta^c)$, over 5000 simulations for our IV and the `naive' IV (where available) for simulations with 150 groups.}
\end{figure}

\begin{figure}[!htbp]
	\vspace{-0.2cm}
	{\centering
		\input{./R/figures/simulations/theta2_e_Pars11.tex}
		\input{./R/figures/simulations/theta2_n_Pars11.tex}
		\\[-0.35cm]
		\input{./R/figures/simulations/gamma_c_Pars11.tex}
		\input{./R/figures/simulations/delta_c_Pars11.tex}
		\label{fig:FrankDGP_ParSet11}
	}
	\vspace{-1cm}
	\caption{Distribution of the estimates of the spillover terms, $(\gamma,\gamma^n,\gamma^c,\delta^c)$, over 5000 simulations for our IV and the `naive' IV (where available) for simulations with 500 groups.}
\end{figure}
\section{Implementation Details for the Linear Model}
\label{sec:StepByStep}

This appendix provides step-by-step instructions for implementing our estimators from \autoref{sec:estimationinference} in the linear potential outcomes model \eqref{eq:LinearModel}.
For simplicity we assume here that the experiment does not include a zero percent saturation; Appendix \ref{sec:0percent} explains the minor modifications needed to accommodate this case. 
The linear outcome model is given by
\[
  Y_{ig} = \alpha_{ig} + \beta_{ig} D_{ig} + \gamma_{ig} \bar{D}_{ig} + \delta_{ig} D_{ig} \bar{D}_{ig}.
\]
In the notation of \autoref{assump:randcoef}, this corresponds to using the basis functions $\mathbf{f}\left(x \right) = (1, x)'$ with $\boldsymbol{\theta}_{ig} = (\alpha_{ig}, \gamma_{ig})$, and $\boldsymbol{\psi}_{ig} = (\alpha_{ig} + \beta_{ig}, \gamma_{ig} + \delta_{ig})$.
We identify direct and indirect effects for compliers ($D_{ig} = C_{ig} = 1$), aka ``the treated,'' along with indirect effects  for never-takers ($D_{ig} = C_{ig} = 0$), aka ``the untreated,'' and the population as a whole.
Specializing \autoref{thm:expect} to the linear model, our estimands are:
\begin{align*}
  \text{Direct, Compliers:} \quad & \mathbbm{E}[\boldsymbol{\psi}_{ig} - \boldsymbol{\theta}_{ig}|C_{ig} = 1] = \mathbbm{E}\left(\left.\begin{bmatrix} \beta_{ig} \\ \delta_{ig} \end{bmatrix}\right| D_{ig} = 1\right) \equiv \begin{bmatrix}
      \beta^c \\ \delta^c
    \end{bmatrix}\\ 
    \text{Indirect, Compliers:} \quad & \mathbbm{E}[\boldsymbol{\psi}_{ig}|C_{ig} = 1] = \mathbbm{E}\left(\left.\begin{bmatrix} \alpha_{ig} + \beta_{ig}\\ \gamma_{ig} + \delta_{ig} \end{bmatrix}\right| D_{ig} = 1\right) \equiv \begin{bmatrix}
      \alpha^c + \beta^c \\ \gamma^c + \delta^c
    \end{bmatrix}\\ 
    \text{Indirect, Never-takers} \quad & \mathbbm{E}[\boldsymbol{\theta}_{ig}|C_{ig} = 0] = \mathbbm{E}\left(\left.\begin{bmatrix} \alpha_{ig} \\ \gamma_{ig} \end{bmatrix}\right| D_{ig} = 0\right) \equiv \begin{bmatrix}
      \alpha^n \\ \gamma^n
    \end{bmatrix}\\ 
    \text{Indirect, Population:} \quad & \mathbbm{E}[\boldsymbol{\theta}_{ig}] = \mathbbm{E}\begin{bmatrix} \alpha_{ig} \\ \gamma_{ig} \end{bmatrix} \equiv \begin{bmatrix}
      \alpha \\ \gamma
    \end{bmatrix}.
\end{align*}


\paragraph{Step 1:} Let $V_g \equiv (1 - S_g)$. Using knowledge of the experimental design, calculate: 
\[
  \overline{s} \equiv \mathbbm{E}[S_g], \,
  \overline{v} \equiv \mathbbm{E}[V_g], \,
  \overline{sv} \equiv \mathbbm{E}[S_g V_g], \,
  \overline{s^2 v} \equiv \mathbbm{E}[S_g^2 V_g], \,
  \overline{sv^2} \equiv \mathbbm{E}[S_gV_g^2], \,
  \overline{s^2} \equiv \mathbbm{E}[S_g^2], \,
  \overline{s^3} \equiv \mathbbm{E}[S_g^3].
\]

\paragraph{Step 2:} Calculate $\widehat{C}_{ig} \equiv \bar{D}_{ig}/\bar{Z}_{ig}$ where $\bar{D}_{ig} \equiv \sum_{j\neq i} D_{jg}/(N_g - 1)$, $\bar{Z}_{ig} \equiv \sum_{j\neq i} Z_{jg}/(N_g - 1)$.

\paragraph{Step 3:} Construct the matrices $\widehat{\mathbf{Q}}_{0,ig}$ and $\widehat{\mathbf{Q}}_{1,ig}$ as follows
\[
  \widehat{\mathbf{Q}}_{0,ig} = 
  \begin{bmatrix}
    \overline{v} & \overline{sv}\, \widehat{C}_{ig}  \\
    \overline{sv}\,\widehat{C}_{ig}  &
  \overline{s^2v}\,\widehat{C}_{ig}^2  + \overline{sv^2}\,\frac{\widehat{C}_{ig}}{N_g - 1}   \end{bmatrix}, \quad 
  \widehat{\mathbf{Q}}_{1,ig} = 
  \begin{bmatrix}
    \overline{s}  & \widehat{C}_{ig} \, \overline{s^2} \\
    \overline{s^2} \, \widehat{C}_{ig}  &
    \overline{s^3} \, \widehat{C}_{ig}^2   + \overline{s^2v}\, \frac{\widehat{C}_{ig}}{N_g - 1}
  \end{bmatrix}.
\]

\paragraph{Step 4:} To estimate $(\alpha, \gamma)$ along with $(\beta^c, \delta^c)$, run a just-identified instrumental variables regression of $Y_{ig}$ on a constant, $D_{ig}$, $\bar{D}_{ig}$, and $D_{ig} \bar{D}_{ig}$ with instruments $\widehat{\boldsymbol{\mathcal{Z}}}_{ig}$ given by  
\[
  \widehat{\boldsymbol{\mathcal{Z}}}_{ig} \equiv
  \begin{bmatrix}
    \widehat{\mathbf{Q}}_{0,ig}^{-1} & -\widehat{\mathbf{Q}}_{0,ig}^{-1}\\ \\
    -\widehat{\mathbf{Q}}_{0,ig}^{-1} & \widehat{\mathbf{Q}}_{0,ig}^{-1} + \widehat{\mathbf{Q}}_{1,ig}^{-1}
  \end{bmatrix}
  \left(\begin{bmatrix}
    1 \\ Z_{ig}
  \end{bmatrix} \otimes
  \begin{bmatrix}
    1 \\ \bar{D}_{ig} 
\end{bmatrix}\right).
\]
The coefficients on the intercept and $\bar{D}_{ig}$ are $(\widehat{\alpha}, \widehat{\gamma})$; those on $D_{ig}$ and $D_{ig} \bar{D}_{ig}$ are $(\widehat{\beta}_1, \widehat{\delta}_1)$.

\paragraph{Step 5:} To estimate $(\alpha^c + \beta^c, \gamma^c + \delta^c)$, run a just-identified instrumental variables regression of $Y_{ig}$ on an intercept and $\bar{D}_{ig}$ with instruments $\widehat{\boldsymbol{\mathcal{Z}}}_{ig}$ given by 
\[
  \widehat{\boldsymbol{\mathcal{Z}}}_{ig} \equiv D_{ig}\widehat{\mathbf{Q}}_{1,g}^{-1} \begin{bmatrix}
    1 \\ \bar{D}_{ig}
  \end{bmatrix}.
\]

\paragraph{Step 6:} To estimate $(\alpha^n, \gamma^n)$, run a just-identified instrumental variables regression of $Y_{ig}$ on an intercept and $\bar{D}_{ig}$ with instruments $\widehat{\boldsymbol{\mathcal{Z}}}_{ig}$ given by
\[
  \widehat{\boldsymbol{\mathcal{Z}}}_{ig} \equiv Z_{ig}(1 - D_{ig})\widehat{\mathbf{Q}}_{1,g}^{-1} \begin{bmatrix}
    1 \\ \bar{D}_{ig}
  \end{bmatrix}.
\]

\paragraph{Inference:} Inference to accompany the estimates from Steps 4--6 is straightforward: simply report the standard errors provided by your preferred IV package, clustering by group if desired. To carry out inference for $(\alpha^c, \gamma^c)$, proceed as follows: First estimate these parameters by subtracting the estimates of $(\beta^c, \delta^c)$ constructed in Step 4 from those of $(\alpha^c + \beta^c, \gamma^c + \delta^c)$ constructed in Step 5. Save the residuals from Steps 4 and 5 and use them to construct the joint variance-covariance matrix of $(\alpha^c, \gamma^c)$ and $(\alpha^c + \beta^c, \gamma^c + \delta^c)$, call it $\Sigma$. The desired standard errors are the square roots of the diagonal elements of $A\Sigma A'$ where $A$ is a matrix whose elements encode the linear combination that corresponds to subtracting the Step 4 estimates from the Step 5 estimates.

\section{Experiments with a 0\% Saturation}
\label{sec:0percent}
Some randomized saturation designs, including the experiment of \cite{crepon2013}, include a zero percent saturation, also known as a ``pure control'' condition.
Under one-sided non-compliance $S_g = 0$ implies $Z_{ig} = D_{ig} = \bar{D}_{ig} = 0$ for all $1 \leq i \leq N_g$.
Accordingly, we cannot estimate the share of compliers $\widehat{C}_{ig}$ from \eqref{eq:Chat} for groups assigned a saturation of zero.
The easiest solution to this problem is simply to drop observations for any zero saturation groups.
Under Assumptions \ref{assump:saturations}--\ref{assump:Bernoulli} and \ref{assump:exclusion} this has no effect on our identification or large-sample results provided that we replace $\mathbf{Q}, \mathbf{Q}_0$ and $\mathbf{Q}_1$ with expectations that condition on $S_g > 0$, namely 
\begin{align*}
  \widetilde{\mathbf{Q}}(\bar{c}, n) &\equiv \mathbbm{E}\left[ \mathbf{W}_{ig}\mathbf{W}_{ig}'|\bar{C}_{ig} = \bar{c}, N_g = n, S_g > 0 \right]\\
  \widetilde{\mathbf{Q}}_0(\bar{c}, n) &\equiv \mathbbm{E}\left[ (1 - Z_{ig}) \mathbf{f}(\bar{D}_{ig})\mathbf{f}(\bar{D}_{ig})'|\bar{C}_{ig} = \bar{c}, N_g = n, S_g > 0 \right]\\
  \widetilde{\mathbf{Q}}_1(\bar{c}, n) &\equiv \mathbbm{E}\left[ Z_{ig} \mathbf{f}(\bar{D}_{ig})\mathbf{f}(\bar{D}_{ig})'|\bar{C}_{ig} = \bar{c}, N_g = n, S_g > 0 \right]
\end{align*}
Zero percent saturation groups, however, \emph{are} informative: they pin down the value of $\mathbbm{E}[Y_{ig}(0, 0)]$ and hence can be used to improve estimates of $\mathbbm{E}\left[ \boldsymbol{\theta}_{ig} \right]$. 
To exploit this information, we replace the instrument vectors from parts (i) and (iv) of \autoref{thm:expect} with
\[
  \widetilde{\boldsymbol{\mathcal{Z}}}_{ig}^W \equiv \begin{bmatrix}
  \mathbbm{1}\left\{ S_g > 0 \right\} \widetilde{\mathbf{Q}}(\bar{C}_{ig}, N_g)^{-1}\mathbf{W}_{ig}\\
  \mathbbm{1}\left\{ S_g = 0 \right\}
  \end{bmatrix}
  \text{,} \ \ \ \ \
  \widetilde{\boldsymbol{\mathcal{Z}}}^0_{ig} \equiv \begin{bmatrix}
    \mathbbm{1}\left\{ S_g > 0 \right\} \widetilde{\mathbf{Q}}_0(\bar{C}_{ig}, N_g)^{-1} \mathbf{f}(\bar{D}_{ig})\\
    \mathbbm{1}\left\{ S_g = 0 \right\}
  \end{bmatrix}  
\]
Calculations similar to those in the proof of \autoref{thm:expect} establish that these are valid and relevant instruments.
Because the dimensions of $\widetilde{\boldsymbol{\mathcal{Z}}}_{ig}^W$ and $\widetilde{\boldsymbol{\mathcal{Z}}}^0_{ig}$ exceed those of the parameters for which they instrument by one, they provide over-identifying information. 
As such, the just-identified IV moment condition from parts (i) and (iv) of \autoref{thm:expect} must be replaced with a linear GMM moment equation.
Subject to this small change, estimation and inference can proceed almost exactly as in \autoref{sec:estimationinference}: we merely substitute $\widehat{C}_{ig}$ for $\bar{C}_{ig}$ in $\widetilde{\mathbf{Q}}$ and $\widetilde{\mathbf{Q}}_0$ to yield a feasible GMM estimator, e.g.\ two-stage least squares. 
With minor notational modifications, our large-sample results continue to apply.

\section{Extending the Definition of \texorpdfstring{$\mathbf{Q}$}{Q}}
\label{sec:Qextend}

Technically, the conditional expectations in \eqref{eq:Qdef}--\eqref{eq:Q1def} are only well-defined when $n \bar{c}$ is a positive integer, whereas \autoref{assump:R} requires the functions $\mathbf{Q}, \mathbf{Q}_0$, and $\mathbf{Q}_1$ to be defined over a continuous range of values for $\bar{c}$.
This problem is easily solved by \emph{extending} the definitions of $\mathbf{Q}_0$ and $\mathbf{Q}_1$.
In many cases, the natural extension will be obvious.
In the linear potential outcomes model, for example, \eqref{eq:Q0LinearBernoulli} and \eqref{eq:Q1LinearBernoulli} agree with \eqref{eq:Q0def} and \eqref{eq:Q1def} when these conditional expectations are well-defined and satisfy all the conditions of \autoref{assump:R} 

More generally, we can always \emph{construct} extended definitions of $\mathbf{Q}_0$ and $\mathbf{Q}_1$ to satisfy these regularity conditions.
Here we provide a construction based on \emph{linear interpolation}.
To begin, let 
\[
  \bar{c}_{\ell}(\bar{c}, n) \equiv \frac{\lfloor (n-1) \bar{c} \rfloor}{n-1}, \quad
  \bar{c}_{u}(\bar{c}, n) \equiv \frac{\lceil (n-1) \bar{c} \rceil}{n-1}.
\]
By construction, $(n-1) \bar{c}_u(\bar{c},n)$ and $(n-1)\bar{c}_{\ell}(\bar{c}, n)$ are non-negative integers.
Now let
\begin{align*}
  \mathbf{Q}_z^{\ell}(\bar{c}, n) &\equiv \mathbbm{E}\left[ \left.\mathbbm{1}(Z_{ig} = z) \mathbf{f}(\bar{D}_{ig})\mathbf{f}(\bar{D}_{ig})' \right| \bar{C}_{ig} = \bar{c}_{\ell}(\bar{c}, n), N_g = n\right]\\
  \mathbf{Q}_z^{u}(\bar{c}, n) &\equiv \mathbbm{E}\left[ \left.\mathbbm{1}(Z_{ig} = z) \mathbf{f}(\bar{D}_{ig})\mathbf{f}(\bar{D}_{ig})' \right| \bar{C}_{ig} = \bar{c}_{u}(\bar{c}, n), N_g = n\right]
\end{align*}
for $z = 0,1$.
Notice that $\mathbf{Q}_0^{\ell}, \mathbf{Q}_1^{\ell}$ and $\mathbf{Q}_0^{u}, \mathbf{Q}_{1}^u$ are well-defined regardless of whether $(n-1) \bar{c}$ is an integer.
From these ingredients, we construct generalizations $\mathbf{Q}_0^*$ and $\mathbf{Q}_1^*$ of $\mathbf{Q}_0, \mathbf{Q}_1$ as
\[
  \mathbf{Q}_z^*(\bar{c}, n) = \left[ 1 - \omega(\bar{c}, n)  \right] \mathbf{Q}_z^{\ell}(\bar{c}, n) + \omega(\bar{c}, n) \mathbf{Q}_z^u (\bar{c}, n); \quad
  \omega(\bar{c}, n) \equiv \frac{\bar{c} - \bar{c}_{\ell}(\bar{c}, n)}{\bar{c}_u(\bar{c}, n) - \bar{c}_{\ell}(\bar{c}, n)} \in [0,1]
\]
for $z = 0,1$.
Since both $\mathbf{Q}_z^{\ell}$ and $\mathbf{Q}_z^{u}$ are symmetric and positive definite, their convex combination $\mathbf{Q}_z^*$ is as well.
To show that this construction satisfies \autoref{assump:R} (iii), define
\begin{equation}
  \mathbf{Q}_{0}^{\infty}(\bar{c}) \equiv  \mathbb{E}\left[(1-S_g)\mathbf{f}(\bar{c}S_g)\mathbf{f}(\bar{c} S_g)'\right], \quad
  \mathbf{Q}_{1}^{\infty}(\bar{c}) \equiv \mathbb{E}\left[S_g\mathbf{f}(\bar{c}S_g)\mathbf{f}(\bar{c}S_g)'\right].
  \label{eq:Q0_Q1_infty}
\end{equation}
Recall that $0 \leq S_g \leq 1$ a discrete random variable with finite support, $\bar{c}$ is a real number between zero and one, and $\mathbf{f}$ is a $K$-vector of Lipschitz-continuous functions, all of which are bounded on $[0,1]$.
It follows that both $\mathbf{Q}_0^{\infty}$ and $\mathbf{Q}_1^{\infty}$ are bounded and Lipschitz-continuous on $[0,1]$.
Accordingly, by \autoref{lem:DbarDist}, Jensen's inequality, and the triangle inequality we can show that
\[
  \left \Vert \mathbf{Q}_z^{\ell}(\bar{c}, n) - \mathbf{Q}_z^{\infty}\left(\bar{c}_{\ell}(\bar{c}, n)\right) \right \Vert \leq \frac{L}{\sqrt{n - 1}}, \quad
  \left \Vert \mathbf{Q}_z^{u}(\bar{c}, n) - \mathbf{Q}_z^{\infty}\left(\bar{c}_{u}(\bar{c}, n)\right) \right \Vert  \leq \frac{L}{\sqrt{n - 1}}
\]
where $L$ denotes an arbitrary, finite, positive constant.  
Similarly,
\[
  \left \Vert \mathbf{Q}_z^{\infty}(\bar{c}) - \mathbf{Q}_z^{\infty}\left(\bar{c}_{\ell}(\bar{c}, n)\right) \right \Vert \leq \frac{L}{n - 1}, \quad
  \left \Vert \mathbf{Q}_z^{\infty}(\bar{c}) - \mathbf{Q}_z^{\infty}\left(\bar{c}_{u}(\bar{c}, n)\right) \right \Vert \leq \frac{L}{n - 1}.
\]
Combining these inequalities an applying the triangle inequality, it follows that
\[
  \left \Vert \mathbf{Q}_z^{u}(\bar{c}, n) - \mathbf{Q}_z^{\ell}(\bar{c},n) \right \Vert \leq \frac{L}{\sqrt{n - 1}}, \quad
  \left \Vert \mathbf{Q}_z^{u}(\bar{c}, n) - \mathbf{Q}_z^{\infty}(\bar{c}) \right \Vert  \leq \frac{L}{\sqrt{n - 1}}
\]
and as a consequence
\[
\left \Vert \mathbf{Q}_z^{\ell}(\bar{c}, n) - \mathbf{Q}_z^{\infty}(\bar{c}) \right \Vert \leq \frac{L}{\sqrt{n - 1}}
\]
where, again, $L$ is an arbitrary, finite, positive constant.
Thus,
\begin{align}
\begin{split}
  \left \Vert \mathbf{Q}_z^{*}(\bar{c}, n) - \mathbf{Q}_{z}^{\infty}(\bar{c}) \right \Vert &\leq \left \Vert \mathbf{Q}_z^{*}(\bar{c}, n) - \mathbf{Q}_{z}^{\ell}(\bar{c}, n) \right \Vert + \left \Vert \mathbf{Q}_z^{\ell}(\bar{c}, n) - \mathbf{Q}_{z}^{\infty}(\bar{c}) \right \Vert \\ 
  &\leq \left \Vert \mathbf{Q}_z^{*}(\bar{c}, n) - \mathbf{Q}_{z}^{\ell}(\bar{c}, n) \right \Vert + \frac{L}{\sqrt{n - 1}}\\
  &= \omega(\bar{c}, n) \left \Vert \mathbf{Q}_z^{u}(\bar{c}, n) - \mathbf{Q}_{z}^{\ell}(\bar{c}, n) \right \Vert + \frac{L}{\sqrt{n - 1}}\\
  &\leq \frac{L}{\sqrt{n-1}}
\end{split}
\label{eq:starVsInfty}
\end{align}
using the definitions of $\mathbf{Q}_z^{*}$ and $\omega(\bar{c}, n)$ from above.
Combining the preceding inequalities, 
\[
  \left \Vert \mathbf{Q}_z^*(\widehat{C}_{ig}, N_g) - \mathbf{Q}_z^*(\bar{C}_{ig}, N_g) \right \Vert \leq L \left\{ \frac{1}{\sqrt{\underline{n} - 1}} + \left| \widehat{C}_{ig} - \bar{C}_{ig} \right| \right\}
\]
since $\underline{n} \leq N_g$ and $\mathbf{Q}_z^{\infty}$ is Lipschitz-continuous.

\section{Include Fewer Basis Functions than Saturations}
\label{sec:BasisFunctionsRank}

Assumption \ref{assump:rank} requires $\mathbf{Q}_z(\bar{c},n)$ to be full rank. 
This condition is crucial for point identifying the coefficients of interest in \autoref{thm:identification}. 
In this section we show that, for large group sizes, the number of saturations in the experimental design constrains the rank of $\mathbf{Q}_z(\bar{c}, n)$.
To ensure point identification in the large-group limit, researchers should not include more basis functions than there are saturations in the design.
The following discussion relies on notation and results from \autoref{sec:Qextend} above, in particular the ``extended'' definition of $\mathbf{Q}_z(\bar{c}, n)$, namely $\mathbf{Q}_z^*(\bar{c},n)$, and its large-group limit $\mathbf{Q}_z^{\infty}(\bar{c})$.

At any $(\bar{c},n)$ where $\mathbf{Q}_z(\bar{c},n)$ is well-defined, $\mathbf{Q}_z(\bar{c}, n) = \mathbf{Q}_z^*(\bar{c},n)$.  
And by \eqref{eq:starVsInfty}, $\mathbf{Q}_z^*(\bar{c}, n)$ is arbitrarily close to $\mathbf{Q}_{z}^{\infty}(\bar{c})$ for large $n$.
For this reason, we begin by considering the rank of $\mathbf{Q}_z^{\infty}(\bar{c})$.
If the number of saturations $|\mathcal{S}|$ in the experimental design is finite, then
\begin{align*}
	\mathbf{Q}_z^{\infty}(\bar{c}) &=  \sum_{s \in \mathcal{S}} s^z(1-s)^{1-z} \mathbf{f}(\bar{c}s)\mathbf{f}(\bar{c}s)' \mathbb{P}(S_g = s).
\end{align*}
Because the right-hand side of this expression is a sum of $|\mathcal{S}|$ rank one matrices, the rank of $\mathbf{Q}_z^{\infty}(\bar{c})$ cannot exceed $|\mathcal{S}|$. 
It follows that $\mathbf{Q}_{z}^{\infty}(\bar{c})$ will be rank deficient when $|\mathcal{S}|$ is less than $K$, the number of basis functions and dimension of $\mathbf{f}$.\footnote{It will also be rank deficient when $|\mathcal{S}| \geq K$ if the basis functions are linearly dependent.}
If $\mathbf{Q}_z^{\infty}$ is rank deficient, it must have at least one eigenvalue equal to zero.
And because the eigenvalues of a matrix are a continuous function of its entries \citep[Theorem D.2]{horn2013matrix}, it follows from \eqref{eq:starVsInfty} that at least one eigenvalue of $\mathbf{Q}^*(\bar{c},n)$ can be made arbitrarily close to zero by increasing $n$.
Hence, to ensure point identification in the large-group limit, researchers should include fewer basis functions that the experimental design has saturations.
All else equal, experiments with more saturations can identify more flexible outcome models.

\section{Testable Implications of IOR}
\label{sec:IOR}

Under one-sided non-compliance and IOR, 
Assumptions \ref{assump:onesided}--\ref{assump:IOR}, $D_{ig} = C_{ig} Z_{ig}$ where $C_{ig}$ is the indicator that person $(i,g)$ is a complier.
Under IOR, $C_{ig}$ can be treated as an unobserved individual characteristic that is \emph{predetermined} at the time of randomization. 
Thus, $S_g$ and $Z_{ig}$ are jointly independent of $C_{ig}$ under the randomized saturation design.
If we assume that the randomization was carried out faithfully, this provides a testable implication of IOR:
\[
  \mathbbm{E}[D_{ig}|Z_{ig} = 1, S_g] = \mathbbm{E}[C_{ig}Z_{ig}|Z_{ig} = 1, S_g] = \mathbbm{E}[C_{ig}|Z_{ig} = 1, S_g] = \mathbbm{E}[C_{ig}].
\]
Thus, if $\mathbbm{E}[D_{ig}|Z_{ig}, S_g = s]$ varies with $s$, we must either conclude that IOR fails, that the saturations were not in fact randomly assigned, or both.
This observation yields a simple regression-based test of IOR.
Suppose that the experimental design features $J$ saturations $\{s_1, s_2, \dots, s_J\}$, \emph{excluding} the zero percent saturation, if present.
For the subset of individuals with $Z_{ig} = 1$, run the regression 
\[
  D_{ig} = \alpha + \sum_{j = 1}^{J-1} \beta_j \mathbbm{1}\{S_g = s_j\} + \varepsilon_{ig}, \quad
  (i,g) \text{ such that } Z_{ig} = 1.
\]
The coefficient $\alpha$ equals the take-up rate among offered individuals in groups with $S_g = s_J$. 
The coefficients $\beta_j$ equal the \emph{difference in take-up rates} for offered individuals in groups with $S_g = s_j$ relative to those with $S_g = s_J$.
Under IOR, the hypothesis $H_0\colon \beta_1 = \beta_2 = \cdots = \beta_{J-1} = 0$ must hold.
Before applying the methods developed in this paper, we recommend that applied researchers test this linear restriction, ideally using a cluster-robust variance matrix. 
If IOR is \emph{a priori} reasonable in their application and the test does not reject, they can proceed with relative confidence to apply our estimators.

\begin{figure}[htbp]
	\centering
\begin{tikzpicture}[x=1pt,y=1pt]
\definecolor{fillColor}{RGB}{255,255,255}
\path[use as bounding box,fill=fillColor,fill opacity=0.00] (0,0) rectangle (289.08,289.08);
\begin{scope}
\path[clip] ( 61.20, 49.20) rectangle (275.88,245.88);
\definecolor{drawColor}{RGB}{190,190,190}

\path[draw=drawColor,line width= 0.4pt,dash pattern=on 4pt off 4pt ,line join=round,line cap=round] ( 61.20,119.87) -- (275.88,119.87);
\definecolor{fillColor}{RGB}{0,0,0}

\path[fill=fillColor] ( 69.15,132.48) circle (  3.15);

\path[fill=fillColor] (135.41,120.62) circle (  3.15);

\path[fill=fillColor] (201.67,117.14) circle (  3.15);

\path[fill=fillColor] (267.93,120.62) circle (  3.15);
\end{scope}
\begin{scope}
\path[clip] (  0.00,  0.00) rectangle (289.08,289.08);
\definecolor{drawColor}{RGB}{0,0,0}

\path[draw=drawColor,line width= 0.4pt,line join=round,line cap=round] ( 69.15, 49.20) -- (267.93, 49.20);

\path[draw=drawColor,line width= 0.4pt,line join=round,line cap=round] ( 69.15, 49.20) -- ( 69.15, 43.20);

\path[draw=drawColor,line width= 0.4pt,line join=round,line cap=round] (135.41, 49.20) -- (135.41, 43.20);

\path[draw=drawColor,line width= 0.4pt,line join=round,line cap=round] (201.67, 49.20) -- (201.67, 43.20);

\path[draw=drawColor,line width= 0.4pt,line join=round,line cap=round] (267.93, 49.20) -- (267.93, 43.20);

\node[text=drawColor,anchor=base,inner sep=0pt, outer sep=0pt, scale=  0.90] at ( 69.15, 27.60) {0.25};

\node[text=drawColor,anchor=base,inner sep=0pt, outer sep=0pt, scale=  0.90] at (135.41, 27.60) {0.50};

\node[text=drawColor,anchor=base,inner sep=0pt, outer sep=0pt, scale=  0.90] at (201.67, 27.60) {0.75};

\node[text=drawColor,anchor=base,inner sep=0pt, outer sep=0pt, scale=  0.90] at (267.93, 27.60) {1.00};

\path[draw=drawColor,line width= 0.4pt,line join=round,line cap=round] ( 61.20, 56.48) -- ( 61.20,238.60);

\path[draw=drawColor,line width= 0.4pt,line join=round,line cap=round] ( 61.20, 56.48) -- ( 55.20, 56.48);

\path[draw=drawColor,line width= 0.4pt,line join=round,line cap=round] ( 61.20, 92.91) -- ( 55.20, 92.91);

\path[draw=drawColor,line width= 0.4pt,line join=round,line cap=round] ( 61.20,129.33) -- ( 55.20,129.33);

\path[draw=drawColor,line width= 0.4pt,line join=round,line cap=round] ( 61.20,165.75) -- ( 55.20,165.75);

\path[draw=drawColor,line width= 0.4pt,line join=round,line cap=round] ( 61.20,202.17) -- ( 55.20,202.17);

\path[draw=drawColor,line width= 0.4pt,line join=round,line cap=round] ( 61.20,238.60) -- ( 55.20,238.60);

\node[text=drawColor,rotate= 90.00,anchor=base,inner sep=0pt, outer sep=0pt, scale=  0.90] at ( 46.80, 56.48) {0.0};

\node[text=drawColor,rotate= 90.00,anchor=base,inner sep=0pt, outer sep=0pt, scale=  0.90] at ( 46.80, 92.91) {0.2};

\node[text=drawColor,rotate= 90.00,anchor=base,inner sep=0pt, outer sep=0pt, scale=  0.90] at ( 46.80,129.33) {0.4};

\node[text=drawColor,rotate= 90.00,anchor=base,inner sep=0pt, outer sep=0pt, scale=  0.90] at ( 46.80,165.75) {0.6};

\node[text=drawColor,rotate= 90.00,anchor=base,inner sep=0pt, outer sep=0pt, scale=  0.90] at ( 46.80,202.17) {0.8};

\node[text=drawColor,rotate= 90.00,anchor=base,inner sep=0pt, outer sep=0pt, scale=  0.90] at ( 46.80,238.60) {1.0};

\path[draw=drawColor,line width= 0.4pt,line join=round,line cap=round] ( 61.20, 49.20) --
	(275.88, 49.20) --
	(275.88,245.88) --
	( 61.20,245.88) --
	( 61.20, 49.20);
\end{scope}
\begin{scope}
\path[clip] ( 61.20, 49.20) rectangle (275.88,245.88);
\definecolor{drawColor}{RGB}{0,0,0}

\path[draw=drawColor,line width= 0.4pt,line join=round,line cap=round] ( 69.15,106.29) -- ( 69.15,158.67);

\path[draw=drawColor,line width= 0.4pt,line join=round,line cap=round] ( 65.54,106.29) --
	( 69.15,106.29) --
	( 72.76,106.29);

\path[draw=drawColor,line width= 0.4pt,line join=round,line cap=round] ( 72.76,158.67) --
	( 69.15,158.67) --
	( 65.54,158.67);

\path[draw=drawColor,line width= 0.4pt,line join=round,line cap=round] (135.41,112.64) -- (135.41,128.60);

\path[draw=drawColor,line width= 0.4pt,line join=round,line cap=round] (131.80,112.64) --
	(135.41,112.64) --
	(139.02,112.64);

\path[draw=drawColor,line width= 0.4pt,line join=round,line cap=round] (139.02,128.60) --
	(135.41,128.60) --
	(131.80,128.60);

\path[draw=drawColor,line width= 0.4pt,line join=round,line cap=round] (201.67,108.76) -- (201.67,125.52);

\path[draw=drawColor,line width= 0.4pt,line join=round,line cap=round] (198.06,108.76) --
	(201.67,108.76) --
	(205.28,108.76);

\path[draw=drawColor,line width= 0.4pt,line join=round,line cap=round] (205.28,125.52) --
	(201.67,125.52) --
	(198.06,125.52);

\path[draw=drawColor,line width= 0.4pt,line join=round,line cap=round] (267.93,115.28) -- (267.93,125.97);

\path[draw=drawColor,line width= 0.4pt,line join=round,line cap=round] (264.32,115.28) --
	(267.93,115.28) --
	(271.54,115.28);

\path[draw=drawColor,line width= 0.4pt,line join=round,line cap=round] (271.54,125.97) --
	(267.93,125.97) --
	(264.32,125.97);
\end{scope}
\begin{scope}
\path[clip] (  0.00,  0.00) rectangle (289.08,289.08);
\definecolor{drawColor}{RGB}{0,0,0}

\node[text=drawColor,anchor=base,inner sep=0pt, outer sep=0pt, scale=  1.00] at (168.54,  3.60) {Saturation};

\node[text=drawColor,rotate= 90.00,anchor=base,inner sep=0pt, outer sep=0pt, scale=  1.00] at ( 22.80,147.54) {Estimated average share of compliers};
\end{scope}
\begin{scope}
\path[clip] (  0.00,  0.00) rectangle (289.08,289.08);
\definecolor{drawColor}{RGB}{0,0,0}

\path[draw=drawColor,line width= 0.4pt,line join=round,line cap=round] ( 61.20, 49.20) --
	(275.88, 49.20) --
	(275.88,245.88) --
	( 61.20,245.88) --
	( 61.20, 49.20);
\end{scope}
\end{tikzpicture}
	\caption{Regression-based test of IOR. The estimated share of compliers is given by the dot and its 95\% confidence interval is given by the bars for each of our four saturation bins. The horizontal dotted line gives the estimated share of compliers across the whole sample.}
	\label{fig:IORtest}
\end{figure}
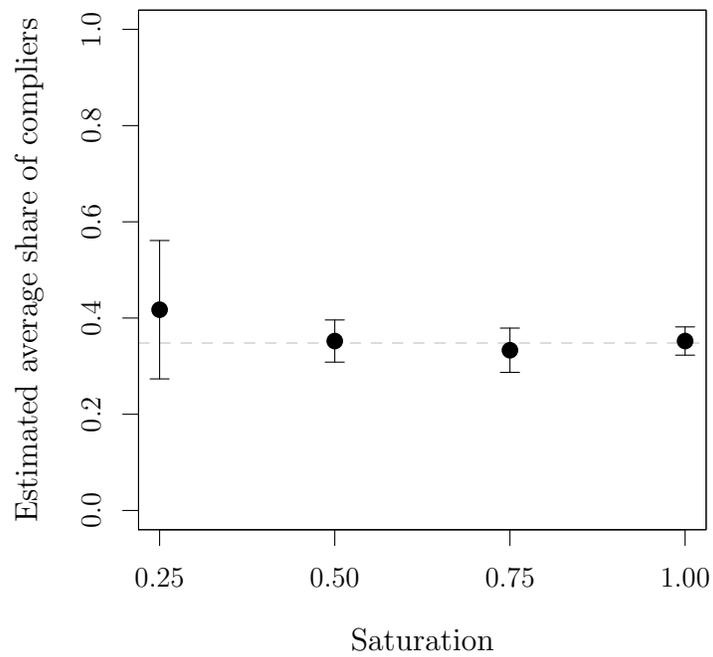

Applying this test to the data from 
\cite{crepon2013} that we use in our empirical example gives a p-value of $0.62$, clustering by city.
This bolsters our confidence that IOR is a reasonable assumption in this application.
An alternative way of viewing this procedure is as a test of whether the share of compliers is constant across saturations.
This follows because, under IOR, the take-up rate among offered individuals is an unbiased estimate of the share of compliers.
\autoref{fig:IORtest} depicts this alternative interpretation of the test.
We find no evidence against IOR in our empirical example.

\section{Extension to Completely Randomized Designs}
\label{sec:CompletelyRandomized}

Our main identification result, \autoref{thm:expect} does not require \autoref{assump:Bernoulli}. 
It only requires Assumptions \ref{assump:randcoef}--\ref{assump:IOR} and \ref{assump:rank} along with $(Z_{ig}, \bar{D}_{ig}) \indep (\mathbf{B}_{ig}, C_{ig})|(\bar{C}_{ig}, N_g)$.
In the body of the paper, we establish this conditional independence relationship by appealing to \autoref{thm:conditional_indep} which \emph{does} require \autoref{assump:Bernoulli}.
In this appendix, we provide an alternative proof of \autoref{thm:conditional_indep} that applies in a \emph{completely randomized} experimental design, in which the number of treatment offers made to a given group is fixed conditional on the realization of $S_g$.
In this case \autoref{assump:Bernoulli} is replaced by the following condition.

\begin{assump}[Completely Randomized Design]
  \label{assump:randomized}
  \[
  \mathbb{P}(\boldsymbol{Z}_{g} = \boldsymbol{z}| N_g = n, S_g = s) =
      \left\{\begin{array}{ll}
        \displaystyle{\binom{n}{\lfloor ns\rfloor}}^{-1}, &\text{ if } \sum_{i} z_{i} = \lfloor ns \rfloor\\ 
    0, & \text{otherwise}\end{array}\right.
\]
where $\lfloor x \rfloor$ denotes the greatest integer less than $x$.
\end{assump}

Under a Bernoulli Design, treatment offers within a group are iid Bernoulli draws: the saturation determines only the probability of making an offer, not the fraction of offers made.
Under a completely randomized design, on the other hand, the number of treatment offers is fixed at $\lfloor ns\rfloor$ given the assigned saturation $s$ and group size $n$.
Offers are still made at random--each individual has the same probability of treatment--but are no longer independent: if Alice is offered treatment, this makes it less likely that Bob will be.
When \autoref{assump:randomized} replaces \autoref{assump:Bernoulli}, \autoref{lem:DbarDist} is replaced by the following result.

\begin{lem}
  \label{lem:DbarDistRandomized}
  Let $\bar{c}$ be a value in $[0,1]$ such that $(n-1)\bar{c}$ is a non-negative integer.  
  Under Assumptions \ref{assump:saturations}, \ref{assump:onesided}--\ref{assump:exclusion} and \ref{assump:randomized}, and conditional on $(N_g = n, S_g = s, \boldsymbol{C}_g = \boldsymbol{c}, \bar{C}_{ig} = \bar{c}, Z_{ig} = z)$, $(n-1)\bar{D}_{ig}$ follows a $\text{Hypergeometric}\left(n-1,\, (n-1)\bar{c},\, \lfloor ns \rfloor - z \right)$ distribution.
\end{lem}

\small
\begin{proof}[Proof of \autoref{lem:DbarDistRandomized}]

Applying \autoref{cor:example} and the  Decomposition property to \autoref{assump:exclusion}(ii) yields $\boldsymbol{Z}_g \indep (\boldsymbol{C}_g, \bar{C}_{ig})|(N_g,S_g)$.
  By the definition of conditional independence, it follows that the distribution of $\boldsymbol{Z}_g|(N_g, S_g, \boldsymbol{C}_g,\bar{C}_{ig})$ is the same as that of $\boldsymbol{Z}_g|(N_g, S_g)$:
\begin{equation}
  \label{eq:design_indep2_append}
  \mathbbm{P}(\boldsymbol{Z}_g=\boldsymbol{z}|N_g = n, S_g = s, \boldsymbol{C}_g, \bar{C}_{ig}) = \mathbbm{P}(\boldsymbol{Z}_g = \boldsymbol{z}|N_g = n, S_g = s).
\end{equation}
Now, define the shorthand $A \equiv \left\{ N_g = n, S_g = s, \boldsymbol{C}_g = \boldsymbol{c}, \bar{C}_{ig} = \bar{c}\right\}$ and let $\mathcal{C}(i)$ be the indices of all non-zero components of $\boldsymbol{c}$, \emph{excluding} the $i$th component, i.e.\ $\mathcal{C}(i) \equiv \left\{ j\neq i\colon c_j = 1 \right\}$.
By the definition of $\bar{D}_{ig}$, the event $\left\{ \bar{D}_{ig} = d  \right\}$ is equivalent to $\left\{ \sum_{j\neq i} C_{jg}Z_{jg} = d (N_g-1) \right\}$.
Consequently, 
\[
  \mathbbm{P}(\bar{D}_{ig} = d|A, Z_{ig}) = \mathbbm{P}\left(\left.\left[\sum_{j\neq i} C_{jg} Z_{jg}\right] = d(n - 1)\right|A,Z_{ig}\right) = \mathbbm{P}\left(\left. \left[\sum_{j\in \mathcal{C}(i)} Z_{jg}\right] = d(n-1) \right|A,Z_{ig}\right)
\]
where the first equality uses the fact that $A$ implies $N_g = n$, and the second uses the fact that $A$ implies $\boldsymbol{C}_g = \boldsymbol{c}$, so we know precisely which of the indicators $C_{jg}$ equal zero and which equal one. 

It remains to calculate the probability that $\sum_{j\in \mathcal{C}(i)}  Z_{jg} = d(n-1)$ given $A$ and $Z_{ig} = z$, under \autoref{assump:randomized}.
By the definition of $\mathcal{C}(i)$ this is simply the probability that exactly $d(n-1)$ of the $(n-1)\bar{c}$ compliers (excluding person $i$) are offered treatment, conditional on $A$ and the treatment offer $z$ made to person $i$.
Now, under \autoref{assump:randomized}, we see that \eqref{eq:design_indep2_append} implies 
\[
  \mathbb{P}(\boldsymbol{Z}_{g} = \boldsymbol{z}| A) =
      \left\{\begin{array}{ll}
        \displaystyle{\binom{n}{\lfloor ns\rfloor}}^{-1}, &\text{ if } \sum_{i} z_{i} = \lfloor sn \rfloor\\ 
    0, & \text{otherwise.}\end{array}\right.
\]
Hence, conditional on $A$, the allocation of treatment offers is equivalent to drawing $\lfloor ns \rfloor$ balls without replacement from an urn containing $n$ balls in total.
Conditioning on $Z_{ig}$ is equivalent to removing one ball in advance, leaving only $n-1$ in the urn.
Of the remaining balls $(n-1)\bar{c}$ are red, corresponding to the compliers, and $(n-1)(1 - \bar{c})$ are white, corresponding to the never-takers.
This follows from our definition of $\bar{C}_{ig}$, which \emph{excludes} person $(i,g)$. 
Conditional on $A$ and $Z_{ig} = z$, the sum $\sum_{j \in \mathcal{C}(i)} Z_{jg}$ is simply the number of red balls that we draw from the urn.
If $z = 0$, then person $(i,g)$ was not offered treatment so we make $\lfloor ns \rfloor$ draws from the urn; if $z = 1$, then person $(i,g)$ was offered treatment, so we make only $\lfloor ns \rfloor - 1$ draws from the urn.
Hence, conditional on $(A,Z_{ig}=z)$, the sum $\sum_{j \in \mathcal{C}(i)} Z_{jg}$ is a Hypergeometric$(N,K,r)$ random variable with $N = n-1$, $K = (n-1)\bar{c}$, and $r = \lfloor ns \rfloor - z$ draws.
In other words,
\[
  \mathbbm{P}(\bar{D}_{ig} = d|N_g = n, S_g = s, \mathbf{C}_g = \mathbf{c}, \bar{C}_{ig} = \bar{c}, Z_{ig} = z) = \frac{\displaystyle{\binom{(n-1)\bar{c}}{(n-1)d}}{\binom{(n-1)(1-\bar{c})}{\lfloor ns \rfloor -z-(n-1)d}}}{\displaystyle {\binom{n-1}{\lfloor ns \rfloor - z}}}.
\]
Because the right hand side of this expression does not depend on $\mathbf{c}$, we have shown that $\bar{D}_{ig}$ is conditionally independent of $\mathbf{C}_g$ given $(N_g,\bar{C}_{ig}, S_g, Z_{ig})$, as required.
\end{proof}

\normalsize

Having established \autoref{lem:DbarDistRandomized}, we now show how to adapt the proof of \autoref{thm:conditional_indep} so that it applies under \autoref{assump:randomized}.
Inspection of the proof of \autoref{thm:conditional_indep} reveals that \autoref{lem:DbarDist} is used only once: to establish \eqref{eq:DbarCindep}, namely
\[
  \bar{D}_{ig} \indep \boldsymbol{C}_{-ig} | (N_g, \bar{C}_{ig}, S_g, Z_{ig}).
\]
But this conditional independence relation \emph{also follows} immediately from \autoref{lem:DbarDistRandomized}.
Therefore, \autoref{thm:conditional_indep} still holds when \autoref{assump:Bernoulli} is replaced by \autoref{assump:randomized}, and hence our main identification result, \autoref{thm:expect} still holds when treatment offers are made according to a completely randomized design.
Note that under this design the matrices $\mathbf{Q}(\bar{c},n)$, $\mathbf{Q}_0(\bar{c},n)$ and $\mathbf{Q}_1(\bar{c},n)$ from \autoref{sec:identification} should be computed using \autoref{lem:DbarDistRandomized} rather than \autoref{lem:DbarDist}.

\section{More Potential Applications of Our Methods}
\label{sec:MoreApplications}

Below we describe five recent empirical studies that appear to satisfy the conditions required to apply our methods.
Each of these papers uses a randomized saturation design and features some degree of non-compliance.
For each paper, we describe the topic and headline result, along with the study population and the definition of the groups/clusters. 
We then discuss the extent to which the paper satisfies the conditions required to use our methods: (1) one-sided non-compliance, (2) many large groups, (3) anonymous interactions, and (4) IOR.  

\paragraph{\cite{abebe2021anonymity}}

This experiment offered a job application workshop and transport subsidy to job seekers in Addis Ababa, Ethiopia. 
The authors show that both the workshop and subsidy improve labour market outcomes: the probability of having a formal job. 
They use a randomized saturation design for the transport intervention to estimate spillover effects, which are described in Section A.3 of the Online Appendix to the paper.
Geographic clusters are drawn from the list of Ethiopian Central Agency enumeration areas, which typically consist of 150-200 housing units (see footnote 24) and ``rarely exceed 300m in diameter'' (see appendix A.3). 
This study features one-sided non-compliance because only those offered the transport subsidy can use it. 
Seventy-four clusters are offered the transport treatment (18 at 20\%, 15 at 40\%, 15 at 75\%, 26 at 90\%) and the overall sample size is 1274. 
The authors of this paper assume anonymous interactions in their analysis.
IOR appears plausible given the size of the clusters and the way in which treatment offers were made to individuals.

\paragraph{\cite{baird2011}}

This experiment randomly offered unconditional (UCTs) or conditional cash transfers (CCTs) to schoolgirls in Malawi. 
The authors show that CCTs reduce school drop-out and grades by more than UCTs, but UCTs reduce marriage and fertility among school dropouts. 
While they do not make use of this variation in the published paper, the experiment uses a randomized saturation design in each treatment arm, with saturations of 0\%, 33\%, 66\%, and 100\%
The sample contains 176 enumeration areas (88 control, 44 UCT, 44 CCT). 
An enumeration area (EA) consists of approximately 250 households (see footnote 10). 
If compliance is defined as as actually receiving the CCT, this is one-sided. 
Anonymous interactions seem plausible in this setting, as there are 250 households per EA and that a large share of the sample is urban or peri-urban (29 are urban, 119 are within 16km of Zomba city, 28 are rural).
It is unclear whether IOR holds in this setting but, as described above, this can be tested.

\paragraph{\cite{banerjee2012}}

This paper tests 4 interventions in 162 police stations in Rajasthan, India. 
Two of these interventions (police training and a freeze on transfers of police staff) improved police effectiveness and public satisfaction. 
Other interventions had no effect, possibly because of a lack of robust implementation.  
Police training was assigned using a randomized saturation design with saturations of 0\%, 25\%, 50\%, 75\%, or 100\%.
There were 162 police stations in the sample, out of 711 in the state of Rajasthan, with an average of 100 officers per station (70,767 officers in 711 stations).
Non-compliance is one-sided since only officers selected for training can receive it (88\% of those offered training accepted.)
In their analysis, the authors assume that interactions are anonymous. 
It seems plausible that the outcomes they study (measures of police effectiveness and public satisfaction) only depend on the share of police treated, not their identity. 
IOR is somewhat dubious in this application but, again, can be tested.

\paragraph{\cite{bursztyn2021}}

This paper randomly incentivized Hong Kong university students to join an anti-authoritarian protest. 
The authors found that ``incentives to attend one protest within a political movement increase subsequent protest attendance, but only when a sufficient fraction of an individual’s social network is also incentivized to attend the initial protest.''
The experiment contains 97 major-cohort cells within a university and the proportion treated in each cell was assigned via a randomized saturation design with saturations of 0\%, 10\%, 50\%, and 75\% and a total of 849 individuals in the sample.
Non-compliance in this setting is two-sided, since unincentivized students can still attend the protest. 
As we outline in our conclusion, extending the methods from this paper to the two-sided non-compliance setting should be relatively straightforward.
The authors implicitly assume anonymous interactions by estimating their spillover regression using percent assigned to treatment.
IOR appears to hold in this experiment, based on the results presented in the paper (the authors find no affect of saturation on year-1 protest attendance).

\paragraph{\cite{callen2019headwaters}}

In this experiment ``a Sri Lankan bank used mobile Point-of-Service (POS) terminals to collect deposits directly from households each week.''
The key result is that when offered a formal savings option, households work more to save more. 
The authors are concerned that formal savings may crowd out informal savings, so they randomize the intensity of treatment within existing informal savings groups (``seetus'' or ROSCAs).
They find that, if anything, formal savings are a complement to informal savings. 
The experiment features 84 informal savings groups. 
Of these 45 are controls, 13 have 20\% treated, 13 have 40\% treated, and 13 have 60\% treated.
The average size of a seetu is approximately 49; the final sample contains 829 individuals.
Non-compliance is one-sided: only people offered the formal savings accounts could use them. 
The authors assume anonymous interactions when they estimate their spillovers regression using percent assigned to treatment. 
It seems likely that IOR holds in this experiment: the treatment offer is individual access to a formal savings technology.

\end{document}